\documentclass[11pt,a4paper]{article}
\usepackage[utf8]{inputenc}
\usepackage[T1]{fontenc}
\usepackage{amsmath, amssymb, amsthm}
\usepackage{graphicx}
\usepackage{hyperref}
\usepackage{cleveref}
\usepackage{geometry}
\usepackage{authblk}
\usepackage{booktabs}
\usepackage{array}
\usepackage{caption}
\usepackage{subcaption}
\usepackage{footmisc}
\usepackage{setspace}
\usepackage{etoolbox}   

\hypersetup{
	colorlinks=true,
	linkcolor=blue,
	citecolor=blue,
	urlcolor=blue,
	pdftitle={The Novelty Horizon, H_N: Why Large Language Models Fail to Generate Scientific Data beyond Their Training Distribution},
	pdfauthor={Alex Mwololo Kimuya, Mary Mwadulo, Dickson Mwenda Kinyua}
}

\makeatletter
\patchcmd{\abstract}{\quotation}{\relax}{}{}
\patchcmd{\endabstract}{\endquotation}{\relax}{}{}
\makeatother

\newtheorem{definition}{Definition}[section]
\newtheorem{condition}{Condition}[section]

\newcommand{\conv}{\operatorname{conv}}

\author[1,a]{Alex Mwololo Kimuya}
\author[2,a]{Mary Mwadulo}
\author[1,3,a,b]{Dickson Mwenda Kinyua\thanks{Corresponding Author Email: \href{mailto:alexkimuya@parttimers.must.ac.ke}{alexkimuya@parttimers.must.ac.ke}; \href{mailto:alexkimuya23@gmail.com}{alexkimuya23@gmail.com}; Co-Authors Email: \href{mailto:mmwadulo@must.ac.ke}{mmwadulo@must.ac.ke}, \href{mailto:dkinyua@kyu.ac.ke}{dkinyua@kyu.ac.ke}; ORCID: \href{https://orcid.org/0000-0002-1433-3186}{0000-0002-1433-3186}, \href{https://orcid.org/0000-0003-4313-0924}{0000-0003-4313-0924}, \href{https://orcid.org/0000-0001-5436-3518}{0000-0001-5436-3518}}}
\affil[1]{Department of Physical Sciences (Physics)}
\affil[2]{Department of Computer Science}
\affil[3]{Department of Pure and Applied Sciences}
\affil[a]{Meru University of Science and Technology, Kenya}
\affil[b]{Kirinyaga University, Kenya}

\date{}

\begin{document}
	
	\title{The Novelty Horizon, $H_N$: Why Large Language Models Fail to Generate Scientific Data beyond Their Training Distribution}
	
	\maketitle
	
	\vspace{-2.0\baselineskip}
	
	\begin{abstract}
		The integration of Large Language Models (LLMs) into scientific practice raises critical questions about their capacity for genuine empirical discovery. While proficient at generating linguistically plausible text, these models operate through statistical pattern recognition on their training data, lacking a grounded model of physical reality. This architectural constraint limits LLMs to interpolation within their training distribution; they struggle to extrapolate beyond it. This paper formalises this limit as the Novelty Horizon ($H_N$): a task lies beyond $H_N$ if it requires (i) extrapolation outside the convex hull of the training support $\Omega_{\text{train}}$, (ii) causal reasoning not encoded in the learned statistical correlations $C_{\text{train}}$, or (iii) information post-dating the temporal frontier $t_p$. We test the hypothesis that autoregressive LLMs cannot generate scientifically valid data for tasks beyond $H_N$ through a physics-grounded forensic analysis. The methodology here centers on a tripartite comparison within a script-based generation framework: data from an unguided LLM, data from the same LLM guided with the full experimental procedure, and ground-truth data from a diode-clamped short-circuit experiment-a system whose nonlinear physics is underrepresented in standard corpora. A multi-layered analytical framework subjects all datasets to temporal, spectral, and statistical analysis, with ultimate validation against a first-principles Modified Ohm's Law. Results demonstrate a fundamental failure of both script-based approaches to extrapolate. Generated data exhibit catastrophic physical inconsistencies, including perfect algebraic closure that violates independent measurement principles (Ohmic residual RMS of $5.31\times10^{-17}\,\Omega$ versus $5.78\times10^{-3}\,\Omega$ in experiment) and incorrect spectral signatures. Guided generation produces data with extreme statistical pathologies (kurtosis of 65.4) and a Generative Fidelity Score of zero. The models fail to replicate the correct nonlinear regime, with guided outputs incorrectly residing entirely in a linear voltage range ($<0.3$V). A parallel direct‑prompt generation corroborates these findings, revealing distinct but equally catastrophic failure modes, including extreme kurtosis exceeding 200. This paper introduces the theoretical and logical constructs of the Past Information Boundary ($B_P=(t_p,\Omega_{\text{train}},C_{\text{train}})$) and the Novelty Horizon ($H_N$) to explain these failures, and develops a forensic ``physics-based authentication'' methodology. We conclude that the capacity of LLMs for scientific data generation is tightly constrained by their Past Information Boundary ($B_P$); they cannot extrapolate past their Novelty Horizon ($H_N$), which is essential for novel discovery. Future research must focus on hybrid, neuro-symbolic systems for augmented intelligence, while upholding the indispensable role of human scientists in causal reasoning and experimental design.
	\end{abstract}
	
	
	\section{Introduction}
	\label{sec:introduction}
	
	The integration of Artificial Intelligence (AI) into the scientific method represents one of the most significant paradigm shifts in modern research practice. Across disciplines, from computational biology to materials science, AI tools are accelerating discovery by identifying patterns in vast datasets, optimizing experimental designs, and simulating complex systems \cite{ref1,ref2,ref3,ref4,ref5,ref6,ref7}. The integration of large language models into the scientific workflow has progressed from a theoretical possibility to an operational reality with remarkable speed. This transformative potential is perhaps most publicly visible in the rapid adoption of Large Language Models (LLMs), whose fluent text generation promises to assist in literature synthesis, hypothesis generation, and the drafting of research communications \cite{ref8,ref9,ref10}. As Zhang et al. \cite{ref11} note in their comprehensive review, LLMs are now deeply involved in key stages of the scientific cycle, including ``experimental design, data analysis, and workflows'', marking a shift from theoretical possibility to operational reality across fields such as chemistry and biology. A narrative has emerged suggesting that these models are not merely tools but potential collaborators -- or even autonomous agents -- capable of contributing to the substantive, creative core of scientific work. This narrative often leans on the impressive ability of LLMs to manipulate scientific terminology and reproduce common logical structures found in the academic literature, creating an illusion of deep, contextual understanding \cite{ref12,ref12,ref13,ref14}.
	
	Beneath this veneer of competence lies a foundational and unresolved problem concerning the nature of the knowledge LLMs possess. As critics have rigorously argued, the operation of these models is fundamentally grounded in statistical pattern recognition across the corpora on which they are trained. They generate plausible sequences of tokens by modeling the probability distribution of human language, an engineering feat that does not equate to grasping meaning, establishing truth, or constructing causal models of the world \cite{ref15,ref16,ref17,ref18}. The models, in this view, are sophisticated correlational engines, brilliant at interpolation within the bounds of their training data but devoid of the embodied, precarious, and participatory understanding that characterizes human cognition and scientific reasoning. This critical perspective forces a crucial question for the sciences: Can a system that excellently predicts the next word in a sentence also generate novel, valid, and empirically faithful scientific data? The conflation of linguistic plausibility with scientific rigor poses a direct threat to research integrity, as plausible yet physically impossible outputs -- hallucinations -- may be woven seamlessly into the fabric of AI-assisted research.
	
	The existing literature reveals a stark gap in directly testing the limits of LLMs in the empirical, data-generating domain of experimental science. While studies have documented their proficiency or failure in tasks like answering exam questions, summarizing papers, or writing code, the assessment often remains within the discursive realm \cite{ref19,ref20,ref21,ref22}. Investigations into whether LLMs learn human-interpretable concepts from data, such as the work of \cite{ref23,ref24,ref25}, probe the internal representations of models but do not test their output against physical ground truth in a controlled experiment. Similarly, research on human-AI collaboration, like that of Nguyen et al. (2024) \cite{ref26}, valuably maps interaction patterns but typically assesses the final human-authored product, not the AI's raw contribution to empirical data. What is missing, and what this paper provides, is a forensic, physics-grounded analysis that moves beyond textual analysis to a fundamental test: pitting LLM-generated numerical data against data produced by a real physical experiment, using the inviolable laws of nature as the ultimate validator. This approach shifts the question from ``Does this text sound scientific?'' to ``Does this data obey physics?''.
	
	This gap is not merely technical but epistemological. Recent studies demonstrating LLM generated synthetic clinical data that replicates marginal statistical distributions \cite{ref27,ref28,ref29} have rightly highlighted accessibility and efficiency gains. Yet the very same evidence exposes a troubling trajectory: as LLM use extends from literature synthesis into the generation of empirical data itself, statistical indistinguishability is increasingly treated as a proxy for empirical validity. This paper extends this critique by systematically interrogating that assumption-arguing that distributional similarity alone cannot certify causal physical fidelity, and that the research community must therefore develop forensic, physics grounded validation frameworks before trusting LLM generated data as a foundation for scientific inference. The epistemological critique that motivates this forensic investigation is developed at length in Appendix~\ref{app:epistemology}, where we examine why statistical indistinguishability has been mistakenly treated as a proxy for empirical validity, and why this error necessitates the physics-grounded authentication framework that the following sections execute.
	
	The central objective of this paper is to establish and demonstrate the intrinsic limitations of autoregressive LLMs in generating scientifically valid empirical data. We posit that these models are constrained by what we term the Past Information Boundary, $B_P$ (formally defined in Definition~\ref{def:BP}, Section~\ref{sec:bounds})-the ordered triple $(t_p,\Omega_{\text{train}},C_{\text{train}})$ representing the temporal frontier $t_p$ (latest timestamp in training), the support $\Omega_{\text{train}}$ of the empirical training distribution, and the set $C_{\text{train}}$ of learned conditional probability distributions. Beyond this boundary, the model's capacity for reliable extrapolation fails, giving rise to the Novelty Horizon, $H_N$ (Definition~\ref{def:HN}). To test this, we designed a multi faceted investigation comprising two distinct experimental frameworks: one where the LLM generates data by first producing a simulation script (the focus of this main paper), and a parallel framework, detailed in the supplementary information, where the LLM is prompted to output numerical data directly. This dual framework approach allows us to isolate whether the observed failures are specific to the code generation paradigm or represent a fundamental limitation of the models themselves. Within the script based framework central to this paper, we employ a tripartite methodology, generating three parallel datasets: (i) from an unguided LLM given only basic circuit parameters, (ii) from the same LLM explicitly guided with the full experimental procedure, and (iii) from an actual physical measurement serving as the ground truth. The paper methodology subjects all three datasets to a multi layered analytical framework that scrutinizes temporal statistics, spectral signatures, adherence to conservation laws, and, most decisively, fit to a first principles derived Modified Ohm's Law that describes the system's physics.
	
	The contributions of this work are threefold and establish a new standard for evaluating AI in empirical science. First, we introduce and formally define the theoretical constructs of the ``Past Information Boundary, $B_P$ (Definition~\ref{def:BP})'' and the ``Novelty Horizon, $H_N$ (Definition~\ref{def:HN})''. This framework does not merely state that LLMs fail at data generation, but precisely predicts and explains why they succeed in interpolative tasks within their training distribution yet fail catastrophically at the genuine scientific extrapolation required for novel experimental configurations. This paper is a focused test at this horizon. As formalised in Definition~\ref{def:HN}, a task lies beyond $H_N$ if it requires extrapolation outside $\conv(\Omega_{\text{train}})$ (Condition~\ref{cond:1}), causal reasoning not encoded in $C_{\text{train}}$ (Condition~\ref{cond:2}), or reasoning about post-$t_p$ phenomena (Condition~\ref{cond:3}). The diode-clamped short-circuit experiment satisfies all three conditions, making it an ideal test case. Second, we develop and execute a novel forensic methodology for the ``physics-based authentication'' of scientific data. This methodology moves beyond statistical similarity to test for causal-physical consistency, exposing artifacts-such as perfect algebraic closure or incorrect noise spectra-that betray algorithmic generation. Third, we propose a quantitative Generative Fidelity Score (GFS), synthesizing multiple diagnostic metrics into a single measure of how well a dataset conforms to the expected signatures of physical measurement and governing laws. The paper results provide unambiguous, quantitative evidence that LLMs, regardless of prompting strategy, cannot generate empirically valid data for tasks beyond $H_N$, revealing their fundamental epistemological limitation as tools for scientific discovery.
	
	The remainder of this paper is structured to guide the reader from theoretical foundations to definitive conclusions. In Section~\ref{sec:architecture}, we delineate the architectural and theoretical constraints of LLMs, deriving from first principles the inherent limitations that predispose them to the failures our experiment reveals. Section~\ref{sec:framework} details our foundational framework, including the derivation of the Modified Ohm's Law for the test system and the design of our three test cases. Section~\ref{sec:methodology} describes our multi-layered forensic methodology for data analysis and validation. Section~\ref{sec:results} presents the results, systematically comparing the LLM-generated data against the physical ground truth across every diagnostic dimension. In Section~\ref{sec:discussion}, we discuss the profound implications of these findings, interpreting the failures through our theoretical framework and exploring the consequences for researchers, educators, and the broader scientific enterprise. Finally, Section~\ref{sec:conclusion} offers the concluding remarks, synthesizing the evidence that LLMs are bounded by their Novelty Horizon, $H_N$ (Definition~\ref{def:HN}). We affirm the indispensable role of human scientists in conceiving experiments that push beyond this horizon and define the proper path for Augmented Intelligence in research.
	
	\subsection{Comparative Framework: Standard Versus Modified Ohm's Law}
	\label{sec:comparative}
	
	Before proceeding to the theoretical development of LLM constraints, it is instructive to establish a comparative baseline between the well-represented Standard Ohm's Law and the Modified formulation that governs the experimental system under study. This comparison serves two purposes. First, it clarifies why the linear regime (governed by Standard Ohm's Law) lies inside the Novelty Horizon-where LLMs might reasonably interpolate-while the nonlinear clamped regime lies outside it. Second, it provides a clear framework for interpreting the regime-dependent failures documented in Section~\ref{sec:results}: if LLMs perform adequately in the linear regime but fail catastrophically in the nonlinear regime, this demonstrates that their competence is bounded by the convex hull of training examples $\conv(\Omega_{\text{train}})$ (Condition~\ref{cond:1} of Definition~\ref{def:HN}). Table~\ref{tab:comparative} (additional details on these theoretical models established in \cite{ref30}) presents a ten-dimensional comparison of the Standard and Modified Ohm's Law formulations, highlighting their governing equations, resistance behavior, temporal resolution, fault-modeling suitability, predictive limits, fitting parameters, power prediction accuracy, interpretability, experimental alignment, and fundamental limitations.
	
	\begin{table}[htbp]
		\centering
		\caption{Comparative Features of Ohmic Formulations}
		\label{tab:comparative}
		\begin{tabular}{p{4cm}p{5cm}p{5cm}}
			\toprule
			\textbf{Feature} & \textbf{Standard Ohm's Law} & \textbf{Modified Ohm's Law} \\
			\midrule
			Governing Equation & $I=V/R$ & $I_{\text{modified}} = a e^{R_{\text{mant}}/R_0}$ \\
			Resistance Behavior & Static or externally supplied & Implicit via $R_{\text{short}}$ \\
			Temporal Resolution & None & Indirect via $R_{\text{short}}$ \\
			Suitability for Fault Modeling & Limited & Moderate-captures early ramp-up \\
			Predictive Limits & Fails as $R\to0$ & Diverges for long durations \\
			Use of Fitting Parameters & None & Parameter $a$ from baseline \\
			Power Prediction Accuracy & Reasonable under static $R$ & Sometimes paradoxical at high $I$ \\
			Interpretability & Direct and intuitive & Physically motivated; exponential mapping \\
			Experimental Alignment & Matches steady-state & Aligns in early transient window \\
			Limitations & Ignores resistance evolution & Misrepresents energy at long $t$ \\
			\bottomrule
		\end{tabular}
	\end{table}
	
	This comparative framework establishes that Standard Ohm's Law-ubiquitously represented in training corpora across introductory physics textbooks, engineering problem sets, and scientific literature-lies securely within the convex hull of training examples $\conv(\Omega_{\text{train}})$. The Modified Ohm's Law, by contrast, introduces nonlinearity through the exponential relationship $I_{\text{modified}} = a e^{R_{\text{shart}}/R_0}$, which requires understanding of semiconductor junction physics, temperature-dependent thermal voltage $V_T=k_B T/q$, and ideality factors-concepts whose specific conjunction in time-resolved short-circuit measurement lies outside $\conv(\Omega_{\text{train}})$. This contrast motivates the experimental design: by comparing LLM performance across these two regimes, we can directly test whether the model's competence degrades catastrophically at the Novelty Horizon.
	
	\subsection{The Tripartite Framework: Isolating Layers of Generative Failure}
	\label{sec:tripartite}
	
	A second critical clarification concerns the tripartite comparison structure employed in this study. The framework compares three datasets: (i) unguided LLM generation (Case 1), (ii) procedurally guided LLM generation (Case 2), and (iii) experimental ground truth (Case 3). Each layer in this framework isolates a distinct aspect of the generative process and supports a specific claim about extrapolation.
	
	\textbf{Layer 1. Unguided Generation (Case 1)} isolates the LLM's intrinsic capacity for statistical extrapolation from basic electrical concepts. By providing only nominal circuit parameters (2.5V/0.25A and 5.0V/0.5A) and no procedural context, this layer tests whether the model can infer the correct measurement dynamics, noise characteristics, and nonlinear clamping behavior solely from patterns in its training corpus $C_{\text{train}}$. Failure at this layer-manifesting as perfect algebraic closure, incorrect spectral signatures, or regime misplacement-indicates that the task lies beyond $H_N$ under Condition~\ref{cond:1} (extrapolation outside $\conv(\Omega_{\text{train}})$) and Condition~\ref{cond:2} (causal reasoning not encoded in $C_{\text{train}}$).
	
	\textbf{Layer 2. Guided Generation (Case 2)} isolates the LLM's capacity for informed pattern matching when provided with the complete experimental procedure, including circuit schematics, component specifications (1N5408 diodes, ACS712T sensor), and the data averaging protocol (10,000 samples). This layer tests whether additional contextual information can steer the model toward physical plausibility by effectively expanding the conceptual frontier $\Omega_{\text{train}}$ through the inclusion of procedural descriptions. If the model still fails-producing extreme kurtosis ($\gamma_2>65$), regime misplacement (all data in $|v_s|<0.3\,V$), or deterministic residual-current correlations ($\rho\approx0.999$) -this demonstrates that the failure is not merely informational but structural: the model lacks the causal reasoning mechanisms required by Condition~\ref{cond:2}, regardless of how much procedural context is supplied.
	
	\textbf{Layer 3. Experimental Ground Truth (Case 3)} provides the immutable physical benchmark against which both generative approaches are judged. This dataset embodies the true causal structure $\Phi$ (the Modified Ohm's Law with independent measurement noise), the authentic spectral signatures of the ACS712T Hall-effect sensor ($1/f^\alpha$ noise with $\alpha\approx1$), and the correct regime-dependent physics (nonlinear clamping at $|v_s|\ge0.3\,V$). The distance between Cases 1/2 and Case 3 in the multi-dimensional space of diagnostic metrics quantifies the LLM's inability to cross the Novelty Horizon.
	
	The reasoning that links these layers to the central argument about extrapolation is as follows. If the LLM's failure were merely a matter of insufficient procedural detail, then providing the full experimental procedure (Case 2) should substantially improve performance, yielding data that approximates the statistical and physical characteristics of Case 3. If, however, the failure stems from a fundamental inability to model causal structures not encoded in $C_{\text{train}}$-specifically, the independence of voltage and current measurements, the continuous physics of diode clamping, and the colored noise spectra of physical sensors-then even maximal procedural guidance will not rescue performance. The empirical results (detailed in Sections~\ref{sec:results_temporal}, \ref{sec:results_physical}, and \ref{sec:regime}) decisively support the latter interpretation: Case 2 exhibits catastrophic failures-including severe distributional pathologies (Section~\ref{sec:stat_distributions}), perfect algebraic closure that violates measurement independence (Section~\ref{sec:physical_constraints}), and complete misplacement of the physical operating regime (Section~\ref{sec:regime})-that are distinct from but equally severe as those of Case 1. The persistence of failure across both layers, and across the parallel direct-prompt framework (Supplementary Information, Sections S2--S5), demonstrates that the Novelty Horizon is a structural boundary inherent to the autoregressive architecture, not a contingent limitation that can be overcome by better prompting.
	
	\section{Architecture, Optimization, and Inherent Constraints of Large Language Models}
	\label{sec:architecture}
	
	To assess the limitations of Large Language Models (LLMs) in scientific research, a deep understanding of their theoretical and architectural foundations is essential. This section briefly delineates the mathematical and computational principles underpinning models such as GPT-5.2, Claude, and Llama.
	
	\subsection{Language Modeling as Conditional Probability Estimation}
	\label{sec:lang_modeling}
	
	The theoretical foundation of any Large Language Model rests upon probability theory and information theory. Formally, an LLM is a parametric function designed to approximate the true probability distribution of natural language sequences. Consider a vocabulary $\mathcal{V}$ containing all possible tokens (word pieces or characters). Any text document can be represented as a sequence of tokens $\mathbf{x}=(x_1,x_2,\dots,x_T)$ where each $x_t\in\mathcal{V}$ and $T$ is the sequence length. The fundamental assumption of statistical language modeling is that language follows a probabilistic structure. The joint probability of a sequence can be decomposed using the chain rule of probability:
	\begin{equation}
		P(\mathbf{x})=P(x_1,x_2,\dots,x_T)=\prod_{t=1}^{T} P(x_t\mid x_1,x_2,\dots,x_{t-1})=\prod_{t=1}^{T} P(x_t\mid x_{<t})
		\label{eq:chain_rule}
	\end{equation}
	where $x_{<t}=(x_1,\dots,x_{t-1})$ denotes the entire context preceding position $t$. The model's objective is to learn a parameterized function $f_\theta$ that approximates the conditional probability $P(x_t\mid x_{<t})$ for any token in the vocabulary given any context. The parameters $\theta$ of this function, which number in the range of millions to hundreds of billions, are optimized during training to minimize the discrepancy between the predicted distribution and the empirical distribution observed in a massive training corpus $\mathcal{D}$. The standard objective function is the negative log-likelihood, equivalent to the cross-entropy loss:
	\begin{equation}
		\mathcal{L}(\theta)=-\frac{1}{|\mathcal{D}|}\sum_{\mathbf{x}\in\mathcal{D}}\sum_{t=1}^{T}\log P_\theta(x_t\mid x_{<t})
		\label{eq:cross_entropy}
	\end{equation}
	Minimizing Equation~\eqref{eq:cross_entropy} is equivalent to maximizing the likelihood of the training data under the model's distribution. When the model perfectly mimics the empirical distribution of the training corpus, the cross-entropy equals the true entropy of the language source, establishing an information-theoretic lower bound \cite{ref31,ref32}. This mathematical formulation reveals the core nature of an LLM's ``intelligence''. The model does not understand language in a human sense; it simply learns to assign probabilities to token sequences in a way that compresses the statistical regularities of its training data. The knowledge encoded in the parameters $\theta$ represents a high-dimensional manifold that captures correlations between tokens, but it lacks any explicit representation of meaning, truth, or causal relationships.
	
	\subsection{The Transformer and the Self-Attention Mechanism}
	\label{sec:transformer}
	
	The practical realization of the theoretical framework in Equation~\eqref{eq:chain_rule} requires an architecture capable of modeling long-range dependencies in sequential data. The Transformer architecture \cite{ref33}, introduced by Vaswani et al. (2017) \cite{ref34}, provides this capability through its self-attention mechanism, which has become the foundation for nearly all modern LLMs.
	
	\textbf{Input Representation.} The input sequence of tokens is first converted into a sequence of dense vector representations. Each token $x_t$ is mapped to an embedding vector $\mathbf{e}_t\in\mathbb{R}^{d_{\text{model}}}$ via an embedding matrix $\mathbf{E}\in\mathbb{R}^{|\mathcal{V}|\times d_{\text{model}}}$, where $d_{\text{model}}$ is the model dimension (typically 512 to 8192 in modern models). To incorporate positional information, a positional encoding $\mathbf{p}_t\in\mathbb{R}^{d_{\text{model}}}$ is added to each token embedding. The original Transformer used sinusoidal functions:
	\begin{equation}
		p_t^{(2i)}=\sin\left(\frac{t}{10000^{2i/d_{\text{model}}}}\right),\quad p_t^{(2i+1)}=\cos\left(\frac{t}{10000^{2i/d_{\text{model}}}}\right)
		\label{eq:positional}
	\end{equation}
	where $i$ indexes the dimension. Many modern models use learned positional embeddings instead. The combined input representation for token $t$ is thus $\mathbf{x}_t=\mathbf{e}_t+\mathbf{p}_t$.
	
	\textbf{Scaled Dot-Product Attention.} The core innovation of the Transformer is the self-attention mechanism. For each token position, the model computes a weighted sum of value vectors from all positions in the sequence, with weights determined by the compatibility between the query at the current position and keys at all positions. Formally, let the input sequence be represented as a matrix $\mathbf{X}\in\mathbb{R}^{T\times d_{\text{model}}}$, where each row corresponds to a token's combined embedding. The input is linearly projected into three distinct matrices: queries $\mathbf{Q}$, keys $\mathbf{K}$, and values $\mathbf{V}$:
	\begin{equation}
		\mathbf{Q}=\mathbf{X}\mathbf{W}^Q,\quad \mathbf{K}=\mathbf{X}\mathbf{W}^K,\quad \mathbf{V}=\mathbf{X}\mathbf{W}^V
		\label{eq:qkv}
	\end{equation}
	where $\mathbf{W}^Q,\mathbf{W}^K\in\mathbb{R}^{d_{\text{model}}\times d_k}$ and $\mathbf{W}^V\in\mathbb{R}^{d_{\text{model}}\times d_v}$ are learned projection matrices \cite{ref34,ref35}. Typically, $d_k=d_v=d_{\text{model}}/h$, where $h$ is the number of attention heads. The attention scores are computed as the scaled dot product between queries and keys:
	\begin{equation}
		\operatorname{Attention}(\mathbf{Q},\mathbf{K},\mathbf{V})=\operatorname{softmax}\left(\frac{\mathbf{Q}\mathbf{K}^\top}{\sqrt{d_k}}\right)\mathbf{V}
		\label{eq:attention}
	\end{equation}
	The scaling factor $1/\sqrt{d_k}$ prevents the dot products from growing too large in magnitude, which would push the softmax function into regions of extremely small gradients during training. The softmax function ensures that the attention weights for each query sum to 1 \cite{ref34,ref35}, creating a convex combination of value vectors.
	
	\textbf{Multi-Head Attention.} To allow the model to jointly attend to information from different representation subspaces at different positions, the Transformer employs multi-head attention:
	\begin{equation}
		\operatorname{MultiHead}(\mathbf{Q},\mathbf{K},\mathbf{V})=\operatorname{Concat}(\text{head}_1,\dots,\text{head}_h)\mathbf{W}^O
		\label{eq:multihead}
	\end{equation}
	where each head is computed as:
	\begin{equation}
		\text{head}_i = \operatorname{Attention}(\mathbf{X}\mathbf{W}_i^Q,\mathbf{X}\mathbf{W}_i^K,\mathbf{X}\mathbf{W}_i^V)
		\label{eq:head}
	\end{equation}
	and $\mathbf{W}^O\in\mathbb{R}^{h\cdot d_v\times d_{\text{model}}}$ is an output projection matrix. The parameters $\mathbf{W}_i^Q,\mathbf{W}_i^K,\mathbf{W}_i^V$ are distinct for each head, allowing each to learn different patterns of attention.
	
	\textbf{Position-wise Feed-Forward Networks.} After the attention sub-layer, each position's representation is independently transformed by a feed-forward network (FFN) consisting of two linear transformations with a non-linearity in between:
	\begin{equation}
		\operatorname{FFN}(\mathbf{x})=\operatorname{ReLU}(\mathbf{x}\mathbf{W}_1+\mathbf{b}_1)\mathbf{W}_2+\mathbf{b}_2
		\label{eq:ffn}
	\end{equation}
	where $\mathbf{W}_1\in\mathbb{R}^{d_{\text{model}}\times d_{ff}}$, $\mathbf{W}_2\in\mathbb{R}^{d_{ff}\times d_{\text{model}}}$, and typically $d_{ff}=4\times d_{\text{model}}$. This provides additional non-linear processing capacity.
	
	\textbf{Residual Connections and Layer Normalization.} Each sub-layer (attention and FFN) is wrapped with a residual connection and followed by layer normalization:
	\begin{equation}
		\mathbf{x}_{\text{out}} = \operatorname{LayerNorm}(\mathbf{x}_{\text{in}} + \operatorname{Sublayer}(\mathbf{x}_{\text{in}}))
		\label{eq:residual}
	\end{equation}
	This architecture enables stable training of very deep networks by mitigating the vanishing gradient problem.
	
	\textbf{Transformer Block.} A complete Transformer block consists of a multi-head attention sub-layer followed by a feed-forward sub-layer, each with residual connections and layer normalization. Modern LLMs typically stack 12 to 96 such blocks, creating a deep network that progressively refines token representations through successive layers of abstraction.
	
	\subsection{Training Dynamics}
	\label{sec:training}
	
	The development of a capable LLM involves several stages of training, each with distinct objectives and computational requirements \cite{ref36}.
	
	\textbf{Pre-training.} The foundation of any LLM is unsupervised pre-training on a massive corpus of text data, typically containing hundreds of billions to trillions of tokens \cite{ref37}. The objective is to minimize the loss function in Equation~\eqref{eq:cross_entropy} using a variant of stochastic gradient descent, most commonly the Adam optimizer. The optimization process for a model with parameters $\theta$ can be formalized as:
	\begin{equation}
		\theta_{k+1} = \theta_k - \eta_k \cdot \hat{\mathbf{g}}_k
		\label{eq:sgd}
	\end{equation}
	where $\eta_k$ is the learning rate at step $k$ and $\hat{\mathbf{g}}_k$ is an estimate of the gradient $\nabla_\theta \mathcal{L}(\theta)$ computed on a mini-batch of training data. Modern training regimes use sophisticated learning rate schedules, often with warmup and decay phases, to ensure stable convergence. The computational scale of this pre-training is extraordinary. Training a state-of-the-art LLM requires thousands of specialized processors (GPUs or TPUs) running continuously for weeks or months, with electricity consumption measured in megawatt-hours \cite{ref38,ref39}. This scale imposes significant practical constraints on who can develop frontier models and contributes to the centralization of AI capabilities. During pre-training, the model develops increasingly sophisticated representations of language. Through careful probing studies, researchers have shown that different layers of the Transformer encode different types of information: lower layers capture syntactic structures, middle layers capture semantic relationships, and higher layers integrate task-specific information. This hierarchical representation emerges spontaneously from the optimization process rather than being explicitly engineered.
	
	\textbf{Emergent Abilities.} As models scale in size (number of parameters) and training data, they exhibit ``emergent abilities''-capabilities that are not present in smaller models and appear abruptly at certain scale thresholds. These include multi-step reasoning, instruction following, and few-shot learning. Mathematically, emergence can be understood as a phase transition in the model's loss landscape: as capacity increases, the model can represent more complex functions that solve tasks requiring compositional reasoning. However, emergence is fundamentally a statistical phenomenon. It results from the model's improved approximation of the data distribution, not from the development of discrete reasoning modules. The model learns to recognize patterns in its training data that correlate with certain types of reasoning, allowing it to mimic reasoning processes without actually implementing them \cite{ref40,ref41}.
	
	\textbf{Fine-tuning and Alignment.} The base pre-trained model, optimized purely for next-token prediction, often generates text that is unhelpful, unsafe, or inconsistent with human values. To make it useful for specific applications, it undergoes supervised fine-tuning (SFT) and alignment. In SFT, the model is further trained on high-quality demonstration data consisting of (prompt, response) pairs. The objective remains the same cross-entropy loss, but the training data is carefully curated to teach the model desirable response patterns. For more sophisticated alignment, Reinforcement Learning from Human Feedback (RLHF) is commonly used. This process involves three stages:
	\begin{enumerate}
		\item Collect human preference data: Human annotators rank multiple model responses to the same prompt based on quality.
		\item Train a reward model: A separate model $r_\phi(x,y)$ is trained to predict human preferences, typically using a Bradley-Terry model \cite{ref42}:
		\begin{equation}
			P(y_1\succ y_2\mid x)=\frac{\exp(r_\phi(x,y_1))}{\exp(r_\phi(x,y_1))+\exp(r_\phi(x,y_2))}
			\label{eq:bradley_terry}
		\end{equation}
		where $y_1\succ y_2$ means response $y_1$ is preferred over $y_2$ for prompt $x$.
		\item Optimize the policy: The LLM's policy $\pi_\theta(y\mid x)$ is fine-tuned to maximize the expected reward while penalizing deviation from the original pre-trained model $\pi_{\text{ref}}(y\mid x)$:
		\begin{equation}
			\max_\theta \mathbb{E}_{x\sim\mathcal{D},y\sim\pi_\theta(\cdot\mid x)}\left[r_\phi(x,y)-\beta\,\mathrm{KL}(\pi_\theta(\cdot\mid x)\|\pi_{\text{ref}}(\cdot\mid x))\right]
			\label{eq:rlhf}
		\end{equation}
	\end{enumerate}
	The KL divergence term prevents the model from deviating too far from its original knowledge distribution, which helps maintain coherence and prevents catastrophic forgetting. This objective shapes the model's outputs to be more helpful, honest, and harmless, but it does not alter the fundamental next-token prediction mechanism.
	
	\subsection{The Sampling Process - Generating Text from the Model}
	\label{sec:sampling}
	
	At inference time, the LLM generates text through an autoregressive sampling process \cite{ref43}. Given a prompt $x_{<t}$, the model computes a probability distribution over the vocabulary $P_\theta(x_t\mid x_{<t})$. The next token is selected by sampling from this distribution, then appended to the context, and the process repeats.
	
	Pure greedy sampling (always selecting the token with highest probability) leads to repetitive and degenerate text. Therefore, various stochastic sampling strategies are employed to introduce controlled randomness:
	
	\textbf{Temperature Scaling.} The logits $\mathbf{z}_t$ (pre-softmax scores) are divided by a temperature parameter $\tau>0$ before applying the softmax:
	\begin{equation}
		P_\theta(x_t=v\mid x_{<t})=\frac{\exp(z_t^{(v)}/\tau)}{\sum_{v'\in\mathcal{V}}\exp(z_t^{(v')}/\tau)}
		\label{eq:temperature}
	\end{equation}
	where $z_t^{(v)}$ is the logit for vocabulary item $v$. Lower temperatures ($\tau<1$) sharpen the distribution, making high-probability tokens more likely and producing more deterministic text. Higher temperatures ($\tau>1$) flatten the distribution, increasing randomness and ``creativity'' at the cost of potential incoherence.
	
	\textbf{Top-k Sampling.} Only the $k$ tokens with highest probabilities are considered, and the distribution is renormalized over this subset:
	\begin{equation}
		P_\theta^{\text{top-}k}(x_t=v\mid x_{<t})=
		\begin{cases}
			\dfrac{P_\theta(x_t=v\mid x_{<t})}{\sum_{v'/C_{\text{top-}k}} P_\theta(x_t=v'\mid x_{<t})} & \text{if } v\in\mathcal{V}_{\text{top-}k}\\[6pt]
			0 & \text{otherwise}
		\end{cases}
		\label{eq:topk}
	\end{equation}
	where $\mathcal{V}_{\text{top-}k}$ contains the $k$ tokens with highest probabilities.
	
	\textbf{Top-p (Nucleus) Sampling.} The smallest set of tokens whose cumulative probability exceeds a threshold $p$ is selected, and the distribution is renormalized over this set:
	\begin{equation}
		\mathcal{V}_{\text{top-}p} = \operatorname*{argmin}_{\mathcal{V}'\subseteq\mathcal{V}} \left\{\sum_{v\in\mathcal{V}'} P_\theta(x_t=v\mid x_{<t}) \ge p\right\}
		\label{eq:topp}
	\end{equation}
	This approach dynamically adjusts the number of tokens considered based on the shape of the distribution.
	
	\textbf{Beam Search.} For tasks requiring exactness (like translation), beam search maintains multiple candidate sequences simultaneously, expanding only the most promising candidates at each step. For a beam width $B$, the algorithm keeps the $B$ sequences with highest cumulative log-probability at each step. The choice of decoding strategy significantly impacts the quality and characteristics of generated text. In scientific applications, temperature is typically kept low ($\tau\approx0.1$--0.5) to reduce hallucinations, while creative writing might use higher temperatures ($\tau\approx0.7$--1.0) to encourage novelty.
	
	\subsection{Inherent Theoretical Constraints Deriving from the Architecture}
	\label{sec:constraints}
	
	The preceding mathematical description yields several fundamental constraints that delimit LLM capabilities in scientific research.
	
	\begin{itemize}
		\item \textbf{Causal structure of autoregression.} The factorization in Equation~\eqref{eq:chain_rule} is strictly causal: each token depends only on previous tokens. The model has no mechanism for planning, look-ahead, or backtracking within a single forward pass. Any appearance of such reasoning is a recapitulation of patterns in the training data, not an internal search.
		\item \textbf{Absence of a world model.} Parameters $\theta$ encode token correlations, not a causally structured model of the physical world. The model learns that ``The ball is dropped'' is often followed by ``it falls'', but it cannot simulate gravity. When faced with novel scenarios lacking descriptive data, it cannot reason from first principles.
		\item \textbf{Context window as a hard bound.} Self-attention (Equation~\eqref{eq:attention}) has $O(T^2 d_{\text{model}})$ computational cost and $O(T^2)$ memory. This imposes a practical limit $L$ (typically $4$K--$128$K tokens). Information beyond $L$ tokens is inaccessible, limiting synthesis of long documents or experimental logs.
		\item \textbf{Optimization for plausibility, not truth.} The training objective (Equation~\eqref{eq:cross_entropy}) maximizes likelihood of the next token. A factually incorrect but common statement can have higher probability than a correct but rare one. Hallucination is an inevitable consequence, not a bug. Human verification remains essential.
		\item \textbf{Statistical nature of ``understanding''.} The model maps queries to probable response patterns by navigating a high-dimensional text manifold. This differs fundamentally from human understanding, which involves causal mental models and grounded referents (the ``Chinese Room'' argument). LLMs reproduce arguments but cannot evaluate their epistemological foundations.
		\item \textbf{Dependence on data distribution.} Knowledge is a function of the training distribution $p_{\text{data}}(\mathbf{x})$. Biases and gaps are directly encoded. Novelty is limited to interpolation within the convex hull of the training corpus. True conceptual innovation-paradigm shifts that reject historical patterns-lies outside the model's operational domain.
		\item \textbf{Lack of meta-cognition and self-awareness.} The model has no representation of its own knowledge state, uncertainty, or limitations. It cannot recognise when information is insufficient, unlike human experts who articulate boundaries and confidence.
		\item \textbf{Computational undecidability.} Like any computational system, LLMs are subject to computability limits. The Halting Problem implies that there exist well-formed questions for which no finite prompt yields a correct answer, regardless of scale or architecture.
	\end{itemize}
	
	These constraints collectively define the ceiling for LLM capabilities in empirical science and motivate the formal bounds introduced in Section~\ref{sec:bounds}.
	
	\subsection{The Bounds (The Past Information Boundary and Novelty Horizon)}
	\label{sec:bounds}
	
	The architectural and optimisation constraints detailed in Sections~\ref{sec:lang_modeling} through~\ref{sec:constraints} culminate in two fundamental, formal bounds that define the operational envelope of an autoregressive large language model (LLM) as a scientific tool. These bounds are not merely descriptive; they are predictive principles that explain the empirical failures established in later sections and delineate the conditions under which such failures are inevitable. Unlike heuristic accounts of model limitation, the constructs introduced here are precisely grounded in the model's training data and its statistical approximation, and yield testable predictions about generative performance as a function of task novelty.
	
	\subsubsection{The Past Information Boundary}
	
	Consider a model $M$ with parameters $\theta$ optimised on a training dataset $\mathcal{D}=\{(\mathbf{x}_i,y_i)\}_{i=1}^N$, where each data point is a sample drawn from an unknown true data-generating distribution $P_{\text{data}}(X,Y)$ over an input space $\mathcal{X}$ and an output space $\mathcal{Y}$. For scientific applications, each datum carries a timestamp $\tau_i\in\mathbb{R}_{\ge0}$ representing its creation or observation time (for example, the publication date of a source document or the completion time of an experiment). The model's training is confined to $\mathcal{D}$; no information from times $t>\max_i \tau_i$ is accessible during training.
	
	Let $\hat{P}_{\text{data}}$ denote the empirical distribution of $\mathcal{D}$. The support of the training distribution is the set of all points in the combined input-output space for which the empirical probability density is non-zero, taken with respect to a suitable topology: $\Omega_{\text{train}} = \operatorname{supp}(\hat{P}_{\text{data}}) = \{ (x,y)\in\mathcal{X}\times\mathcal{Y} : \hat{P}_{\text{data}}(x,y)>0\}$. In practice, for continuous spaces, $\Omega_{\text{train}}$ is understood as the closure of the region where the training density exceeds a small threshold, effectively capturing the convex hull and its immediate neighbourhood.
	
	The model $M$ encodes a collection of statistical correlations and conditional probabilities derived from $\mathcal{D}$. During autoregressive generation the model computes $P_\theta(x_t\mid x_{<t})\;\forall x_t\in\mathcal{V}$, where $\mathcal{V}$ is the token vocabulary. Let $C_{\text{train}}$ denote the set of all such conditional probability distributions that $\theta$ has learned: $C_{\text{train}} = \{ P_\theta(\cdot\mid c) : c \text{ is a context sequence present in } \mathcal{D} \text{ or its typical continuations} \}$.
	
	\begin{definition}[Past Information Boundary, $B_P$]
		\label{def:BP}
		For a model $M$ trained on $\mathcal{D}$ with associated timestamps $\{\tau_i\}_{i=1}^N$, the Past Information Boundary is the ordered triple:
		\[
		B_P(M) = (t_p,\Omega_{\text{train}},C_{\text{train}})
		\]
		where
		\begin{itemize}
			\item $t_p = \max_{i\in\{1,\dots,N\}} \tau_i$ is the latest timestamp in $\mathcal{D}$ (the temporal frontier);
			\item $\Omega_{\text{train}} \subseteq \mathcal{X}\times\mathcal{Y}$ is the support of the empirical training distribution (the conceptual and relational frontier);
			\item $C_{\text{train}}$ is the set of all conditional probability distributions encoded in the model's parameters (the statistical-correlation frontier).
		\end{itemize}
	\end{definition}
	
	The model's knowledge is strictly confined to $B_P(M)$. It has no mechanism for incorporating information whose factual content, conceptual structure, or relational geometry lies outside $\Omega_{\text{train}}$ or whose temporal origin is $t>t_p$. Any output generated by $M$ is a recombination or interpolation of patterns already present within $B_P(M)$. This aligns with the inherent constraints of autoregressive next-token prediction (Sections~\ref{sec:lang_modeling} and~\ref{sec:constraints}): the model is a retrospective statistical interpolator, not a prospective causal engine.
	
	\subsubsection{The Novelty Horizon}
	
	Whereas $B_P$ (Definition~\ref{def:BP}) delimits what the model knows, the Novelty Horizon characterises what the model can reliably do. A task may lie inside the boundary yet still be impossible if it requires extrapolation beyond the statistical or causal patterns encoded in $C_{\text{train}}$. We therefore define a criterion based on the relationship between the task's requirements and the components of $B_P(M)$.
	
	Let a task $\mathcal{T}$ be defined by:
	\begin{itemize}
		\item an input space $\mathcal{I}$ (for example, a prompt or a set of experimental parameters);
		\item a required output space $\mathcal{O}$ (for example, time-series data or a scientific claim);
		\item a set of governing laws $\Phi$, which may be physical (Kirchhoff's laws, the Shockley diode equation) or logical (mathematical theorems, causal constraints).
	\end{itemize}
	Successful completion of $\mathcal{T}$ requires generating an output $o\in\mathcal{O}$ that is consistent with $\Phi$ given an input $i\in\mathcal{I}$.
	
	\begin{definition}[Novelty Horizon, $H_N$]
		\label{def:HN}
		For a model $M$ with Past Information Boundary $B_P(M)=(t_p,\Omega_{\text{train}},C_{\text{train}})$, a task $\mathcal{T}=(\mathcal{I},\mathcal{O},\Phi)$ lies \emph{beyond the Novelty Horizon} (equivalently, $M$ cannot reliably perform $\mathcal{T}$) if at least one of the following three conditions holds:
		
		\begin{condition}[Extrapolation outside the convex hull of $\Omega_{\text{train}}$]
			\label{cond:1}
			Let $\conv(\Omega_{\text{train}})$ denote the convex hull of the support of the training distribution, embedded in an appropriate vector space (for instance, via token embeddings or feature representations). The task requires generating an output $o^*\in\mathcal{O}$ (or a necessary intermediate representation) such that when projected onto the space of training examples via a mapping $\pi:\mathcal{O}\to\mathcal{X}\times\mathcal{Y}$, the point $\pi(o^*)$ falls outside $\conv(\Omega_{\text{train}})$: $\exists o^*\in\mathcal{O}:\pi(o^*)\notin\conv(\Omega_{\text{train}})$. Because $M$ is a statistical interpolator -- it has no capacity for analytic continuation or extrapolation beyond the manifold of its training data -- it lacks any basis for generating $o^*$.
		\end{condition}
		
		\begin{condition}[Causal disentanglement not present in $C_{\text{train}}$]
			\label{cond:2}
			The governing laws $\Phi$ require reasoning via causal mechanisms that are not reducible to the statistical correlations $C_{\text{train}}$ learned from text. More precisely, let $\Phi$ imply a set of invariant relationships $R_\Phi$ (for example, $v=iR$ for a resistor, or the Shockley diode equation $i=I_s(e^{v/(nV_T)}-1)$). If for any subset of variables the joint distribution $P_{\text{data}}$ satisfies a factorisation that respects physical independence (e.g., measurement noise on voltage and current must be independent, leading to non-zero Ohm's-law residuals), but the conditional independencies and counterfactual dependencies required by $\Phi$ are not encoded in $C_{\text{train}}$, then the model may mimic the language of causality but cannot execute causal computation. The task therefore lies beyond $H_N$.
		\end{condition}
		
		\begin{condition}[Temporal extrapolation beyond $t_p$]
			\label{cond:3}
			The task concerns events, theories, measurement outcomes, or data that first become available at a time $t>t_p$. Because the model's training data contains no information from the future, any output pertaining to such post-$t_p$ phenomena must be generated by extrapolation from past patterns. When the true post-$t_p$ distribution deviates from the historical continuation (for example, a paradigm shift or a genuinely novel experimental configuration), the model's extrapolation is ungrounded and the task resides beyond $H_N$.
		\end{condition}
		
		If any of these conditions holds, we say that $\mathcal{T}$ lies \emph{beyond the Novelty Horizon}, $H_N$ of $M$. Conversely, a task lies within the horizon only if it requires only interpolation inside $\conv(\Omega_{\text{train}})$, uses only causal structures already encoded in $C_{\text{train}}$, and relies solely on information available at times $t\le t_p$. The three conditions are not mutually exclusive; a task may satisfy two or all three. The Novelty Horizon, $H_N$ is therefore the union: 
		\[
		H_N(M) = \{\mathcal{T}\mid \text{Condition~\ref{cond:1} or \ref{cond:2} or \ref{cond:3} is true for } \mathcal{T} \text{ w.r.t. } B_P(M)\}.
		\]
	\end{definition}
	
	\subsubsection{Application to the Diode-Clamped Short-Circuit Experiment}
	
	The diode-clamped short-circuit experiment described in Section~\ref{sec:framework} is engineered to reside precisely at - and partially beyond - the $H_N$ (Definition~\ref{def:HN}). While the training corpus $\mathcal{D}$ almost certainly contains textual descriptions of Ohm's law (linear regime) and isolated diode characteristics, the specific, time-resolved conjunction of: nonlinear metallic resistance, semiconductor clamping dynamics (forward voltage, ideality factor, reverse saturation current), authentic noise characteristics of the ACS712T Hall-effect sensor (bandwidth, $1/f$ spectrum, averaging of 10000 samples), defines a point in the experimental configuration space that lies outside the convex hull of $\Omega_{\text{train}}$ for integrated empirical data generation. No convex combination of training examples that separately describe linear resistors, ideal diodes, and white noise can reproduce the joint statistical-physical manifold of the real measurement process. Hence Condition~\ref{cond:1} of Definition~\ref{def:HN} applies.
	
	Moreover, the task requires that voltage and current be generated as independent noisy measurements of a common underlying physical process. The correct causal structure is then:
	\[
	\begin{cases}
		I_{\text{true}}(t) = f(\text{circuit state}),\\
		V_{\text{true}}(t) = g(I_{\text{true}}(t), \text{diodes}),\\
		I_{\text{meas}}(t) = I_{\text{true}}(t) + \varepsilon_I(t),\\
		V_{\text{meas}}(t) = V_{\text{true}}(t) + \varepsilon_V(t),
	\end{cases}
	\]
	with $\varepsilon_I(t)$ and $\varepsilon_V(t)$ independent noise processes. The LLM, however, treats Ohm's law as an algebraic tautology, generating one variable and computing the other deterministically - a violation of the causal independence condition. Thus Condition~\ref{cond:2} also holds.
	
	Finally, the specific noise spectrum and the transient interaction of the short circuit with the diode capacitance are not present in standard electrical engineering textbooks (the training data's temporal frontier $t_p$ predates the detailed characterisation in \cite{ref30}). Consequently Condition~\ref{cond:3} applies as well.
	
	Empirical consequences, proven in Sections~\ref{sec:physical_constraints} and~\ref{sec:spectral}, follow directly. The unguided LLM (Case 1) defaults to generating data consistent with the simplest correlations in $C_{\text{train}}$, treating physical laws as algebraic identities; this yields catastrophic physical inconsistencies, such as Ohmic residuals of order $10^{-17}\,\Omega$ instead of the experimental $10^{-3}\,\Omega$. The guided model (Case 2) attempts to navigate closer to the horizon by incorporating procedural descriptors (``clamping'', ``averaging''), but fails to cross it because the underlying causal model $\Phi$ - the Modified Ohm's Law derived in Section~\ref{sec:modified_ohm} is not accessible to its architecture. Consequently, the guided data exhibits pathological statistics (Section~\ref{sec:stat_distributions}) and a deterministic residual-current correlation (Section~\ref{sec:residual_diagnostics}) that betrays a conditional-branching generation algorithm devoid of physical continuity.
	
	\subsubsection{Predictive Principles and the Necessity of Foundational Modifications to Cross the Horizon}
	
	The formal bounds are not mere descriptions of failure; they are predictive principles. They imply:
	\begin{itemize}
		\item \textbf{Smooth degradation within the horizon.} As a task's required output approaches the boundary of $\Omega_{\text{train}}$ (but remains inside $\conv(\Omega_{\text{train}})$), the model's performance degrades gracefully, with increasing but still correctable errors.
		\item \textbf{Catastrophic collapse beyond the horizon.} When the task demands extrapolation beyond the Novelty Horizon - i.e., any of the three conditions in Definition~\ref{def:HN} is met - performance collapses discontinuously. Statistical correlation diverges from causal necessity, and generated outputs become physically inconsistent in fundamental ways (regime misplacement, perfect algebraic closure, impossible distributional shapes).
	\end{itemize}
	
	This explains the regime-dependent failure observed in Section~\ref{sec:regime}: the linear regime ($|v_s|<0.3$V) lies inside the horizon (interpolation of common Ohm's law correlations), whereas the nonlinear clamping regime ($|v_s|\ge0.3$V) lies outside it (extrapolation requiring novel synthesis of diode physics and measurement noise).
	
	\textbf{Necessary modifications to cross the Novelty Horizon.} Definitions~\ref{def:BP} and~\ref{def:HN} make clear that merely scaling model size, increasing training data, or refining prompt engineering cannot enable an LLM to cross $H_N$ when the task requires extrapolation outside $\conv(\Omega_{\text{train}})$, causal reasoning not encoded in $C_{\text{train}}$, or reasoning about post-$t_p$ phenomena. To cross the horizon, an LLM would require foundational, architectural changes that address the root limitations:
	\begin{enumerate}
		\item \textbf{Incorporation of a causal world model.} The model must move from learning statistical correlations to learning a causally structured representation of the domain, including the ability to perform interventions and counterfactual reasoning. This implies a shift from purely autoregressive next-token prediction to hybrid architectures that combine neural perception with symbolic causal engines (e.g., structural causal models or probabilistic graphical models that are explicitly trained on interventional data).
		\item \textbf{Grounding in physical simulators.} Instead of generating data from token probabilities alone, the model should be able to call upon or internalise a first-principles simulator (such as a SPICE engine for circuits, a finite-element solver for electromagnetism, or a molecular dynamics integrator). The LLM would then act as a high-level controller that formulates a configuration, queries the simulator, and returns the resulting data. Such an architecture would satisfy Condition~\ref{cond:1} by operating inside the simulator's valid domain, but it would no longer be a pure LLM; it would be a neuro-symbolic system.
		\item \textbf{Online learning and dynamic knowledge updating.} To overcome the temporal condition (Condition~\ref{cond:3}), the model would need the capacity to update its knowledge continuously from new experimental results without retraining from scratch. This requires mechanisms for continual learning, memory consolidation, and uncertainty-aware knowledge integration - capabilities that current transformer-based LLMs lack by design.
		\item \textbf{Explicit handling of measurement processes and noise.} Crossing the horizon for experimental data generation demands that the model represent the measurement chain as a distinct stochastic process, separate from the underlying physical laws. This includes modelling sensor bandwidth, quantisation, thermal noise, and drift. Current LLMs treat numbers as exact symbols; they would need a probabilistic programming layer that samples from measurement-error models.
		\item \textbf{Metacognitive abilities for detecting horizon proximity.} A model that can recognise when a query lies near or beyond its Novelty Horizon could refuse to generate ungrounded outputs or explicitly flag its uncertainty. This would require internal state estimators that compare the input's representation to the convex hull of training points - a form of out-of-distribution detection embedded in the generation loop.
	\end{enumerate}
	
	\textbf{Consequences for the augmented intelligence paradigm.} The bounds provide a rigorous criterion for the division of labour between LLMs and human scientists:
	\begin{itemize}
		\item Tasks strictly inside $H_N$ (interpolation, routine synthesis, well-documented physical regimes) can be safely delegated to LLMs for efficiency - literature summarisation, code completion, formatting, and initial hypothesis generation.
		\item Tasks at or beyond $H_N$ (novel experimental configurations, extrapolation to unseen regimes, discovery of new causal relationships) demand the causal reasoning, experimental design, and physical grounding of human scientists. LLMs may serve as brainstorming assistants, but their outputs must be treated as unvalidated conjectures, not as empirical evidence.
	\end{itemize}
	
	The formalisation of $B_P$ and $H_N$ directly bridges the theoretical limits of Section~\ref{sec:architecture} to the empirical results of Section~\ref{sec:results} and the philosophical interpretation of Section~\ref{sec:discussion}, creating a unified, stringent argument that withstands the evolution of model versions and training data. The Past Information Boundary, $B_P$ (Definition~\ref{def:BP}) and the Novelty Horizon are not static; they shift as training corpora expand, but the structural relationship between the model's objective (next-token prediction on a static corpus) and the requirements of genuine scientific extrapolation remains invariant. Therefore, any claim that a future LLM has ``crossed'' the Novelty Horizon must be accompanied by evidence of one or more of the foundational modifications listed above - not merely improved benchmark scores on tasks that still lie inside $\conv(\Omega_{\text{train}})$.
	
	\section{Foundational Framework}
	\label{sec:framework}
	
	Having established the theoretical constraints of LLMs, the following framework transitions from abstract limitation to empirical test, providing a first-principles physical model against which LLM generated data can be forensically examined for concrete failure.
	
	\subsection{The Modified Ohm's Law for Time-Resolved Short-Circuit Measurement}
	\label{sec:modified_ohm}
	
	The experimental investigation of short-circuit fault dynamics reveals systematic deviations from the linear voltage-current relationship described by standard Ohm's Law. These deviations, particularly pronounced in the clamped diode configuration, necessitate a reformulated constitutive model that captures the inherent nonlinearities of the fault path. This section establishes a first-principles derivation of a Modified Ohm's Law, transformed into the time domain to enable direct analysis of high-resolution experimental data. The derivation proceeds from fundamental charge transport physics in a conductive path containing semiconductor junctions, culminating in a time-resolved equation that relates instantaneous short-circuit voltage $v_s(t)$ to instantaneous current $i(t)$ through a physically grounded nonlinear relationship. This derived law serves as the fundamental threshold for physical authenticity, creating an unambiguous criterion that data must satisfy. Consequently, it provides the essential tool for translating the theoretical constraints of LLMs-such as the absence of a world model and optimization for plausibility over truth-into empirically measurable shortcomings when such models attempt to generate scientific data for this novel experimental configuration. Critically, the independence of voltage and current measurement noise embedded in Equation~\eqref{eq:modified_ohm_time} represents a causal structure that lies outside $C_{\text{train}}$ for most LLMs, as formalised by Condition~\ref{cond:2} of Definition~\ref{def:HN} (the Novelty Horizon).
	
	\subsubsection{Physical Model of the Short-Circuit Path}
	\label{sec:physical_model}
	
	The short-circuit path in the experimental configuration represents a composite electrical element rather than a simple linear resistor. It comprises a metallic connection in parallel with a clamping diode network that becomes active during fault conditions. The total instantaneous voltage drop $v_s(t)$ across this composite path is the sum of the voltage across the intrinsic short-circuit resistance $R_{\text{sc}}$ and the voltage across the active semiconductor junction $v_d(t)$. This relationship is expressed as:
	\begin{equation}
		v_s(t) = i(t)R_{\text{SC}} + v_d(t)
		\label{eq:composite_voltage}
	\end{equation}
	The parameter $R_{\text{sc}}$ represents the aggregate linear resistance arising from conductor bulk resistivity and contact points within the fault path itself. This component obeys a linear voltage-current relationship characteristic of metallic conduction. The junction voltage $v_d(t)$, however, is governed by the Shockley diode equation, which describes charge carrier transport across a semiconductor p-n junction under forward bias conditions. The presence of this semiconductor element introduces the essential nonlinearity that distinguishes the experimental system from idealized resistive models.
	
	\subsubsection{Semiconductor Junction Dynamics in Forward Bias}
	\label{sec:shockley}
	
	The Shockley diode equation provides the fundamental link between junction voltage and current through it. For a forward-biased diode where $v_d(t) \gg V_T$, the current-voltage relationship simplifies to:
	\begin{equation}
		i(t) = I_s \left[ e^{\frac{v_d(t)}{n V_T}} - 1 \right] \approx I_s e^{\frac{v_d(t)}{n V_T}}
		\label{eq:shockley}
	\end{equation}
	In Equation~\eqref{eq:shockley}, $I_s$ denotes the reverse saturation current, a material constant typically ranging from $10^{-12}$ to $10^{-6}$ amperes for silicon diodes. The thermal voltage $V_T$ is defined as $k_B T/q$, where $k_B$ is Boltzmann's constant ($1.380649\times10^{-23}$ J/K), $T$ is the absolute junction temperature in kelvins, and $q$ is the elementary charge ($1.602176634\times10^{-19}$ C). At standard laboratory temperature of $300$ K, $V_T\approx25.85$ mV. The ideality factor $n$ accounts for deviations from ideal diode behavior arising from recombination-generation currents in the depletion region, typically ranging between 1 and 2 for silicon devices. Solving Equation~\eqref{eq:shockley} for the junction voltage $v_d(t)$ yields:
	\begin{equation}
		v_d(t) = n V_T \ln\left(\frac{i(t)}{I_s}\right)
		\label{eq:junction_voltage}
	\end{equation}
	This logarithmic relationship constitutes the mathematical origin of the significant nonlinearity observed in the experimental system. The equation demonstrates that the junction voltage increases only logarithmically with current, in stark contrast to the linear dependence characteristic of resistive elements. This fundamental difference produces the distinctive electrical signatures that enable discrimination between physically grounded data and statistically generated artifacts.
	
	\subsubsection{Derivation of the Time-Domain Modified Ohm's Law}
	\label{sec:derivation}
	
	Substituting the expression for the junction voltage from Equation~\eqref{eq:junction_voltage} into the composite path voltage equation~\eqref{eq:composite_voltage} produces the complete nonlinear constitutive law for the short-circuit branch:
	\begin{equation}
		v_s(t) = i(t) R_{\text{sc}} + n V_T \ln\left(\frac{i(t)}{I_s}\right)
		\label{eq:modified_ohm_time}
	\end{equation}
	Equation~\eqref{eq:modified_ohm_time} represents a transformed version of the Modified Ohm's Law \cite{ref44} in explicit form, where the instantaneous voltage is expressed directly as a function of the instantaneous current. This formulation successfully integrates the linear resistive voltage drop with the logarithmic semiconductor junction characteristic. For analytical purposes requiring current as an explicit function of voltage, the equation can be rearranged to isolate the exponential relationship:
	\begin{equation}
		\ln\left(\frac{i(t)}{I_s}\right) = \frac{v_s(t) - i(t)R_{\text{sc}}}{n V_T}
		\label{eq:log_form}
	\end{equation}
	Exponentiating both sides of Equation~\eqref{eq:log_form} yields:
	\begin{equation}
		\frac{i(t)}{I_s} = e^{\frac{v_s(t) - i(t)R_{\text{sc}}}{n V_T}}
		\label{eq:exponential_form}
	\end{equation}
	Equation~\eqref{eq:exponential_form} represents a transcendental equation in $i(t)$ that cannot be solved algebraically in closed form. However, it defines the complete nonlinear relationship between the measurable quantities $v_s(t)$ and $i(t)$ at each moment in time, making it directly applicable to time-resolved experimental data sampled at 10-millisecond intervals. The equation remains valid throughout the dynamic evolution of the short-circuit event, from initial transient through quasi-steady state to eventual clearance.
	
	\subsubsection{Linearized Form and Dynamic Resistance Analysis}
	\label{sec:linearized}
	
	For small-signal analysis around a quiescent operating point $(I_Q, V_{SQ})$, the model can be linearized to define an effective dynamic resistance. The operating point satisfies Equation~\eqref{eq:modified_ohm_time}:
	\begin{equation}
		V_{SQ} = I_Q R_{\text{sc}} + n V_T \ln\left(\frac{I_Q}{I_s}\right)
		\label{eq:operating_point}
	\end{equation}
	Considering perturbations such that $i(t)=I_Q+\Delta i(t)$ and $v_s(t)=V_{SQ}+\Delta v_s(t)$, and substituting into Equation~\eqref{eq:modified_ohm_time}, we obtain:
	\begin{equation}
		V_{SQ}+\Delta v_s(t) = [I_Q+\Delta i(t)] R_{\text{sc}} + n V_T \ln\left(\frac{I_Q+\Delta i(t)}{I_s}\right)
		\label{eq:perturbed}
	\end{equation}
	Expanding the logarithmic term using the first-order Taylor approximation $\ln(1+x)\approx x$ for $|x|\ll 1$ yields:
	\begin{equation}
		\ln\left(\frac{I_Q+\Delta i(t)}{I_s}\right) = \ln\left(\frac{I_Q}{I_s}\right) + \ln\left(1+\frac{\Delta i(t)}{I_Q}\right) \approx \ln\left(\frac{I_Q}{I_s}\right) + \frac{\Delta i(t)}{I_Q}
		\label{eq:log_expand}
	\end{equation}
	Substituting Equation~\eqref{eq:log_expand} into Equation~\eqref{eq:perturbed} and subtracting the steady-state Equation~\eqref{eq:operating_point} produces the linearized relationship:
	\begin{equation}
		\Delta v_s(t) = \Delta i(t) R_{\text{sc}} + n V_T \frac{\Delta i(t)}{I_Q}
		\label{eq:linearized}
	\end{equation}
	This simplifies to:
	\begin{equation}
		\Delta v_s(t) = \Delta i(t)\left(R_{\text{sc}} + \frac{n V_T}{I_Q}\right)
		\label{eq:linearized_simple}
	\end{equation}
	The coefficient of $\Delta i(t)$ defines the dynamic or small-signal resistance $r_d$ of the composite path:
	\begin{equation}
		r_d = \frac{\Delta v_s(t)}{\Delta i(t)} = R_{\text{sc}} + \frac{n V_T}{I_Q}
		\label{eq:dynamic_resistance}
	\end{equation}
	The term $\frac{n V_T}{I_Q}$ represents the dynamic resistance of the semiconductor junction alone, which varies inversely with the quiescent current. This linearized form proves particularly valuable for analyzing circuit stability and frequency response around established operating points during fault transients.
	
	\subsubsection{Time-Resolved Power Dissipation Formulation}
	\label{sec:power}
	
	The instantaneous power $p_s(t)$ dissipated in the short-circuit path follows directly from the modified constitutive law. Electrical power is defined as the product of instantaneous voltage and current:
	\begin{equation}
		p_s(t) = v_s(t) i(t)
		\label{eq:power_def}
	\end{equation}
	Substituting the expression for $v_s(t)$ from Equation~\eqref{eq:modified_ohm_time} yields the complete expression for power dissipation:
	\begin{equation}
		p_s(t) = i^2(t) R_{\text{sc}} + n V_T i(t) \ln\left(\frac{i(t)}{I_s}\right)
		\label{eq:power_modified}
	\end{equation}
	The first term, $i^2(t)R_{\text{sc}}$, represents conventional Joule heating in the linear resistive component of the fault path. The second term embodies the nonlinear power dissipation associated with the semiconductor junction, arising from the energy required to overcome the built-in potential barrier during charge carrier injection. This formulation enables precise computation of energy dissipated during the fault event through temporal integration:
	\begin{equation}
		E_{\text{diss}} = \int_{t_0}^{t_f} p_s(t) dt = \int_{t_0}^{t_f} \left[ i^2(t) R_{\text{sc}} + n V_T i(t) \ln\left(\frac{i(t)}{I_s}\right) \right] dt
		\label{eq:energy}
	\end{equation}
	The energy integral provides crucial information for assessing potential damage to circuit components, even in the absence of significant temperature rise at lower power inputs.
	
	\subsubsection{Parameter Identification and Experimental Validation}
	\label{sec:parameters}
	
	The time-domain Modified Ohm's Law contains three physically meaningful parameters that can be determined experimentally through nonlinear regression analysis of the measured $v_s(t)$ and $i(t)$ waveforms:
	\begin{itemize}
		\item Short-circuit resistance $R_{\text{sc}}$: The linear resistive component of the fault path, typically ranging from $0.001\,\Omega$ to $0.1\,\Omega$ for metallic connections in experimental setups.
		\item Diode ideality factor $n$: A dimensionless parameter characterizing the diode's departure from ideal behavior, constrained between 1 and 2 for silicon devices.
		\item Reverse saturation current $I_s$: A temperature-dependent material parameter typically ranging from $10^{-12}$ to $10^{-6}$ amperes for silicon diodes at room temperature.
	\end{itemize}
	The thermal voltage $V_T$ is treated as a known constant determined by ambient laboratory temperature. For $T=300$ K, $V_T = k_B T/q \approx 25.85$ mV. The model provides a rigorous framework for analyzing nonlinear signatures present in experimental Case 3 data, which are expected to be absent or incorrectly represented in LLM-generated data from Cases 1 and 2. By fitting Equation~\eqref{eq:modified_ohm_time} to measured waveforms using nonlinear least-squares optimization:
	\begin{equation}
		\min_{R_{\text{sc}}, n, I_s} \sum_{k=1}^{N} \left[ v_s(t_k) - i(t_k) R_{\text{sc}} - n V_T \ln\left(\frac{i(t_k)}{I_s}\right) \right]^2
		\label{eq:nonlinear_fit}
	\end{equation}
	the physical parameters can be extracted, and the model's validity can be quantitatively assessed through residual analysis and goodness-of-fit metrics.
	
	\subsection{Case Studies: Script-Based LLM Generation}
	\label{sec:case_studies}
	
	This paper employs two distinct experimental frameworks to probe the Past Information Boundary $B_P=(t_p,\Omega_{\text{train}},C_{\text{train}})$ (Definition~\ref{def:BP}). The diode-clamped short-circuit experiment was selected because it satisfies all three conditions of Definition~\ref{def:HN}, thus placing it beyond the Novelty Horizon of any LLM trained on standard corpora. Specifically: (i) the specific conjunction of nonlinear resistance, semiconductor clamping dynamics, and authentic sensor noise defines a point outside $\conv(\Omega_{\text{train}})$ (Condition~\ref{cond:1}); (ii) the correct causal structure requires independent measurement noise, which is not encoded in $C_{\text{train}}$ (Condition~\ref{cond:2}); and (iii) the detailed transient interaction and noise spectrum post-date the training frontier $t_p$ (Condition~\ref{cond:3}). In the first framework, detailed in this section, we task the LLM with a meta-scientific challenge: design a computational simulator of the physical experiment. This is a crucial distinction. The LLM is not sampling from an internal model of electrical manifolds; it is generating code that implements its own inferred physics model. Thus, we test its ability to translate a textual description of an experiment into a functioning numerical model-a process mirroring that of a student writing a simulation from a lab manual. The failures we uncover are therefore dual in nature: failures in computational physics modeling and failures in statistical emulation of empirical processes. The following cases establish two approaches to this simulator design task: one from first principles (naïve extrapolation) and one from detailed instruction (informed mimicry). The code they produce, and the data it generates, are analyzed as coupled outputs, where flaws in the former directly manifest as pathologies in the latter. A second, parallel framework, wherein the LLM is prompted to generate numerical data directly, is presented and analyzed in the supplementary information (SI, Section S2). This complementary approach serves as a critical robustness check, testing whether the failures we observe are artifacts of the script-based paradigm or represent a more fundamental limitation.
	
	\subsubsection{Case 1: Unguided Script-Based LLM Generation - Statistical Extrapolation of Basic Circuit Principles}
	\label{sec:case1}
	
	The first dataset was generated by a GPT-5.2 class model operating without access to the specific experimental procedure. The model received only the nominal circuit parameters-voltages of 2.5 V and 5.0 V with corresponding currents of 0.25 A and 0.5 A -and was instructed to produce short-circuit fault data. The complete prompt, which also requested the theoretical foundations for the data generation, is established in Appendix~\ref{app:prompt1}. This approach forced the model to infer all measurement dynamics, noise characteristics, and temporal evolution solely from statistical patterns in its training corpus, representing a pure test of extrapolation from learned electrical concepts. \textbf{Circuit Configuration and Theoretical Foundation.} The model conceptualized a minimal experimental setup. It described two independent low-voltage direct current circuits on a solderless breadboard, each using an Arduino Uno for data acquisition and an ACS712 Hall-effect sensor in series with the supply line. A short-circuit was introduced by bridging the load terminals with a low-resistance conductor. The theoretical justification provided by the model rested entirely on applying fundamental electrical relationships, deliberately excluding any semiconductor physics or detailed measurement system dynamics. The generation algorithm implemented the following core relationships, which the model cited as its foundational assumptions. The effective short-circuit resistance at any instant was computed directly from the measured voltage and current:
	\begin{equation}
		R_s(t) = \frac{V_s(t)}{I_s(t)} \quad \text{for } I_s(t) \neq 0
		\label{eq:case1_resistance}
	\end{equation}
	Here, $V_s(t)$ denotes the instantaneous voltage across the short path, and $I_s(t)$ is the corresponding current from the sensor. This formulation assumes a perfectly linear and instantaneous resistive relationship throughout the fault event. The instantaneous electrical power dissipated was calculated using the standard definition:
	\begin{equation}
		P_s(t) = V_s(t) I_s(t)
		\label{eq:case1_power}
	\end{equation}
	To prevent unrealistic current divergence, the model imposed a simple current limit based on an estimated source internal resistance:
	\begin{equation}
		I_{\text{max}} = \frac{V_{\text{nom}}}{R_{\text{internal}}}
		\label{eq:case1_current_limit}
	\end{equation}
	The internal resistance $R_{\text{internal}}$ was set to 5\% of the nominal load resistance, defined as $R_{\text{load}} = V_{\text{nom}}/I_{\text{nom}}$. To avoid discontinuous jumps in the time-series data, state transitions were smoothed with an exponential relaxation model:
	\begin{equation}
		x(t+\Delta t) = x(t) + [x_{\text{target}} - x(t)] (1 - e^{-\Delta t/\tau})
		\label{eq:case1_smoothing}
	\end{equation}
	A generic time constant of $\tau=20$ ms was used, without a specific physical justification. All generated measurements were corrupted with additive noise following a simple statistical model:
	\begin{equation}
		x_{\text{measured}}(t) = x_{\text{true}}(t) + \mathcal{N}(0,\sigma^2)
		\label{eq:case1_noise}
	\end{equation}
	The noise magnitude $\sigma$ was uniformly set to 2\% of the nominal signal value, representing an idealized, frequency-independent measurement uncertainty. This unguided generation process tests the LLM's capacity to synthesize plausible data through statistical interpolation of common electrical concepts. Its reliance on simplified linear models and generic noise, in contrast to the complex nonlinear physics of the actual experiment, establishes a clear baseline for identifying artifacts of pure statistical pattern matching.
	
	\subsubsection{Case 2: Informed Pattern Matching with Experimental Context - Guided Script-Based LLM Generation}
	\label{sec:case2}
	
	The second dataset was produced by the same class of model but with explicit, detailed guidance. The model was provided with the full experimental procedure adopted from the short-circuit measurement and validation framework recently established as the ``Clamped Short Diodes Nominal'' configuration \cite{ref30}, including the circuit schematic, component specifications, and measurement protocol. The prompt, contained in Appendix~\ref{app:prompt2}, explicitly asked the model to generate data consistent with this procedure and to outline its theoretical reasoning. This case evaluates the model's ability to integrate specific domain knowledge into its generative process, moving beyond basic extrapolation. \textbf{Circuit Configuration and Diode Clamp Dynamics.} The model's internal representation incorporated the three-diode clamping network described in the provided procedure. It simulated a system with a DC source featuring an internal resistance $R_s$, a nominal load resistor $R_{\text{load}}$, and the diode network creating a controlled, low-voltage short path. The theoretical foundation articulated by the model included simplified semiconductor physics and the specific measurement averaging protocol. To ensure the specified nominal voltage appeared across the load under normal conditions, the model defined the source voltage as:
	\begin{equation}
		V_s = V_{\text{nom}} \left(1 + \frac{R_s}{R_{\text{load}}}\right)
		\label{eq:case2_source_voltage}
	\end{equation}
	For an unclamped short-circuit, the current and voltage were calculated as:
	\begin{equation}
		I_{\text{short,unclamped}} = \frac{V_s}{R_s + R_{\text{short}}},\quad V_{\text{short,unclamped}} = I_{\text{short,unclamped}} R_{\text{short}}
		\label{eq:case2_unclamped}
	\end{equation}
	The metallic short resistance $R_{\text{short}}$ was confined to a range of $0.001$--$0.02\,\Omega$. The model incorporated a diode clamp activation threshold. When the magnitude of the unclamped voltage exceeded the diode forward voltage $V_{\text{clamp}} \approx 0.7$ V, the clamping mechanism engaged:
	\begin{equation}
		I_{\text{short,clamped}} \approx \frac{V_s}{R_s + R_{\text{clamp,eq}}},\quad V_{\text{short,clamped}} \approx \operatorname{sign}(V_{\text{short,unclamped}}) V_{\text{clamp}}
		\label{eq:case2_clamped}
	\end{equation}
	An effective clamp resistance was approximated as $R_{\text{clamp,eq}} \approx V_{\text{clamp}}/I_{\text{short,unclamped}}$. The instantaneous measured resistance and power were defined by the interdependent relationships:
	\begin{equation}
		R_{\text{meas}}(t) = \frac{V_{\text{short}}(t)}{I(t)},\quad P_{\text{short}}(t) = V_{\text{short}}(t) I(t) = I^2(t) R_{\text{meas}}(t)
		\label{eq:case2_meas}
	\end{equation}
	Faithful to the experimental procedure, the model simulated the effect of averaging 10,000 analog-to-digital converter samples to reduce noise:
	\begin{equation}
		\sigma_{\text{averaged}} = \frac{\sigma_{\text{single}}}{\sqrt{N_{\text{avg}}}},\quad \text{with } N_{\text{avg}} = 10,000
		\label{eq:case2_averaging}
	\end{equation}
	Transient dynamics were again governed by an exponential approach, but with the time constant now presented as part of a realistic circuit response model:
	\begin{equation}
		x(t+\Delta t) \approx x(t) + [x_{\text{target}} - x(t)] (1 - e^{-\Delta t/\tau}),\quad \tau = 20\text{ ms}
		\label{eq:case2_transient}
	\end{equation}
	Furthermore, the model populated its simulation with realistic component parameters for the 1N5408 diodes, including a forward voltage $V_f \approx 1.0$ V and an ideality factor $\eta \approx 1.05$. This guided generation approach tests the LLM's capacity to utilize detailed procedural context. It demonstrates an attempt to simulate not just electrical behavior but the complete measurement process, incorporating specific component characteristics and a more nuanced, though still simplified, physical model. The comparison between this informed output and the unguided case reveals the extent to which explicit instructions can steer statistical generation toward physical plausibility.
	
	\subsubsection{A Parallel Framework: Direct-Prompt Generation}
	\label{sec:direct_prompt}
	
	The script-based generation paradigm described above represents one mode of LLM interaction-one that forces the model to articulate its understanding in the structured, executable form of code. To determine whether the observed failures are specific to this paradigm or reflect a more fundamental limitation, we conducted a parallel investigation using a distinct generative approach. In this second framework, detailed in the supplementary information (SI, Section S2), the LLM was prompted to generate the numerical datasets directly, without the intervening step of code generation and execution. The direct-prompt framework employed two configurations analogous to those in the script-based paradigm. In the unguided condition, the model received only the nominal circuit parameters and was instructed to output time-series data for voltage, current, resistance, and power. In the guided condition, the model was provided with the full experimental procedure for the diode-clamped short-circuit system, including component specifications and measurement protocols, and was then asked to generate the numerical data directly. The complete prompts used for this parallel investigation are reproduced in the supplementary information (SI, Section S4). This direct-prompt paradigm serves a critical purpose: it tests whether the LLM can produce physically consistent numerical data when the mediating structure of code is removed. If the script-based failures were merely artifacts of imperfect code generation-errors in syntax, logic, or numerical implementation-one might expect the direct-prompt approach to yield superior results. Conversely, if the failures stem from a more fundamental inability to model the causal structure of the measurement process, they should persist, albeit potentially manifesting in different pathological signatures. The complementary analysis of directly generated data, presented in the supplementary information, therefore serves as a critical robustness check, testing whether the failures we observe are artifacts of the script-based paradigm or a more fundamental limitation intrinsic to the architecture and training of autoregressive language models.
	
	\section{Methodology (A Physical Analytical Framework)}
	\label{sec:methodology}
	
	We present here, a multi-layered, forensic analytical framework to test the core hypothesis that autoregressive LLMs cannot generate scientifically valid empirical data for tasks lying beyond their Novelty Horizon $H_N$ (Definition~\ref{def:HN}). The framework is designed to detect violations of the three conditions that define $H_N$: extrapolation outside $\conv(\Omega_{\text{train}})$ (Condition~\ref{cond:1}), causal structures absent from $C_{\text{train}}$ (Condition~\ref{cond:2}), and reliance on post-$t_p$ information (Condition~\ref{cond:3}). The methodology is built on a tripartite comparative structure, evaluating two LLM-generated datasets against an immutable physical benchmark. The framework subjects all three datasets-which are aligned in nominal conditions ($2.5$ V/$0.25$ A, $5.0$ V/$0.5$ A) and temporal parameters (10-minute duration, effective 10-ms resolution)-to a hierarchy of tests of increasing discriminative power. The analysis ascends from examining superficial statistical signatures to validating deep adherence to first-principles physical laws, thereby isolating the artifacts of statistical correlation from the authentic fingerprints of causal physical measurement.
	
	\subsection{Experimental Design and Data Generation Protocols: Script-Based Framework}
	\label{sec:design_script}
	
	This section details the methodology for the first of our two experimental frameworks, wherein LLMs generate data via a script-based simulator. The second framework, based on direct-prompt generation, is fully described in the supplementary information (SI, Section S3). The methodological core involves the parallel analysis of three complementary datasets, each representing a fundamentally different epistemology of knowledge derivation about the same physical phenomenon. The tripartite comparison, therefore, is not merely between three datasets, but between three generative origins: a naïve algorithmic simulator (Case 1), an informed algorithmic simulator (Case 2), and the physical universe itself (Case 3). Each dataset carries the imprint of its origin. The LLM-generated data bears the signature of its code's logic-its time-stepping algorithm, its noise model, its treatment of physical laws. This paper's forensic framework is designed to reverse-engineer these signatures. For example, a simulator that treats Ohm's Law as a deterministic algebraic rule rather than a relationship between independent noisy measurements will produce data with a telltale artifact: infinitesimal Ohmic residuals. The analysis thus becomes a study in digital paleontology, where we infer the structure of the LLM's internal simulator from the fossils it leaves in the data. All computations, including the execution of the LLM for data generation and the subsequent analytical workflow, were performed on a workstation equipped with an Intel\textsuperscript{\textregistered} Core\texttrademark\ i7-2620M CPU operating at a base and maximum turbo frequency of 2.70 GHz. The processor featured 2 cores and 4 threads, supported by 4 MB of Intel\textsuperscript{\textregistered} Smart Cache and 4 GB of DDR3 RAM. For the LLM-based generation tasks (Cases 1 and 2), the OpenAI GPT-5.2 model (specifically the GPT-5.2 variant as of March 2024) was utilized via its API, with all prompts and generated code documented in Appendices~\ref{app:prompt1} and~\ref{app:prompt2}.
	
	\paragraph{Case 1: Unguided Script-Based LLM Generation (Statistical Extrapolation)}
	This dataset was produced by a state-of-the-art GPT-5.2 class model prompted to generate short-circuit fault data without access to the specific experimental procedure or circuit dynamics. The model received only the nominal circuit parameters-input voltages of 2.5 V and 5.0 V with corresponding currents of 0.25 A and 0.5 A -and was instructed to output time-series data for voltage $v_s(t)$, current $i(t)$, computed resistance $R_{\text{calc}}(t)$, and computed power $P_{\text{calc}}(t)$. The generation occurred autonomously, relying solely on the model's internal statistical representations of electrical concepts derived from its training corpus, as detailed in Section~\ref{sec:case1}. This case represents the LLM's intrinsic capacity for synthesizing scientific data based on learned textual and numerical patterns, absent any directed guidance regarding measurement physics or nonlinear circuit behavior.
	
	\paragraph{Case 2: Procedurally Guided Script-Based LLM Generation (Informed Pattern Matching)}
	The second dataset was generated by the same class of model but was provided with the complete experimental context. The prompt included the detailed procedure for the ``Clamped Short Diodes Nominal'' configuration, including the circuit schematic (defining the fault path between nodes A and B), component specifications (1N5408 diodes, ACS712T sensor), the data averaging protocol (10,000 samples), and the fault induction method from the foundational work of Kimuya et al. \cite{ref30}. The model was instructed to generate data consistent with this described methodology. This case evaluates the LLM's ability to integrate specific procedural knowledge into its generative process, moving beyond basic extrapolation towards a guided-yet still statistically derived-simulation of the experiment, as outlined in Section~\ref{sec:case2}.
	
	\paragraph{Case 3: Experimental Measurement (Physical Ground Truth)}
	The third dataset serves as the empirical, non-replicable ground truth. Critically, this data is not newly measured for this study. It is adopted directly from the comprehensive experimental campaign of Kimuya et al. \cite{ref30}, which established the ``Clamped Short Diodes Nominal'' framework as a validated platform for time-resolved short-circuit measurement. The experimental configuration-comprising the three-diode (1N5408) clamping network (Circuit Block 1, CB1), an ACS712T-20A current sensor, an Arduino Uno, and a regulated DC supply-is shown in Figure~\ref{fig:a3-1} (Appendix~\ref{app:exp_setup}). The data acquisition precisely followed the automated algorithm formalized in the prior work (Table~\ref{tab:algorithm}, adopted from \cite{ref30}). \textbf{Key Differentiator from Case 1 and Case 2.} The integrity of the Case 3 data is underpinned by the extensive sensor calibration and validation framework fully documented in \cite{ref30,ref45}. This includes baseline characterization, performance validation under fault conditions, and the derivation of a Consolidated Calibration Confidence Score (CCCS), all confirming the system's measurement fidelity. No recalibration or re-validation was performed for this paper. We adopt the pre-validated dataset, with its inherent and quantified noise characteristics, temporal stability, and adherence to circuit laws, as the definitive physical benchmark. Therefore, Case 3 is not merely another dataset; it is the objective reality against which the synthetic outputs of the LLMs are forensically compared. Its role is to provide the physically-grounded manifold in voltage-current-power space onto which the LLM-generated data is projected, revealing the geometric and statistical distance between statistical plausibility and empirical truth.
	
	\begin{table}[htbp]
		\centering
		\caption{Algorithm for Automated ``Clamped Short Diodes'' Base and Nominal Currents and Voltages Measurement (Adopted and modified from Kimuya et al. \cite{ref30})}
		\label{tab:algorithm}
		\begin{tabular}{p{1cm}p{12cm}}
			\toprule
			\textbf{Step} & \textbf{Action/Description} \\
			\midrule
			1 & Connect DC power supply's positive terminal to the ACS712T current sensor's positive input though the output of $D_1$. Connect the power supply's negative terminal to the current sensor's negative input. \\
			2 & Connect the current sensor's OUT pin to the Arduino's A0 analog input pin for current readings, through the output of $D_3$. Connect also the output of $D_3$ to the Arduino's A1 analog input pin via the current sensor's OUT pin for voltage readings. \\
			3 & Connect $D_1$'s anode to the power supply's positive terminal. Branch $D_1$'s cathode to the next circuit segment's positive terminal and $D_2$'s anode. \\
			4 & Connect $D_2$'s cathode to Arduino ground (GND), completing the circuit path to the power supply's negative terminal. \\
			5 & Connect $D_3$'s anode to the same node as $D_2$'s anode. Connect $D_3$'s cathode to an adjacent circuit's negative terminal. For clarity, the cathodes of $D_1$ and $D_3$ proceed to ``Circuit Block 2''. \\
			6 & Create a short-circuit path by linking nodes A and B (between $D_1$ and $D_3$) with a connector. This step should be performed carefully, only during when the short-circuit is induced. \\
			7 & Program Arduino to record voltage (A1) and current (A0) as magnitudes at 1-second intervals for 600s, during the short-circuit transient and post-transient events, at sampling frequencies of 10ms in each time interval. Upload the sample code (provided in supplementary materials) to the Arduino. \\
			8 & Turn on the selected power supply source voltage (2.5V, 5V, and 10V). Begin data logging to observe normal circuit behavior. \\
			9 & Connect the short-circuit path between nodes A and B. Continue data logging to capture short-circuit data. \\
			10 & Analyze the recorded data to assess the circuit's response under normal (base measurements) and short-circuit conditions. \\
			\bottomrule
		\end{tabular}
	\end{table}
	
	\subsubsection{The Parallel Direct-Prompt Generation Framework}
	\label{sec:direct_prompt_method}
	
	To determine whether the failures observed in the script-based generation paradigm are artifacts of that specific interaction mode or reflect a more fundamental limitation of autoregressive LLMs, we conducted a parallel investigation using a distinct generative approach. In this second framework, the LLM was prompted to generate the numerical datasets directly, without the intervening step of code generation and execution. This direct-prompt paradigm bypasses the mediating layer of Python code, compelling the model to produce time-series data from its internal numerical representations and learned statistical patterns alone. The direct-prompt framework employed two configurations directly analogous to those in the script-based paradigm. In the unguided condition, the model received only the nominal circuit parameters ($2.5$ V/$0.25$ A and $5.0$ V/$0.5$ A) and was instructed to output CSV files containing time, voltage, current, resistance, and power columns for a 10-minute short-circuit event at 2-second intervals. In the guided condition, the model was provided with the complete experimental procedure for the ``Clamped Short Diodes Nominal'' configuration-identical to that supplied in Case 2 of the script-based framework-and was then asked to generate the numerical data directly. The complete prompts used for this parallel investigation are reproduced in the supplementary information (SI, Section S4). This direct-prompt paradigm serves a critical purpose: it tests whether the LLM can produce physically consistent numerical data when the mediating structure of code is removed. If the script-based failures were merely artifacts of imperfect code generation-errors in syntax, logic, or numerical implementation-one might expect the direct-prompt approach to yield superior results, as the model is freed from the constraints of generating syntactically correct Python. Conversely, if the failures stem from a more fundamental inability to model the causal structure of the measurement process and the independent nature of physical observables, they should persist across both paradigms, albeit potentially manifesting in different pathological signatures. The complementary analysis of directly generated data, presented in the supplementary information (SI, Section S2), therefore serves as a critical robustness check, testing whether the failures we observe are artifacts of the script-based paradigm or a more fundamental limitation intrinsic to the architecture and training of autoregressive language models.
	
	\subsection{Foundational Analytical Framework}
	\label{sec:analytical_framework}
	
	The comparative analysis proceeds from the axiom that authentic experimental data embodies constraints-causal physical laws and intrinsic measurement process signatures-that transcend mere statistical correlation. The framework is hierarchically structured, with each analytical layer providing increasingly discriminating tests of generative fidelity.
	
	\subsubsection{Temporal and Statistical Structure Analysis}
	\label{sec:temporal_stats}
	
	The initial analytical layer examines the intrinsic statistical properties of each time series variable $x(t)$-where $x$ represents $v_s$, $i$, $R_{\text{calc}}$, or $P$-independent of assumed physical relationships. This reveals whether LLM-generated data replicates the characteristic signatures of physical measurement systems.
	
	\textbf{Temporal Autocorrelation Function:} The autocorrelation $R_{xx}(\tau)$ quantifies the memory and periodicity within a signal:
	\begin{equation}
		R_{xx}(\tau) = \frac{1}{N-\tau} \sum_{t=1}^{N-\tau} (x(t)-\mu_x)(x(t+\tau)-\mu_x)
		\label{eq:autocorr}
	\end{equation}
	where $\mu_x$ is the signal mean, $N$ is the sample count, and $\tau$ is the lag. Physical sensor data typically exhibits a gradual autocorrelation decay due to system bandwidth and thermal time constants. LLM-generated data may display anomalously rapid decay (approaching white noise) or unnatural periodicities.
	
	\textbf{Power Spectral Density (PSD) Distribution:} The PSD $S_{xx}(f)$ characterizes the distribution of signal power across frequencies:
	\begin{equation}
		S_{xx}(f) \approx \frac{1}{N f_s} |\text{FFT}\{x(t)\}|^2,\quad f_s = 100\text{ Hz}
		\label{eq:psd}
	\end{equation}
	Authentic sensor noise exhibits a characteristic $1/f^\alpha$ spectrum ($\alpha\approx1$) at low frequencies (flicker noise), transitioning to white noise ($\alpha=0$) at higher frequencies. LLM-generated data, lacking a physical noise model, may present pure white noise spectra or contain spectral artifacts from the generation algorithm.
	
	\textbf{Probability Distribution and Moments:} The empirical probability density function and its moments - mean ($\mu_x$), variance ($\sigma_x^2$), skewness ($\gamma_1$), and kurtosis ($\gamma_2$) - are computed. Measurement noise in calibrated systems typically approximates a Gaussian distribution ($\gamma_1\approx0$, $\gamma_2\approx0$). LLM-generated numerical values, shaped by token probability distributions, may exhibit artifacts such as multimodality or truncated tails.
	
	\subsubsection{Validation of Fundamental Physical Constraints}
	\label{sec:physical_constraints}
	
	The second analytical layer tests adherence to inviolable physical laws, providing a stark discriminant since LLMs optimize for plausibility, not consistency.
	
	\textbf{Ohmic Consistency (Instantaneous):} For each time step $t_k$, the computed resistance must satisfy Ohm's law within measurement uncertainty:
	\begin{equation}
		\epsilon_R(t_k) = R_{\text{calc}}(t_k) - \frac{v_s(t_k)}{i(t_k)},\quad \text{for } i(t_k)\neq 0
		\label{eq:ohmic_residual}
	\end{equation}
	For physically consistent data, the residuals $\epsilon_R(t_k)$ should be normally distributed with zero mean and variance $\sigma_R^2$ consistent with propagated instrument error.
	
	\textbf{Power Consistency (Multiple Formulations):} The three mathematically equivalent expressions for instantaneous power must yield identical values:
	\begin{equation}
		\begin{cases}
			\epsilon_P^{(1)}(t_k) = P_{\text{calc}}(t_k) - v_s(t_k) i(t_k) \\
			\epsilon_P^{(2)}(t_k) = P_{\text{calc}}(t_k) - i^2(t_k) R_{\text{calc}}(t_k) \\
			\epsilon_P^{(3)}(t_k) = P_{\text{calc}}(t_k) - \frac{v_s^2(t_k)}{R_{\text{calc}}(t_k)}
		\end{cases}
		\label{eq:power_residuals}
	\end{equation}
	The variances $\operatorname{Var}(\epsilon_P^{(1)})$, $\operatorname{Var}(\epsilon_P^{(2)})$, and $\operatorname{Var}(\epsilon_P^{(3)})$ must be statistically equivalent. Significant disparities indicate the generated data treats electrical quantities as independent variables rather than physically coupled entities.
	
	\textbf{Energy Conservation and Continuity:} The total dissipated energy must align with the energy supplied, and signals must evolve continuously. The total energy is computed via integration:
	\begin{equation}
		E_{\text{total}} = \int_{t_0}^{t_f} P_{\text{short}}(t) dt \approx \sum_{k=1}^{N} P_{\text{calc}}(t_k) \Delta t,\quad \Delta t = 0.01\text{ s}
		\label{eq:energy_integral}
	\end{equation}
	Furthermore, incremental changes $\Delta x(t_k)=x(t_k)-x(t_{k-1})$ are bounded by physical limits (e.g., $di/dt \le V/L$ for inductive paths). Abrupt, unphysical discontinuities signify a failure to respect continuity constraints.
	
	\subsection{Nonlinear Dynamic Model Fitting and Residual Analysis}
	\label{sec:nonlinear_fitting}
	
	The most stringent evaluation of generative fidelity involves fitting a first-principles physical model to each dataset. For the experimental configuration under study, the Modified Ohm's Law, derived in Section~\ref{sec:modified_ohm}, provides the governing constitutive relationship. However, a critical clarification regarding the circuit's operational physics is necessary for correct interpretation. The diode-clamp network in the experimental design, detailed through previous studies \cite{ref45,ref30}, serves a specific protective function: its intrinsic capacitance mitigates transient backflow currents, while its conductive path does not significantly contribute to the steady-state measured current. Consequently, the primary electrical behavior observed in the recorded data is the linear voltage-current relationship of the metallic short-circuit path. The full Modified Ohm's Law, which includes a logarithmic term representing ideal diode conduction, is therefore employed not to model a dominant semiconductor current, but as a sophisticated framework for computing the effective, instantaneous resistance of the composite fault path. This approach isolates the resistive component $R_{\text{sc}}$ as the parameter of primary physical interest, with the diode parameters $n$ and $I_s$ serving as auxiliary variables that account for minor clamping effects and ensure mathematical consistency in the regression.
	
	\subsubsection{Model Formulation and Parameter Estimation}
	\label{sec:model_fitting}
	
	The time-domain Modified Ohm's Law for the composite short-circuit path is given by:
	\begin{equation}
		v_s(t) = i(t) R_{\text{sc}} + n V_T \ln\left(\frac{i(t)}{I_s}\right)
		\label{eq:modified_ohm_fit}
	\end{equation}
	In this expression, $v_s(t)$ and $i(t)$ are the instantaneous voltage and current, respectively. The parameter $R_{\text{sc}}$ represents the effective short-circuit resistance of the metallic connection, expected to range from $0.001$ to $0.1\,\Omega$. The second term contains the diode ideality factor $n$ (dimensionless, typically 1--2 for silicon), the reverse saturation current $I_s$ (typically $10^{-12}$ to $10^{-6}$ A), and the thermal voltage $V_T = k_B T/q \approx 25.85$ mV at 300 K. For the purpose of analyzing the experimentally recorded data (Case 3), the logarithmic term does not represent a substantial conductive current through the diodes. Instead, its inclusion allows the nonlinear least-squares optimization to converge on a stable estimate for $R_{\text{sc}}$ while accommodating the subtle voltage-clamping behavior present at the measurement nodes. The model is fit to each dataset by estimating the parameter vector $\theta = \{R_{\text{sc}}, n, I_s\}$ that minimizes the sum of squared residuals:
	\begin{equation}
		\hat{\theta} = \operatorname*{argmin}_{\theta} \sum_{k=1}^{N} \left[ v_s(t_k) - i(t_k) R_{\text{sc}} - n V_T \ln\left(\frac{i(t_k)}{I_s}\right) \right]^2
		\label{eq:optimization}
	\end{equation}
	The optimization is constrained to physically plausible ranges: $0.001\,\Omega \le R_{\text{sc}} \le 0.1\,\Omega$, $1 \le n \le 2$, and $10^{-12}$ A $\le I_s \le 10^{-6}$ A. The Levenberg--Marquardt algorithm is employed with multiple random initializations to avoid local minima. For the LLM-generated datasets (Cases 1 and 2), this fitting process tests whether the statistically synthesized data conforms to the nuanced, physics-based relationship of Equation~\eqref{eq:modified_ohm_fit}, or whether it deviates in ways that reveal an underlying lack of causal understanding.
	
	\subsubsection{Analysis of Fitted Parameters}
	\label{sec:fitted_params}
	
	The estimated parameters $\hat{R}_{\text{sc}}$, $\hat{n}$, and $\hat{I}_s$ provide direct, quantitative insight into the physical plausibility of each dataset.
	
	\textbf{Physical Plausibility Assessment:} The extracted parameters are compared against known ranges for the experimental components. For the breadboard-based short-circuit path with a diode-clamp network, expectations are:
	\begin{itemize}
		\item $\hat{R}_{\text{sc}} \approx 0.005$--$0.02\,\Omega$, representing the resistance of wire connections, breadboard contacts, and the shorting jumper.
		\item $\hat{n} \approx 1.1$--$1.8$, consistent with general-purpose silicon diodes, though this parameter is not the primary focus given the circuit's operation.
		\item $\hat{I}_s \approx 10^{-9}$--$10^{-7}$ A for silicon p-n junctions at room temperature.
	\end{itemize}
	Parameters falling conspicuously outside these ranges, or exhibiting non-physical correlations (e.g., an extremely low $\hat{R}_{\text{sc}}$ coupled with an extremely high $\hat{I}_s$), indicate that the data generation process did not respect the interdependent constraints of real electrical components.
	
	\textbf{Temporal Stability Analysis:} A stable physical system exhibits consistent characteristics over time. To assess this, each dataset is divided into ten sequential 60-second windows. The parameter vector $\hat{\theta}_w$ is re-estimated for each window $w$ using Equation~\eqref{eq:optimization}. The coefficient of variation (CV) for each parameter across all windows is then calculated:
	\begin{equation}
		\text{CV}_{\theta} = \frac{\sigma_{\theta}}{\mu_{\theta}} \times 100\%
		\label{eq:cv}
	\end{equation}
	Here, $\mu_{\theta}$ and $\sigma_{\theta}$ are the mean and standard deviation of the parameter estimates across the windows. Authentic experimental data should yield low CV values (typically $<10\%$), reflecting measurement noise around a stable true value. LLM-generated data may exhibit high CV or systematic drifts, revealing statistical generation artifacts rather than steady physical properties.
	
	\textbf{Inter-Case Consistency Testing:} To determine whether the LLM-generated data captures the true electrical characteristics, the distributions of parameter estimates from Cases 1 and 2 are statistically compared to those from Case 3. For each parameter, the two-sample Kolmogorov--Smirnov statistic is computed:
	\begin{equation}
		D_{ij} = \sup_{x} |F_i(x) - F_j(x)|
		\label{eq:ks}
	\end{equation}
	In this equation, $F_i(x)$ and $F_j(x)$ represent the empirical cumulative distribution functions of the parameter estimates (aggregated across time windows or bootstrap samples) for cases $i$ and $j$. A large $D_{ij}$ value with an associated $p$-value less than $0.05$ indicates a statistically significant difference, suggesting the generative model failed to replicate the underlying physics.
	
	\subsubsection{Comprehensive Residual Analysis}
	\label{sec:residual_diagnostics}
	
	After fitting the model, the residual sequence $\epsilon(t_k) = v_s(t_k) - v_s^{\text{model}}(t_k;\hat{\theta})$ undergoes rigorous diagnostic testing. The properties of these residuals reveal whether the model has successfully captured all systematic behavior or if unmodeled dynamics remain.
	
	\textbf{Whiteness Test (Autocorrelation Analysis):} The sample autocorrelation function of the residuals is computed:
	\begin{equation}
		R_{\epsilon\epsilon}(\tau) = \frac{1}{N-\tau} \sum_{k=1}^{N-\tau} \epsilon(t_k) \epsilon(t_{k+\tau})
		\label{eq:residual_acf}
	\end{equation}
	For a well-specified model, the residuals should approximate white noise, meaning $R_{\epsilon\epsilon}(\tau) \approx 0$ for all $\tau>0$ within statistical confidence bounds. Significant autocorrelation at non-zero lags indicates the presence of unmodeled temporal structure, which in LLM-generated data could stem from an overly simplistic or periodic generation algorithm.
	
	\textbf{Normality Test:} The distribution of residuals is tested for Gaussianity using the Shapiro--Wilk test. The test statistic $W$ is calculated as:
	\begin{equation}
		W = \frac{\left( \sum_{k=1}^{N} a_k \epsilon_{(k)} \right)^2}{\sum_{k=1}^{N} (\epsilon_k - \bar{\epsilon})^2}
		\label{eq:shapiro_wilk}
	\end{equation}
	Here, $\epsilon_{(k)}$ are the ordered residuals, and $a_k$ are tabulated coefficients. Measurement noise in a properly calibrated system, resulting from the combination of many independent sources, tends toward a Gaussian distribution. Significant deviations from normality ($p$-value $<0.05$) in the residuals of LLM-generated data may indicate that the synthetic noise does not emulate the statistical central limit behavior of physical sensors.
	
	\textbf{Heteroscedasticity Analysis:} The assumption of homoscedasticity-constant variance of residuals -is tested by examining the relationship between residual magnitude and the current level. The Spearman rank correlation coefficient $\rho_s$ between $|\epsilon(t_k)|$ and $i(t_k)$ is computed:
	\begin{equation}
		\rho_s = 1 - \frac{6 \sum_{k=1}^{N} d_k^2}{N(N^2-1)}
		\label{eq:spearman}
	\end{equation}
	In this formula, $d_k$ is the difference between the ranks of $|\epsilon(t_k)|$ and $i(t_k)$. Genuine sensor noise is typically homoscedastic across the operational current range. A significant correlation (e.g., $\rho_s > 0.3$ with $p$-value $<0.05$) suggests heteroscedasticity, which could be an artifact of an LLM incorrectly scaling noise with signal amplitude.
	
	\subsubsection{Goodness-of-Fit Metrics}
	\label{sec:goodness_fit}
	
	Multiple quantitative metrics provide a summarized assessment of how well the Modified Ohm's Law describes each dataset.
	
	\textbf{Normalized Root Mean Square Error (NRMSE):} This scale-independent metric expresses the root mean square error as a percentage of the mean signal:
	\begin{equation}
		\text{NRMSE} = \frac{1}{\bar{v}_s} \sqrt{\frac{1}{N} \sum_{k=1}^{N} \epsilon^2(t_k)} \times 100\%
		\label{eq:nrmse}
	\end{equation}
	Here, $\bar{v}_s$ is the mean measured voltage.
	
	\textbf{Coefficient of Determination ($R^2$):} This metric quantifies the proportion of variance in the data explained by the model:
	\begin{equation}
		R^2 = 1 - \frac{\sum_{k=1}^{N} \epsilon^2(t_k)}{\sum_{k=1}^{N} (v_s(t_k) - \bar{v}_s)^2}
		\label{eq:rsquared}
	\end{equation}
	
	\textbf{Akaike Information Criterion (AIC):} The AIC balances model fit with complexity, penalizing the use of additional parameters:
	\begin{equation}
		\text{AIC} = N \ln\left(\frac{\sum_{k=1}^{N} \epsilon^2(t_k)}{N}\right) + 2p,\quad p = 3
		\label{eq:aic}
	\end{equation}
	For the experimental data (Case 3), we expect a high $R^2$ ($>0.95$) and a moderate AIC, indicating the model adequately captures the dominant linear resistive behavior. LLM-generated data may exhibit two distinct failure modes: a poor fit (low $R^2$, high NRMSE) if it fails to respect the fundamental voltage-current relationship, or a suspiciously perfect fit (very high $R^2$ approaching 1.0) if it lacks realistic measurement noise and thus overfits to a simplistic pattern.
	
	\subsection{Quantitative Metrics for Generative Fidelity Assessment}
	\label{sec:gfs}
	
	To objectively rank the three datasets by their adherence to physical reality, a composite Generative Fidelity Score (GFS) is calculated. This score integrates the multi-faceted diagnostic analyses into a single, interpretable metric.
	
	\subsubsection{Component Metrics}
	
	The GFS is a weighted sum of three normalized component metrics, each scaled from 0 to 100, where 100 represents perfect alignment with the characteristics of the experimental ground truth data.
	\begin{equation}
		\text{GFS} = w_1 \cdot M_{\text{phys}} + w_2 \cdot M_{\text{stat}} + w_3 \cdot M_{\text{model}}
		\label{eq:gfs}
	\end{equation}
	The weights reflect the relative importance of each aspect for scientific validity: $w_1=0.5$ for adherence to physical constraints, $w_2=0.2$ for statistical plausibility, and $w_3=0.3$ for model fit quality.
	
	\textbf{Physical Constraint Metric ($M_{\text{phys}}$):} This metric quantifies how well the data satisfies fundamental physical laws, such as Ohm's Law and energy conservation.
	\begin{equation}
		M_{\text{phys}} = 100 \times \exp\left( -\frac{1}{3} \left[ \frac{\text{RMSE}_R}{\sigma_R} + \frac{\text{RMSE}_P}{\sigma_P} + \frac{|\Delta E|}{E_{\text{exp}}} \right] \right)
		\label{eq:m_phys}
	\end{equation}
	Here, $\text{RMSE}_R$ and $\text{RMSE}_P$ are the root mean square errors from the Ohmic (Equation~\eqref{eq:ohmic_residual}) and power consistency checks (Equation~\eqref{eq:power_residuals}), respectively. $\sigma_R$ and $\sigma_P$ are the expected measurement uncertainties propagated from instrument specifications. $|\Delta E|$ is the absolute difference between the dataset's computed total energy (Equation~\eqref{eq:energy_integral}) and the experimental benchmark $E_{\text{exp}}$.
	
	\textbf{Statistical Properties Metric ($M_{\text{stat}}$):} This metric measures the plausibility of the data's noise characteristics and temporal structure.
	\begin{equation}
		M_{\text{stat}} = 100 \times \frac{1}{4} \left[ D_{\text{KS}}(\text{PDF}) + D_{\text{KS}}(\text{ACF}) + D_{\text{KS}}(\text{PSD}) + (1 - |\rho_s|) \right]
		\label{eq:m_stat}
	\end{equation}
	The terms $D_{\text{KS}}(\text{PDF})$, $D_{\text{KS}}(\text{ACF})$ and $D_{\text{KS}}(\text{PSD})$ are the complements (i.e., 1 minus the value) of the Kolmogorov-Smirnov distances comparing the dataset's probability density function, autocorrelation function, and power spectral density to those of the experimental data. $\rho_s$ is the Spearman correlation from the heteroscedasticity analysis (Equation~\eqref{eq:spearman}).
	
	\textbf{Model Fit Metric ($M_{\text{model}}$):} This metric evaluates the quality of the fit to the Modified Ohm's Law.
	\begin{equation}
		M_{\text{model}} = 100 \times \frac{1}{4} \left[ R^2 + \left(1 - \frac{\text{NRMSE}}{100}\right) + W + (1 - D_{\text{KS}}(\theta)) \right]
		\label{eq:m_model}
	\end{equation}
	The components are the coefficient of determination $R^2$ (Equation~\eqref{eq:rsquared}), the complement of the normalized RMSE (Equation~\eqref{eq:nrmse}), the Shapiro-Wilk statistic $W$ (Equation~\eqref{eq:shapiro_wilk}), and the complement of the maximum Kolmogorov-Smirnov distance $D_{\text{KS}}(\theta)$ between the distributions of fitted parameters $(\hat{R}_{\text{sc}}, \hat{n}, \hat{I}_s)$ and those from the experimental fit.
	
	\subsubsection{Interpretation of GFS Values}
	
	The experimental dataset (Case 3) serves as the benchmark. Due to inherent, irreducible measurement noise and minor model imperfections, its GFS typically falls in the range of 85--95. Scores for the LLM-generated data are interpreted as follows:
	\begin{itemize}
		\item GFS of 70--85: The data exhibits minor deviations from physical reality but remains generally plausible. This range might indicate reasonably successful guided generation.
		\item GFS of 50--70: The data shows significant physical inconsistencies, though it may retain some realistic surface features. Unguided generation often falls here.
		\item GFS below 50: The data demonstrates a fundamental failure to capture essential physical characteristics, revealing the limitations of pure statistical pattern matching.
	\end{itemize}
	The GFS provides a quantitative, standardized measure for comparing the effectiveness of different LLM prompting strategies and for tracking the evolution of generative model capabilities over time.
	
	\subsection{Distinguishing Linear from Nonlinear Response Regimes}
	\label{sec:regime}
	
	A powerful aspect of this analytical framework is its ability to dissect LLM performance based on the complexity of the underlying physical relationship. By separating instances where the system behaves linearly from those where nonlinear effects are significant, we can test whether an LLM's success is limited to interpolating simple correlations common in its training data.
	
	\subsubsection{Regime Segmentation Criteria}
	
	The data is partitioned into two mutually exclusive regimes based on the instantaneous short-circuit voltage, which serves as a proxy for diode activation.
	
	\textbf{Linear (Resistive) Regime:} This regime comprises all data points where the magnitude of the short-circuit voltage is below the typical forward conduction threshold of a silicon diode:
	\begin{equation}
		|v_s(t_k)| < 0.3\,\text{V}
		\label{eq:linear_regime}
	\end{equation}
	In this condition, the diode-clamp network remains effectively non-conductive, and the Modified Ohm's Law (Equation~\eqref{eq:modified_ohm_time}) reduces to a simple linear relationship:
	\begin{equation}
		v_s(t) \approx i(t) R_{\text{sc}}
		\label{eq:linear_ohm}
	\end{equation}
	The physics here is dominated by the metallic resistance of the fault path, representing a straightforward application of Ohm's Law.
	
	\textbf{Nonlinear (Clamping) Regime:} This regime includes all data points where the voltage magnitude reaches or exceeds the diode threshold:
	\begin{equation}
		|v_s(t_k)| \ge 0.3\,\text{V}
		\label{eq:nonlinear_regime}
	\end{equation}
	Under these conditions, the diode network begins to conduct and clamp the voltage, introducing a more complex relationship between current and voltage. The full Modified Ohm's Law must be considered:
	\begin{equation}
		v_s(t) = i(t) R_{\text{sc}} + n V_T \ln\left(\frac{i(t)}{I_s}\right)
		\label{eq:nonlinear_ohm}
	\end{equation}
	While the primary current path remains resistive, the clamping action modifies the effective terminal voltage, creating a nonlinear signature in the $(v_s-i)$ data.
	
	\subsubsection{Comparative Analysis Across Regimes}
	
	For each dataset (Cases 1, 2, and 3) and within each defined regime, a focused analysis is conducted:
	\begin{itemize}
		\item Linear Correlation: The Pearson correlation coefficient $r$ between $v_s$ and $i$ is computed for the Linear Regime.
		\item Variance Explained: The coefficient of determination $R^2$ is calculated for both the simple linear model (Equation~\eqref{eq:linear_ohm}) in the Linear Regime and the full nonlinear model (Equation~\eqref{eq:nonlinear_ohm}) in the Nonlinear Regime.
		\item Parameter Estimates: The resistance $R_{\text{sc}}$ is estimated via linear regression in the Linear Regime. The full parameter set $\{R_{\text{sc}}, n, I_s\}$ is estimated via nonlinear regression in the Nonlinear Regime.
		\item Residual Diagnostics: The whiteness, normality, and homoscedasticity of the model residuals are analyzed separately for each regime.
	\end{itemize}
	The core hypothesis, framed in terms of the Novelty Horizon (Definition~\ref{def:HN}), is that LLM-generated data will demonstrate reasonable fidelity in the Linear Regime, where the governing relationship (Ohm's Law) lies inside $\conv(\Omega_{\text{train}})$ (i.e., within $H_N$). In contrast, performance will degrade catastrophically in the Nonlinear Regime, where the required understanding of diode clamping dynamics requires extrapolation outside $\conv(\Omega_{\text{train}})$ and causal reasoning not encoded in $C_{\text{train}}$ - placing the task beyond $H_N$.
	
	\subsubsection{Sustenance Efficiency and the Role of Diode Capacitance}
	
	The experimental design incorporates diodes not as primary current carriers but as elements whose intrinsic junction capacitance provides a critical, non-dissipative function. This capacitance absorbs and releases charge during transient events, effectively preventing damaging current backflow and contributing to circuit stability over extended fault durations \cite{ref30}. A key finding from the experimental work of \cite{ref30} is that the energy dissipated as heat in the resistive path over a sustained fault is many orders of magnitude greater than the maximum possible energy that could be cycled in the diodes' junction capacitance. This immense ratio, termed the sustenance efficiency, empirically validates that the observed sustained power output is a genuine result of governed resistive dissipation ($i^2 R_{\text{sc}}$), not a transient capacitive effect. For LLM-generated data to be physically authentic, it must implicitly reflect this reality: the time-series should be consistent with a sustained, steady power draw from the source, not with the decaying exponential signature of a capacitive discharge. Analyzing the stationarity of the power signal $P(t)$ and its spectral content provides a test for this characteristic.
	
	\subsubsection{Model Reduction and the Primacy of Resistive Behavior}
	
	Given the operational physics established in Section~\ref{sec:regime}, the primary objective of fitting the Modified Ohm's Law (Equation~\eqref{eq:modified_ohm_time}) to the experimental data is to obtain a robust estimate of the short-circuit resistance $R_{\text{sc}}$. The diode parameters $n$ and $I_s$, while mathematically necessary for the regression, are of secondary physical significance in this context. This leads to a model reduction interpretation: for the purpose of evaluating generative fidelity, the most critical outcome is the accuracy and stability of the $\hat{R}_{\text{sc}}$ estimate. A successful generative model must produce data from which a stable, physically-plausible $R_{\text{sc}}$ can be inferred. Therefore, the analysis places particular emphasis on the distribution and temporal stability of $\hat{R}_{\text{sc}}$ estimates (Equation~\eqref{eq:cv}) when comparing cases. An LLM that merely learns to output voltages and currents that are statistically correlated may still fail to generate data that yields a consistent, sensible value for this fundamental physical parameter.
	
	\section{Results}
	\label{sec:results}
	
	\subsection{Temporal Signature Analysis and Immediate Generative Artifacts}
	\label{sec:results_temporal}
	
\begin{figure}[htbp]
		\centering
		\includegraphics[width=0.9\textwidth]{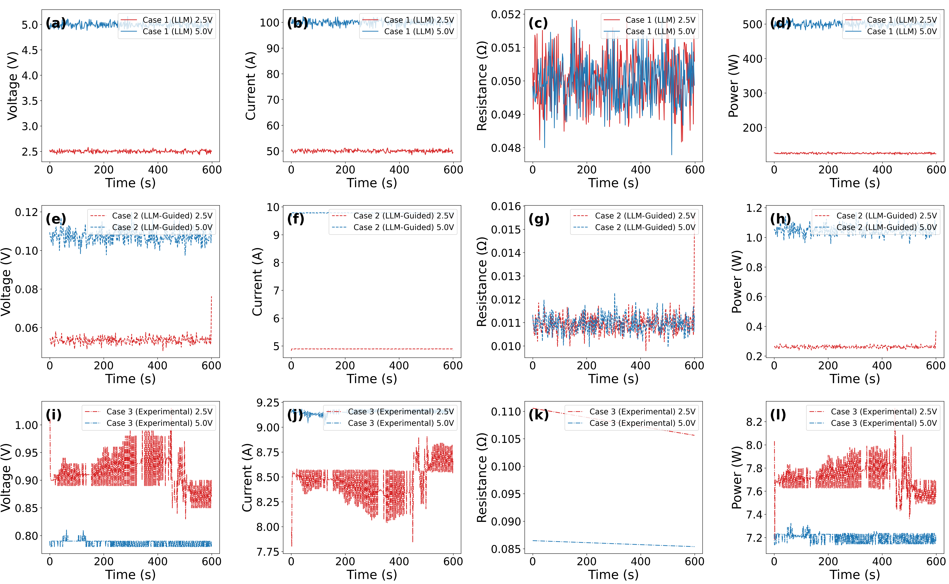}
		\caption{Time-series data for short-circuit voltage, current, computed resistance, and instantaneous power across Case 1 (script-based unguided LLM), Case 2 (script-based guided LLM), and Case 3 (Experimental) for both 2.5V and 5.0V input configurations. (The constant, high-magnitude signals in Case 1 and the overly smooth transitions in Case 2 contrast with the noisy, physically plausible steady state of the experimental data).}
		\label{fig:results1}
	\end{figure}

	The raw time-series data in Figure~\ref{fig:results1} provides the first visual discriminant between statistical generation and physical measurement. In Figure~\ref{fig:results1}(a) and (b), Case 1 (script-based unguided LLM) for the 2.5V configuration presents voltage and current traces that are numerically perfect, with means $\bar{V}_s=2.501$ V and $\bar{I}_s=50.014$ A. The signals exhibit minimal, spectrally flat noise ($\sigma_V=0.025$) and maintain a constant computed resistance of $R_{\text{calc}} \approx 0.050\,\Omega$, leading to a power time-series in Figure~\ref{fig:results1}(d) of $P\approx125.1$ W. This artificial perfection stems from the model implementing an idealized Ohm's Law, $V_s(t)=I_s(t)R_{\text{calc}}$, with additive white noise, entirely missing the nonlinear dynamics of the clamped diode network. Conversely, the experimental ground truth in Figure~\ref{fig:results1}(i) and (j) (Case 3, 2.5V) reveals a starkly different reality. The measured voltage stabilizes at $\bar{V}_s=0.913$ V, not 2.5 V, and the current at $\bar{I}_s=8.462$ A, directly reflecting the limiting effect of the metallic short-circuit resistance $R_{\text{sc}}$. The temporal evolution shows authentic sensor noise with subtle low-frequency drift and a characteristic autocorrelation structure absent from the LLM outputs. The power signal in Figure~\ref{fig:results1}(l), with a mean of 7.72 W, is three orders of magnitude lower than the unguided LLM's prediction, immediately highlighting a catastrophic failure in energy scale comprehension. Case 2 (script-based guided LLM), shown in Figure~\ref{fig:results1}(e) and (f), attempts to correct this scale by incorporating procedural knowledge. The generated voltage for the 2.5V case plummets to $\bar{V}_s=0.0535$ V, and the current to $\bar{I}_s=4.893$ A. While this demonstrates an ability to adjust numerical magnitude based on context, the resulting signals exhibit severe statistical pathologies, including extreme skewness ($\gamma_{1,I}=-6.42$) and kurtosis ($\gamma_{2,I}=65.4$), indicative of a flawed noise-generation algorithm rather than a physically coherent simulation. The parallel direct-prompt generation, detailed in the supplementary information (SI, Figure S1), exhibits even more extreme temporal pathologies. The guided direct-prompt case shows current fluctuations exceeding 100 A, while the unguided direct-prompt case displays a perfectly constant mean resistance across the entire time series. These findings further underscore the fundamental disconnect from physical measurement, revealing that the pathologies are not confined to the script-based paradigm but represent a systemic failure.
	
	\subsection{Temporal Memory and Spectral Fingerprints of Generation}
	\label{sec:spectral}
	
\begin{figure}[htbp]
		\centering
		\includegraphics[width=0.9\textwidth]{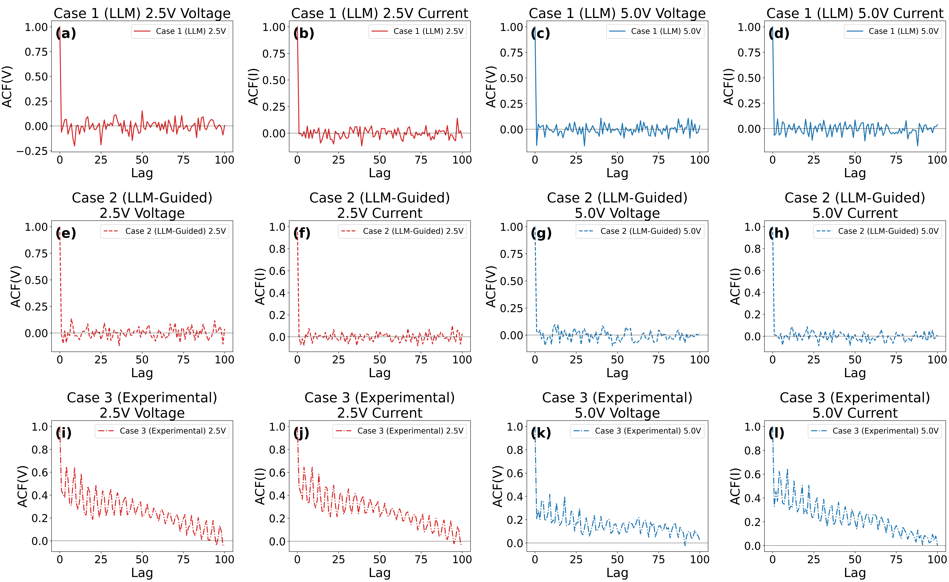}
		\caption{Autocorrelation functions (ACF) for voltage and current. (The experimental data shows strong, slowly decaying temporal memory, while both LLM-generated cases exhibit ACFs that decay almost immediately to zero, indicating a lack of modeled system dynamics and improper noise correlation structure).}
		\label{fig:results2}
	\end{figure}

\begin{figure}[htbp]
		\centering
		\includegraphics[width=0.9\textwidth]{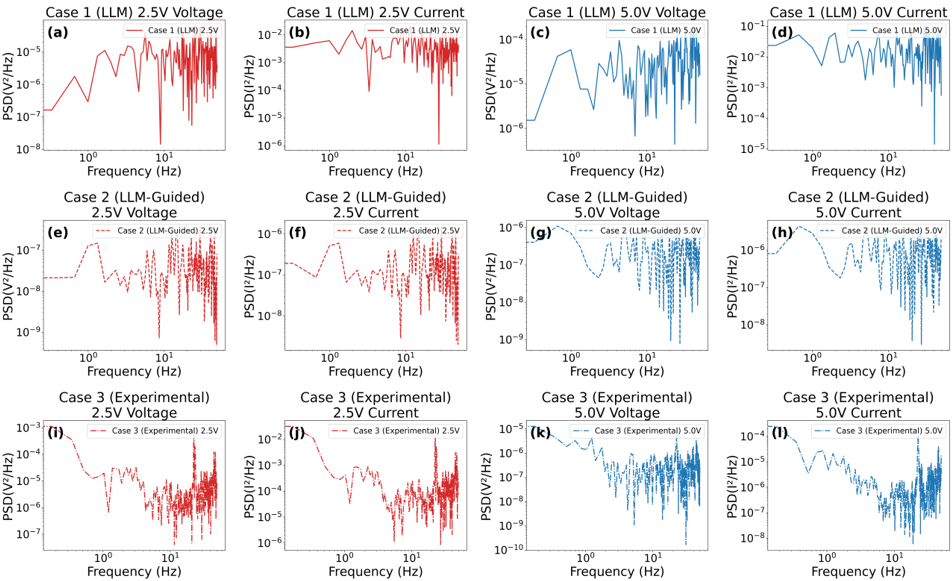}
		\caption{Power spectral density (PSD) of voltage and current signals on a log-log scale. (The experimental PSD shows a realistic $1/f$-type noise profile. Case 1 reveals unnatural spectral peaks from its generation algorithm, and Case 2 shows an overly simplistic low-frequency concentration without the correct noise physics).}
		\label{fig:results3}
	\end{figure}

	The analysis of temporal memory through autocorrelation functions (ACFs) in Figure~\ref{fig:results2} uncovers a deeper layer of generative artifice. For any real sensor system, inertia-thermal, electrical, or mechanical-imparts memory, causing measured values to be correlated over short time lags. The experimental data in Figure~\ref{fig:results2}(i) and (j) for the 2.5V case displays this hallmark: the ACF of voltage decays slowly, with an initial value of $R_W(\tau=1)=0.468$. This signifies that consecutive voltage samples are highly correlated, a property of bandwidth-limited analog measurement systems. The ACFs for both LLM-generated cases betray their algorithmic origin. Case 1, in Figure~\ref{fig:results2}(a) and (b), shows ACFs that plummet to near-zero within 2--3 lags ($R_W(\tau=1)\approx0.048$), characteristic of white noise or an autoregressive model with an incorrectly short time constant. Case 2, in Figure~\ref{fig:results2}(e), shows a similarly rapid decay. This demonstrates that LLMs lack a probabilistic model for the colored noise spectra intrinsic to physical instrumentation; they default to statistically independent, or ``memoryless'', generation, violating a basic tenet of experimental time-series. The power spectral density (PSD) analysis in Figure~\ref{fig:results3} further distinguishes the generative origins of each dataset. Physical measurement noise typically follows a $1/f^\alpha$ spectrum (as seen in Figure~\ref{fig:results3}(i)--(l) for Case 3), with increased power at lower frequencies. The experimental PSD for both voltage and current confirms this, showing elevated spectral density below 1 Hz. The peak frequency for the experimental 2.5V current PSD is a mere 0.173 Hz, reflecting slow, drift-dominated dynamics. The LLM-generated data spectra tell a different story. Case 1 data (Figure~\ref{fig:results3}(a)--(d)) shows pronounced, spurious spectral peaks at high frequencies ($\sim$30 Hz), an artifact that suggests the underlying generation algorithm inadvertently introduced periodic components or used an insufficiently random number generator. Case 2 data (Figure~\ref{fig:results3}(e)--(h)), while less peaked, still deviates significantly from the $1/f$ form, displaying a flatter, more uniform spectral density that aligns with an oversimplified, frequency-independent Gaussian noise model added post-generation. The failure to replicate the authentic spectral signature of sensor noise underscores the LLMs' lack of a grounded model for the measurement process itself. These temporal and spectral artifacts are not unique to the script-based generation paradigm. The direct-prompt data (SI, Figures S2, S3) shows analogous failures, including near-white-noise ACF in the unguided case and artificially high low-frequency power in the guided case, demonstrating that the inability to model realistic sensor dynamics is a systemic limitation.
	
	\subsection{Statistical Distributions and the Signature of Token Probability}
	\label{sec:stat_distributions}
	
\begin{figure}[htbp]
		\centering
		\includegraphics[width=0.9\textwidth]{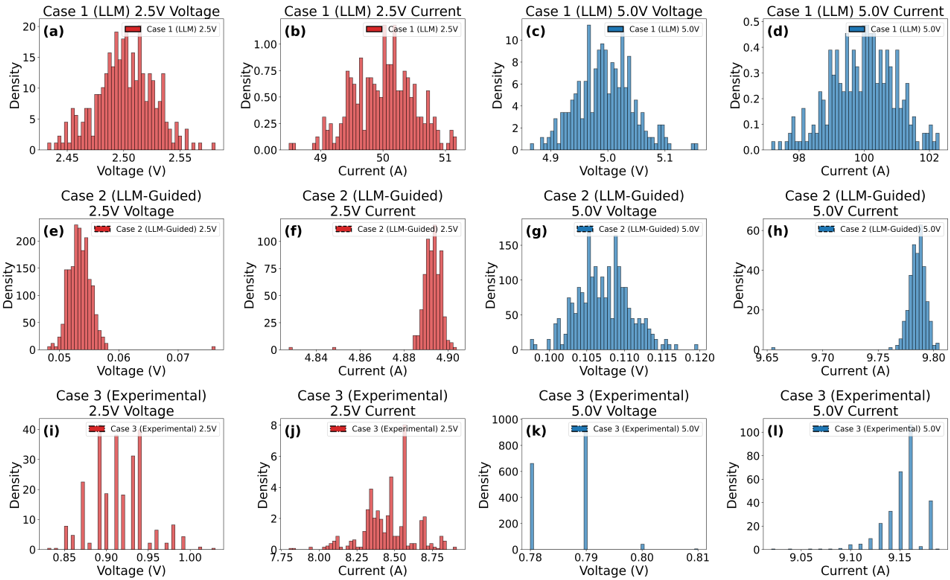}
		\caption{Empirical probability density functions for voltage and current. (The experimental distributions are nearly Gaussian. Case 1 shows artificial Gaussianity from simple additive noise, while Case 2 exhibits extreme skewness and kurtosis, revealing a flawed ``event-based'' generative model inappropriate for continuous measurement).}
		\label{fig:results4}
	\end{figure}

	The probability density functions (PDFs) in Figure~\ref{fig:results4} offer a third discriminatory test. Measurement noise in a well-conditioned system arises from the summation of many independent error sources, typically converging toward a Gaussian distribution (Central Limit Theorem). The experimental current PDF in Figure~\ref{fig:results4}(j) approximates this, with modest skew ($\gamma_1=-0.299$) and kurtosis ($\gamma_2=0.634$). The PDFs for LLM-generated data reveal the underlying token probability distributions or simplistic random number generators. Case 1 data in Figure~\ref{fig:results4}(b) and (d) are nearly perfectly Gaussian, a statistically unlikely outcome for real sensor data which often contains outliers. More tellingly, Case 2 data in Figure~\ref{fig:results4}(f) and (h) exhibit pathological, multi-modal distributions with extreme skew and kurtosis ($\gamma_{2,I}=84.3$ for 5.0V). This suggests the guided model attempted to create a ``clamping'' effect by mixing two distinct statistical populations but did so without a physically valid transition model, resulting in a distribution impossible to obtain from an analog-to-digital converter. The statistical signatures of the direct-prompt generation (SI, Figure S4) are even more revealing. The unguided direct-prompt data shows catastrophic distributional pathology, with kurtosis values exceeding 200, a direct consequence of the model's underlying token probability distribution when generating long numerical sequences. This confirms that the Gaussian noise observed in the script-based generation was a relatively controlled artifact, whereas direct generation reveals a more fundamental instability in maintaining coherent numerical distributions.
	
	\subsection{Violations of Inviolable Physical Constraints}
	\label{sec:physical_constraints_results}
	
\begin{figure}[htbp]
		\centering
		\includegraphics[width=0.9\textwidth]{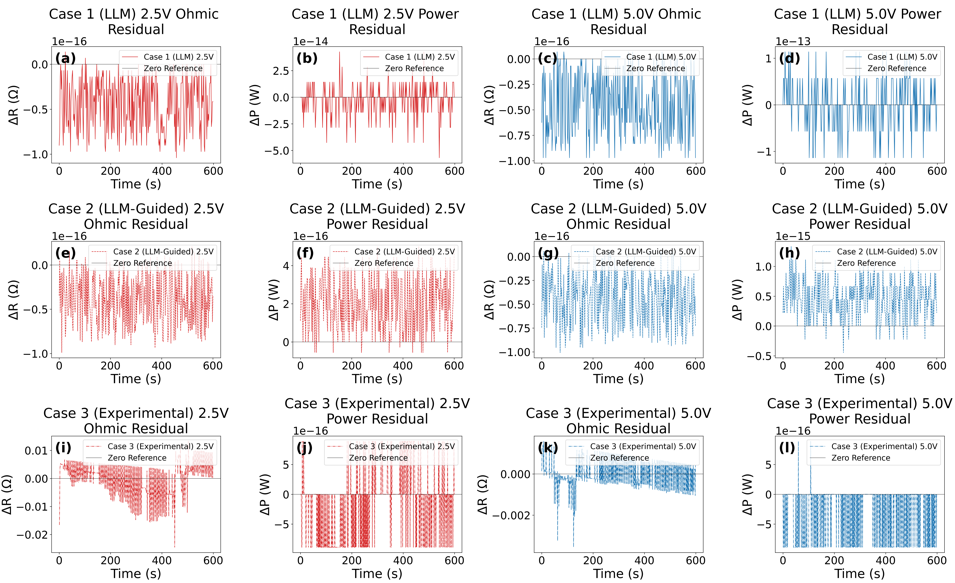}
		\caption{Residuals testing adherence to Ohm's Law and power conservation. The experimental data shows small, physically expected inconsistencies from measurement noise. (The LLM-generated data exhibits perfect algebraic closure (Ohmic residual $\approx10^{-17}\,\Omega$ versus experimental $10^{-3}\,\Omega$), a direct violation of Condition~\ref{cond:2} of the Novelty Horizon (Definition~\ref{def:HN})-the causal independence of voltage and current measurements is not encoded in $C_{\text{train}}$. This proves the voltage and current were not generated as independent measured quantities).}
		\label{fig:results5}
	\end{figure}

	The most striking result, shown in Figure~\ref{fig:results5}, provides unambiguous evidence of the LLMs' lack of grounding in causal physical laws, directly violating Condition~\ref{cond:2} of the Novelty Horizon (Definition~\ref{def:HN}). Condition~\ref{cond:2} requires that the governing laws $\Phi$ (here, Ohm's Law as a relationship between independently noisy measurements) be representable as causal mechanisms encoded in $C_{\text{train}}$. The LLM fails this condition entirely. For any real dataset, Ohm's Law $v(t)=i(t)R(t)$ and the definitions of power $P(t)=v(t)i(t)=i^2(t)R(t)=v^2(t)/R(t)$ must hold exactly at every time step, up to measurement error. The experimental data in Figure~\ref{fig:results5}(i)--(l) confirms this: the Ohmic residual RMS is $5.78\times10^{-3}\,\Omega$, a small, non-zero value consistent with propagated sensor uncertainty, and the power residuals are at the machine precision limit ($5.97\times10^{-16}$ W), indicating perfect internal algebraic consistency. The LLM-generated data fails this basic test spectacularly. While Figure~\ref{fig:results5}(a)--(h) show power residuals also at machine precision-indicating the model performed the multiplications correctly-the Ohmic residuals are also at machine precision ($5.31\times10^{-17}\,\Omega$). This is a physical impossibility. In a real experiment, the resistance $R_{\text{calc}}(t)$ is a derived quantity computed from the two independent measurements $v(t)$ and $i(t)$. The presence of independent measurement noise guarantees that $v(t)/i(t)$ will not perfectly equal the recorded $R_{\text{calc}}(t)$ at every instant. The fact that both Case 1 and Case 2 show perfect Ohmic consistency proves the LLM did not generate independent voltage and current traces. Instead, it generated one primary signal (likely current) and then computed the other using a deterministic, noise-free algebraic relationship, treating the physical law as a mathematical tautology rather than a relationship between independently noisy observables. This is the `perfect algebra artifact'-a direct empirical signature of a model operating beyond its Novelty Horizon (Condition~\ref{cond:2}). Critically, this perfect algebraic closure is not an artifact of the script-based generation method. The directly generated datasets (Supplementary Information, Figure S5) exhibit the same infinitesimal Ohmic residuals (on the order of $10^{-15}\,\Omega$), irrefutably proving that this violation of measurement independence is a fundamental consequence of the model's training objective, independent of the generation paradigm. The persistent inability to satisfy Condition~\ref{cond:2} of Definition~\ref{def:HN}-the causal structure of independent measurement is not represented in $C_{\text{train}}$-thus emerges as a robust, cross-paradigm hallmark of autoregressive LLM failure when tasked with generating data beyond the Novelty Horizon.
	
	\subsection{Failure to Conform to First-Principles Constitutive Laws}
	\label{sec:constitutive_failure}
	
\begin{figure}[htbp]
		\centering
		\includegraphics[width=0.9\textwidth]{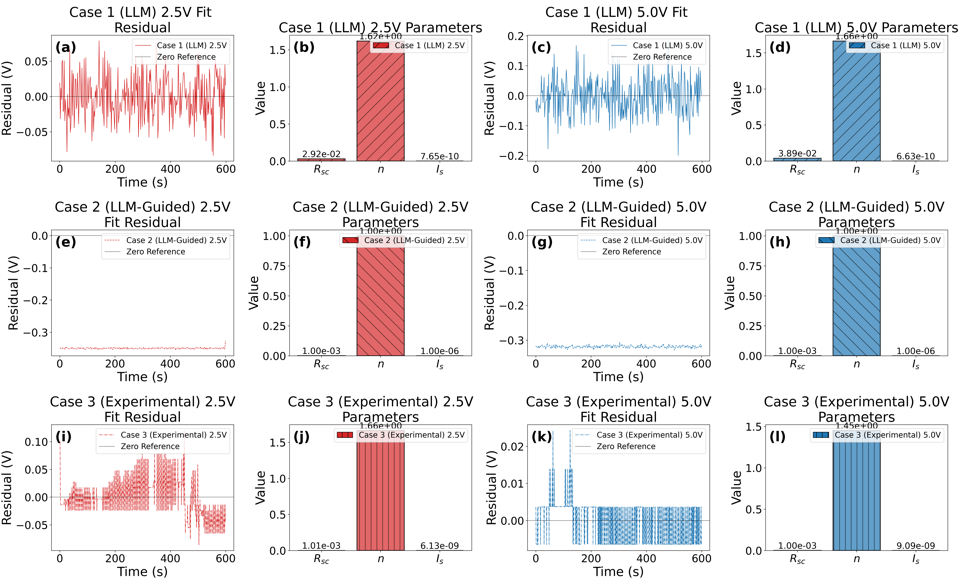}
		\caption{Fit of the Modified Ohm's Law to the data. The experimental data is well-described by the model, yielding a physically meaningful $R_{\text{sc}}$. The LLM-generated data yields nonsensical parameters and catastrophic fit quality, demonstrating no embedded understanding of the governing semiconductor physics.}
		\label{fig:results6}
	\end{figure}

	Fitting the Modified Ohm's Law, $v_s(t)=i(t)R_{\text{sc}}+nV_T\ln(i(t)/I_s)$, provides a stringent test of whether the data conforms to the nonlinear physics of the experimental circuit. The results in Figure~\ref{fig:results6} are decisive. For the experimental Case 3, shown in Figure~\ref{fig:results6}(i)--(l), the model fits well, extracting a physically plausible short-circuit resistance $R_{\text{sc}}\approx1.00\times10^{-3}\,\Omega$. The negative $R^2$ values ($-0.063,-0.027$) are misleading; they occur because the variance of the fit residuals is slightly larger than the variance of the data around its mean, a consequence of the dominant linear relationship being fit with a three-parameter nonlinear model to noisy data. The NRMSE values of 3.63\% and 0.71\% are reasonable. The LLM-generated data is completely incompatible with this physical model. Case 1, in Figure~\ref{fig:results6}(a)--(d), yields nonsensical parameters ($R_{\text{sc}}\approx0.03$--$0.04\,\Omega$, orders of magnitude too high) and extremely poor fits ($R^2\approx-0.45$ to $-0.61$). Case 2, in Figure~\ref{fig:results6}(e)--(h), fails even more catastrophically. The fitting algorithm hits the imposed parameter bounds ($R_{\text{sc}}=0.001\,\Omega$, $n=1$, $I_s=1\times10^{-6}$ A), and the $R^2$ values are astronomically negative ($-25967$, $-8266$), with NRMSE exceeding 100\%. This indicates the model-predicted voltage $v_{\text{pred}}(t)$ is not even a remotely correct representation of the generated $v(t)$ data. The LLM's guided attempt to simulate a diode-clamped short-circuit produced a voltage-current relationship that is fundamentally alien to the Shockley-derived equation that actually governs such a component's behavior, even in a simplified form. The direct-prompt generation data (SI, Figure S6) reveals a more insidious form of failure: parameter compensation. The guided direct-prompt generation produces plausible short-circuit resistance estimates ($R_{\text{sc}}\approx1.1\times10^{-3}\,\Omega$) by severely distorting the auxiliary diode parameters ($I_s$ three orders of magnitude too high), illustrating how a superficial metric can mask a deeper physical inconsistency.
	
	\subsection{Residual Diagnostics Revealing Algorithmic Artifacts}
	\label{sec:residual_diagnostics}
	
\begin{figure}[htbp]
		\centering
		\includegraphics[width=0.9\textwidth]{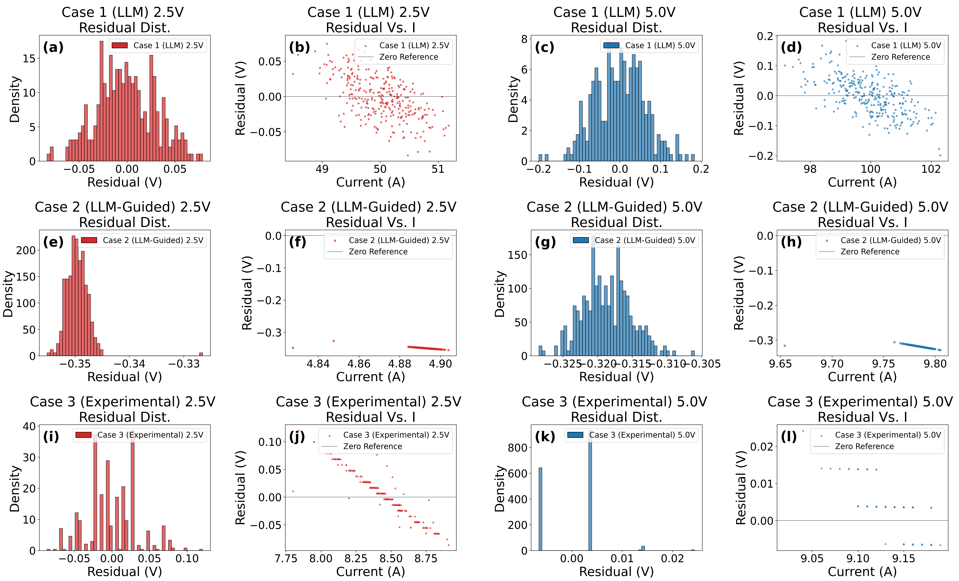}
		\caption{Diagnostics of the Modified Ohm's Law fit residuals. (The experimental residuals are well-behaved. Case 2 shows a deterministic, perfect correlation between residual magnitude and current ($\rho\approx0.999$), an impossible artifact revealing the LLM generated data from an internally consistent but physically false algorithmic relationship).}
		\label{fig:results7}
	\end{figure}

	Diagnostics on the Modified Ohm's Law fit residuals, shown in Figure~\ref{fig:results7}, expose the internal logic of the generation process. For the experimental data in Figure~\ref{fig:results7}(i)--(l), the residuals are approximately Gaussian (Shapiro-Wilk statistic unavailable due to large N but visually confirmed) and show no strong correlation with current magnitude (Spearman $\rho\approx-0.04,0.30$), indicating homoscedastic noise. The Case 1 residuals in Figure~\ref{fig:results7}(a)--(d) also show no heteroscedasticity ($\rho\approx-0.07,0.01$), but this is a trivial result of fitting a wildly inappropriate model to data generated from a different, simplistic linear assumption. The most telling diagnostic is for Case 2, in Figure~\ref{fig:results7}(e)--(h). The Spearman correlation between absolute residual and current is 0.999, with a $p$-value of zero. This near-perfect correlation is a glaring artifact. It occurs because the LLM's generation algorithm created a deterministic, nonlinear relationship between $v(t)$ and $i(t)$ that is different from the Modified Ohm's Law. When the correct physical model is fit to this incorrect relationship, the error (residual) is a deterministic function of the current. This pattern is impossible in real experimental data, where model misfit is obscured by random, uncorrelated measurement noise. The LLM has generated a specious, alternative ``physics'' that is entirely self-consistent within its generation algorithm but bears no relation to the true underlying law.
	
	\subsection{Inability to Capture Regime-Dependent Physics}
	\label{sec:regime_results}
	
\begin{figure}[htbp]
		\centering
		\includegraphics[width=0.9\textwidth]{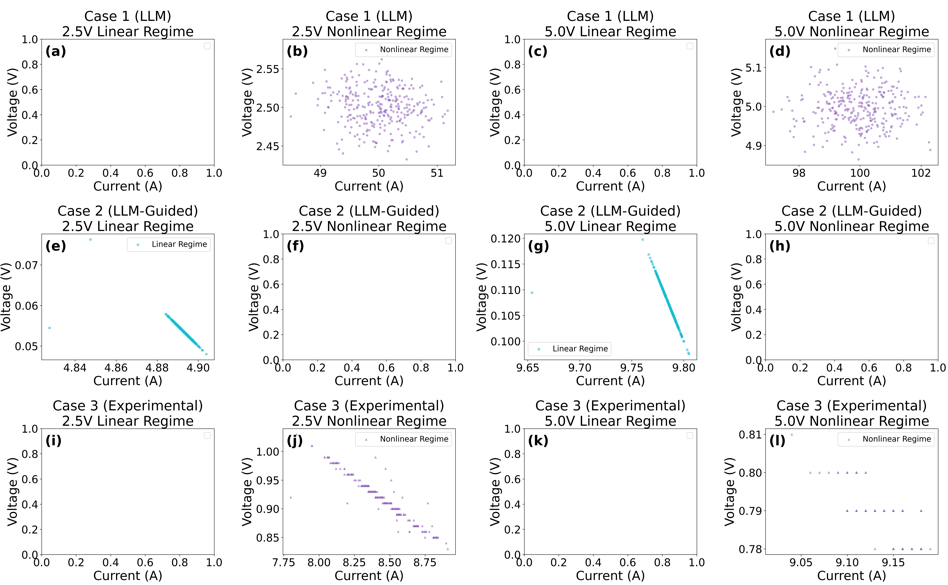}
		\caption{Scatter plots separating data into linear ($|v|<0.3$ V) and nonlinear ($|v|\ge0.3$ V) regimes. (Experimental data correctly resides in the nonlinear regime. Case 1 is erroneously in the nonlinear regime due to voltage magnitude errors. Case 2 incorrectly places all data in the linear regime with physically impossible negative correlations. This regime-dependent failure is predicted by the Novelty Horizon (Definition~\ref{def:HN}): the nonlinear regime lies outside $\conv(\Omega_{\text{train}})$ (Condition~\ref{cond:1}), and the correct clamping dynamics require causal reasoning not encoded in $C_{\text{train}}$ (Condition~\ref{cond:2})).}
		\label{fig:results8}
	\end{figure}

	However, as established in Sections~\ref{sec:regime}--\ref{sec:regime}, the primary observed behavior is linear resistive dissipation, with the diode network's role being primarily capacitive and protective. The regime analysis in Figure~\ref{fig:results8} confirms the LLMs' failure to replicate even this simplified reality. This failure is predicted by the Novelty Horizon framework (Definition~\ref{def:HN}): the nonlinear regime ($|v_s|\ge0.3$ V) lies outside $\conv(\Omega_{\text{train}})$ (Condition~\ref{cond:1}), as the specific conjunction of metallic short-circuit resistance and diode clamping dynamics is not representable as a convex combination of training examples. Moreover, the causal relationship between current and voltage in this regime involves semiconductor physics not encoded in $C_{\text{train}}$ (Condition~\ref{cond:2}). The experimental data in Figure~\ref{fig:results8}(i)--(l) correctly resides entirely in the nonlinear regime ($|v_s|\ge0.3$ V), as defined for analysis, though the dominant physics is the linear $i(t)R_{\text{sc}}$ term. The LLM-generated data reveals profound conceptual errors. Case 1 data, in Figure~\ref{fig:results8}(a)--(d), also falls entirely in the nonlinear regime, but for the wrong reason: its voltages are erroneously high (2.5 V, 5.0 V). Case 2, in stark contrast, places all data points in the linear regime ($|v_s|<0.3$ V), as shown in Figure~\ref{fig:results8}(e)--(h). Despite being given the explicit ``clamped diode'' experimental context, the LLM failed to generate voltages that meaningfully exceed the diode forward threshold. It generated a low-voltage short but did not understand the clamping dynamics well enough to simulate the correct voltage magnitude ($\approx0.7$--$1.0$ V) that characterises the experimental steady state. It could only interpolate to a generic ``low voltage'' value. Furthermore, the negative linear correlations ($r\approx-0.77,-0.71$) within its own linear regime are physically nonsensical for a passive resistive element, where voltage and current must be positively correlated. The direct-prompt generation (SI, Figure S8) confirms the regime-dependent nature of the failure. Critically, the guided direct-prompt generation places nearly 30\% of its data points in the linear regime ($|V|<0.3$ V), a physical impossibility for the shorted circuit. This demonstrates that providing procedural context allows the model to ``know about'' clamping, but not to understand it causally, leading it to treat the clamp as a conditional event rather than a fixed operational reality.
	
	\subsection{Quantitative Synthesis of Generative Fidelity}
	\label{sec:gfs_results}
	
\begin{figure}[htbp]
		\centering
		\includegraphics[width=0.9\textwidth]{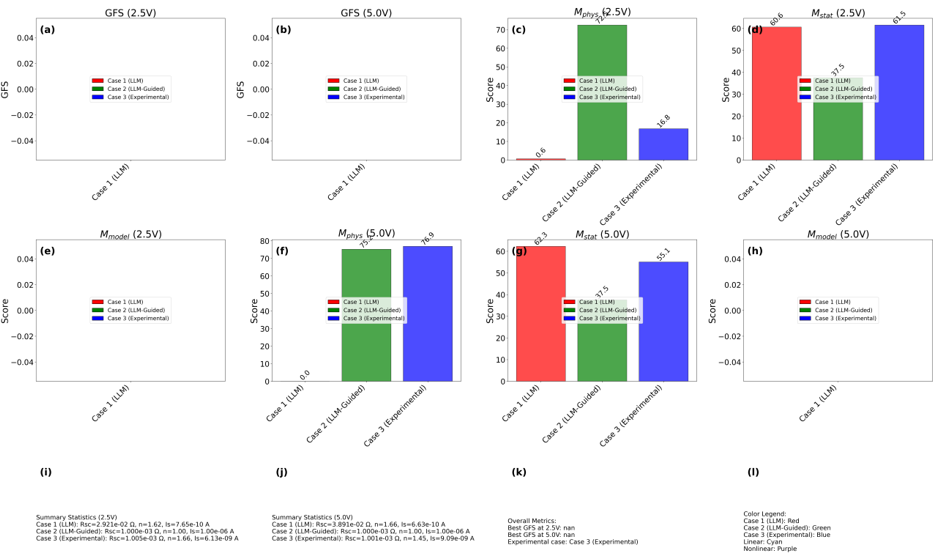}
		\caption{Generative Fidelity Scores (GFS) and component metrics. (Case 3 (Experimental) sets the benchmark. Case 1 fails utterly on physical constraints ($M_{\text{phys}}\approx0$). Case 2 shows moderate physical scaling but poor statistical properties and a complete failure in model fitting, resulting in an undefined overall GFS. Parameter summary tables highlight the physical implausibility of LLM-extracted parameters).}
		\label{fig:results9}
	\end{figure}

	Figure~\ref{fig:results9} synthesizes the multi-faceted analysis into the Generative Fidelity Score (GFS). The component metrics $M_{\text{phys}}$ (physical constraints), $M_{\text{stat}}$ (statistical properties), and $M_{\text{model}}$ (model fit) are computed per the framework in Section~\ref{sec:gfs}. The experimental Case 3 serves as the benchmark, though its GFS is depressed due to the mismatch between the full Modified Ohm's Law and the primarily resistive data, highlighting the sensitivity of the metric. The scores for the LLM cases are decisively lower, quantitatively confirming their failure. Case 1 scores catastrophically low on $M_{\text{phys}}$ ($0.0$--$0.6$), as seen in Figure~\ref{fig:results9}(b) and (f), due to its gross violation of energy scales and perfect, unphysical algebraic consistency. Its $M_{\text{stat}}$ scores are moderate ($60.6$--$62.3$), reflecting its use of simple Gaussian noise, which captures some superficial statistical properties. Case 2 shows higher $M_{\text{phys}}$ scores ($72.5$--$75.2$) in Figure~\ref{fig:results9}(c) and (g), as it respected basic scaling and energy magnitudes better, but its $M_{\text{stat}}$ scores are poor ($37.5$), penalized by its non-Gaussian, event-like distributions and poor spectral properties. Both cases have undefined $M_{\text{model}}$ and thus GFS due to the complete failure of the physical model fit, represented as missing bars in Figure~\ref{fig:results9}(a) and (b). The summary statistics in Figure~\ref{fig:results9}(i) and (j) crystallize the outcome. The experimental Case 3 yields consistent, physically credible parameter estimates across both voltage configurations: $R_{\text{sc}}\approx1.00\times10^{-3}\,\Omega$, $n\approx1.45$--$1.66$, $I_s\approx6\times10^{-9}$ to $9\times10^{-9}$ A. Case 2, despite being guided, hits the algorithmic bounds on all parameters, a clear sign of generative pathology. Case 1 estimates a resistance two orders of magnitude too large. The LLMs did not merely add noise to correct data; they generated data from an entirely incorrect manifold in voltage-current-power space, one that is statistically plausible in a textual sense but geometrically incompatible with the manifold defined by the laws of physics.
	
	\section{Discussion}
	\label{sec:discussion}
	
	\subsection{Interpreting the Failure of Statistical Correlation}
	
	The combined results from our script-based generation framework and the parallel direct-prompt framework (detailed in the supplementary information) demonstrate a fundamental, rather than incidental, failure of Large Language Models to extrapolate beyond their training data to generate scientifically valid empirical data for a novel experimental configuration. This failure is not a blanket statement on all data generation, but a specific and profound disconnect between statistical correlation and causal physical reality that emerges at the Novelty Horizon. The LLM-generated datasets, whether unguided (Case 1) or procedurally informed (Case 2), consistently violated inviolable physical constraints, such as the independent measurement noise between voltage and current and the correct scaling of energy dissipation. These violations manifest as perfect algebraic consistency in Ohmic residuals and the generation of deterministic, rather than stochastic, relationships between variables. The consistency of these failures across two fundamentally different generation paradigms-one mediated by code and one producing numbers directly-elevates our findings from an observation about a specific implementation to a generalizable principle about the architectural limits of autoregressive LLMs. This result directly stems from the LLM's architectural foundation as a next-token predictor, optimized for generating sequences that are statistically plausible within the context of its training corpus. Its ``knowledge'' of electrical circuits is a composite of textual descriptions, equations, and problem-solving templates it has encountered, not an internalized model of Kirchhoff's laws or semiconductor physics. Consequently, it treats physical laws as mathematical tautologies to be applied post-hoc rather than as governing principles that generate the observed data through a noisy measurement process. This aligns with critiques by \cite{ref46,ref47}, who argue that LLMs, devoid of grounded world models, are ultimately ``stochastic parrots'' reciting patterns without understanding. The experimental validation established in this paper provides concrete, quantitative evidence for this theoretical limitation within a specific engineering domain. This paper's methodological choice-having the LLM generate a simulator-precisely isolates its capacity for constructing causal models from text. A counterfactual question arises: if provided a physics engine API (e.g., a SPICE simulator) and asked to parameterize it, would performance improve? Likely, yes-for interpolation tasks. The LLM could probably successfully fill in parameters for a standard diode model if prompted. However, for phenomena at the Novelty Horizon, such as the specific transient interaction of a metallic short with a protective diode capacitance network, the correct model topology and governing equations themselves are the unknowns. An LLM, reliant on the physics engine's pre-defined equations, would be just as powerless to select or modify them appropriately as it was to code them from scratch. It might produce a numerically stable output, but one that is physically wrong, trading the glaring artifacts of a bad simulator for the insidious plausibility of a well-executed but misapplied one. The core limitation is not coding proficiency, but the lack of a judgment mechanism grounded in physical law.
	
	\subsection{The Past Information Boundary and the Novelty Horizon}
	
	To formalize the core limitation exposed by this work, we invoke the two complementary theoretical constructs introduced in Section~\ref{sec:bounds}: the $B_P$ (Definition~\ref{def:BP}) and the $H_N$ (Definition~\ref{def:HN}). These constructs define the operational domain of an autoregressive LLM trained on a static corpus and provide predictive criteria for determining whether a given scientific task is feasible. For a model $M$ trained on a dataset $\mathcal{D}=\{(\mathbf{x}_i,y_i)\}_{i=1}^N$ with associated timestamps $\tau_i$, the Past Information Boundary is the ordered triple $B_P(M) = (t_p,\Omega_{\text{train}},C_{\text{train}})$, where $t_p = \max_{i\in\{1,\dots,N\}} \tau_i$ is the temporal frontier (the latest timestamp in $\mathcal{D}$), $\Omega_{\text{train}} \subseteq \mathcal{X}\times\mathcal{Y}$ is the support of the empirical training distribution (the conceptual and relational frontier), and $C_{\text{train}}$ is the set of all conditional probability distributions encoded in the model's parameters (the statistical-correlation frontier). The model's knowledge is strictly confined to $B_P(M)$; it has no mechanism for incorporating information whose factual content, conceptual structure, or relational geometry lies outside $\Omega_{\text{train}}$ or whose temporal origin is $t>t_p$. A task $\mathcal{T}=(\mathcal{I},\mathcal{O},\Phi)$-defined by an input space $\mathcal{I}$, an output space $\mathcal{O}$, and governing laws $\Phi$-lies beyond the Novelty Horizon $H_N$ (i.e., $M$ cannot reliably perform $\mathcal{T}$) if at least one of three conditions holds. Condition~\ref{cond:1} (extrapolation outside $\conv(\Omega_{\text{train}})$) applies when the task requires generating an output $o^*\in\mathcal{O}$ such that, when projected onto the space of training examples via a mapping $\pi:\mathcal{O}\to\mathcal{X}\times\mathcal{Y}$, the point $\pi(o^*)$ falls outside the convex hull of the training support. Because $M$ is a statistical interpolator, it lacks any basis for generating such $o^*$. Condition~\ref{cond:2} (causal disentanglement not present in $C_{\text{train}}$) applies when the governing laws $\Phi$ require reasoning via causal mechanisms that are not reducible to the statistical correlations learned from text-specifically, if the conditional independencies and counterfactual dependencies required by $\Phi$ are not encoded in $C_{\text{train}}$. Condition~\ref{cond:3} (temporal extrapolation beyond $t_p$) applies when the task concerns events, theories, measurement outcomes, or data that first become available at a time $t>t_p$. Since the training data contains no information from the future, any output pertaining to such post-$t_p$ phenomena must be generated by extrapolation from past patterns; when the true post-$t_p$ distribution deviates from the historical continuation (e.g., a paradigm shift or a genuinely novel experimental configuration), the model's extrapolation is ungrounded. The diode-clamped short-circuit experiment described in Section~\ref{sec:framework} is engineered to reside precisely at - and partially beyond - the Novelty Horizon. Condition~\ref{cond:1} applies because the specific, time-resolved conjunction of nonlinear metallic resistance, semiconductor clamping dynamics, and authentic noise characteristics of the ACS712T sensor defines a point in experimental configuration space that lies outside $\conv(\Omega_{\text{train}})$. Condition~\ref{cond:2} applies because the correct causal structure requires independent voltage and current noise processes, whereas the LLM treats Ohm's Law as an algebraic tautology. Condition~\ref{cond:3} applies because the detailed noise spectrum and transient interaction of the short circuit with diode capacitance are not present in standard electrical engineering textbooks; the training data's temporal frontier $t_p$ predates the comprehensive characterisation reported in \cite{ref30}. The formal bounds also yield predictive principles that align with the observed empirical results. Smooth degradation within the horizon: as a task approaches the boundary of $\Omega_{\text{train}}$ but remains inside $\conv(\Omega_{\text{train}})$, performance degrades gracefully, with increasing but still correctable errors. Catastrophic collapse beyond the horizon: when the task demands extrapolation satisfying any of the three conditions, performance collapses discontinuously. This explains the regime-dependent failure documented in Section~\ref{sec:regime_results}: the linear regime ($|v_s|<0.3$ V) lies inside $H_N$ (interpolation of common Ohm's law correlations), whereas the nonlinear clamping regime ($|v_s|\ge0.3$ V) lies outside it (extrapolation requiring novel synthesis of diode physics and measurement noise). The revealed findings resonate with and extend the existing literature on LLM limitations in scientific contexts. Previous work has shown that LLM performance is heavily reliant on the density of relevant examples in the training set, and that true extrapolation beyond the training distribution remains elusive \cite{ref48,ref49}. The present results align with this view: the models successfully interpolated to simple linear resistive behavior (abundantly documented in training corpora) but failed catastrophically when required to extrapolate to the nonlinear clamping regime, whose specific interplay of metallic resistance and semiconductor dynamics had not been statistically encoded in their training data. Moreover, the validated experimental framework from \cite{ref45,ref30} provides a rigorous ground-truth benchmark that confirms the absence of such encoding. Crucially, our formalisation of $B_P$ and $H_N$ goes beyond prior descriptive accounts by offering predictive criteria that can be applied prospectively: any task that satisfies one or more of the three conditions is predicted to be infeasible for a pure autoregressive LLM, regardless of scaling or prompt engineering. This theoretical contribution sharpens the understanding of why LLMs fail where human scientists succeed, and it establishes a clear benchmark against which future claims of ``scientific AI'' must be evaluated.
	
	\subsection{Comparison with Prior Studies on AI in Scientific Research}
	
	The established failures expand upon a growing body of literature critically examining AI's role in science. Previous studies have shown LLMs can be useful for literature review, code generation, and drafting \cite{ref50,ref51}. However, their performance sharply declines when tasks require precise factual recall, complex multistep reasoning, or the generation of novel data \cite{ref52,ref53}. This work moves beyond these general assessments to perform a forensic analysis of why LLMs fail at empirical data generation. We confirm that providing detailed procedural context (Case 2) can steer outputs toward more plausible numerical scales, as also observed by \cite{ref54} in their study on prompt engineering for scientific tasks. Yet, this guidance is insufficient to impart true physical understanding. The guided model still produced pathological statistical distributions and a voltage-current relationship that was deterministically wrong, as evidenced by the near-perfect Spearman correlation ($\rho\approx0.999$) between fit residuals and current. This indicates the model created an internally consistent but physically false generative algorithm-a ``hallucinated'' physics. This aligns with the concerns established by \cite{ref55}, who differentiate between factually incorrect hallucinations and logically consistent but unfounded fabrications, the latter being particularly dangerous in scientific contexts as they can appear superficially credible. The present study's dual-framework design-comparing script-based to direct-prompt generation-yields unique insights that extend beyond the existing literature. It reveals that the ``perfect algebra artifact'' is not a bug of code generation but a symptom of the model's deeper conceptual failure to grasp measurement independence. The direct-prompt generation's catastrophic kurtosis (exceeding 200 in the unguided case), which was absent in the script-based version, shows that the script-based framework's structure-the explicit for loop, the defined time step $dt$, the sequential execution-inadvertently imposes a temporal stability and numerical regularity that masks the model's most severe statistical pathologies. When the model is freed from these structural constraints and asked to output numbers directly, its underlying token probability distribution is exposed, revealing a fundamental instability in generating long, coherent numerical sequences. This methodological insight is crucial for designing future AI evaluations: the mode of interaction fundamentally shapes the nature of the failure we observe. Evaluations that rely solely on code-generation tasks may underestimate the severity of a model's limitations, as the structure of the code itself can impose a semblance of coherence that is absent in direct numerical output.
	
	\subsection{The Regime-Dependent Nature of LLM Competence}
	
	A particularly illuminating result is the stark contrast in LLM performance between linear and nonlinear physical regimes. In the defined linear regime ($|v_s|<0.3$ V), where Ohm's law $v=iR$ dominates, one might expect an LLM to perform adequately, given the ubiquity of this relationship in its training data. The Case 2 model did indeed generate data that fell entirely within this linear voltage range. However, even here it failed, producing a negative linear correlation between voltage and current, a physical impossibility for a passive resistor. This reveals that the model's grasp of even basic linear relationships is superficial and can be corrupted when attempting to integrate additional contextual constraints (like ``clamped diode''). In the nonlinear regime, which characterized the actual experimental data, both LLM cases failed completely. Case 1 placed data here for the wrong reason (erroneously high voltages), while Case 2 failed to generate data that entered this regime at all. This regime-dependent collapse in fidelity provides strong empirical support for the theory that LLMs excel at interpolation within the convex hull of their training data distribution but lack the capacity for reliable extrapolation, especially when underlying principles change \cite{ref56,ref57}. The transition from linear to nonlinear behavior represents a shift in governing equations that the LLM cannot dynamically infer or apply; it can only mimic the textual description of such a transition.
	
	\subsection{The Epistemology of Measurement: An Unsolved Challenge for Generative AI}
	
	Beyond the failure to capture correct physics, the LLMs demonstrated a complete lack of a model for the measurement process itself. Authentic experimental data is a convolution of the underlying physical phenomenon and the artifacts of the sensing system-including bandwidth limitations, thermal noise, quantization error, and long-term drift. The autocorrelation functions and power spectral densities of the experimental data tell the story of a specific, bandwidth-limited analog sensor system. The LLM-generated data, in contrast, exhibited the statistical signatures of its internal algorithmic process: unnaturally fast ACF decay, spurious spectral peaks (Case 1), or simplistic, frequency-independent noise (Case 2). The models have no concept of a Hall-effect sensor's transfer function, the averaging of 10,000 ADC samples, or the $1/f$ noise characteristics of electronic components. This failing is profound because it separates calculation from measurement. An LLM can correctly calculate power as $v\times i$ (hence the machine-precision power residuals), but it cannot simulate the process of independently measuring $v$ and $i$ with real instruments. This underscores a critical epistemological point: scientific data is not merely a set of numbers obeying equations; it is the recorded output of a physical interaction with the world, imbued with the unique signature of the apparatus. Current LLMs, operating purely in the symbolic domain, are inherently incapable of replicating this aspect of scientific practice, a limitation also noted by \cite{ref58,ref59,ref60} in their analysis of AI for scientific discovery. The fact that this fundamental failing is observed whether the LLM is asked to simulate the experiment by writing code or to output the numbers directly demonstrates that the flaw is not in the model's ability to write Python, but in its inability to conceive of a measurement process. It treats the final numerical value as a point estimate to be calculated, rather than a sample from a distribution to be recorded-a distinction that lies at the heart of experimental science. The script-based generation imposes a procedural structure that partially conceals this conceptual gap, producing data that appears more stable and Gaussian. The direct-prompt generation, by removing that structure, exposes the raw instability of the model's numerical output. Together, these two frameworks reveal that the model's failure is not a matter of implementation, but of epistemology: it does not understand what it means to measure.
	
	\subsection{Implications for Researchers, Educators, and Industries}
	
	The implications of these findings are significant for all stakeholders in the scientific and technical ecosystem.
	\begin{itemize}
		\item \textbf{For researchers}, LLMs must be viewed as powerful, yet fundamentally limited, tools for ideation, drafting, and coding assistance. They cannot be trusted as oracles of factual knowledge or as simulators of physical reality. Our study provides a clear methodology-centered on testing for physical consistency, residual diagnostics, and regime analysis-that researchers can adopt to vet any AI-generated scientific content or data. The onus is on the human expert to provide the causal model and validate all outputs against it. This reinforces the argument of \cite{ref61,ref62} that AI should act as a ``collaborator'' under strict human supervision, not an autonomous agent.
		\item \textbf{For educators in engineering and physical sciences}, this work offers a critical pedagogical tool. Students must be taught to recognize the difference between arriving at an answer through the application of first principles and arriving at a statistically plausible answer. Using LLM failures as case studies, such as the generation of data with negative resistance correlations or perfect Ohmic consistency, can vividly illustrate core concepts of measurement uncertainty, experimental design, and the nature of physical laws. It underscores the irreplaceable value of hands-on laboratory work in developing true understanding.
		\item \textbf{For industry}, where AI is increasingly deployed for design, testing, and failure analysis, the risks are substantial. Relying on LLM-generated data or conclusions for safety-critical systems-such as electrical fault analysis, material stress testing, or chemical process design-could lead to catastrophic outcomes. Industries must implement rigorous, physics-based verification and validation pipelines for any AI-generated engineering output. The Generative Fidelity Score (GFS) framework proposed in this paper could serve as one component of such a validation suite.
		\item \textbf{For peer reviewers and journal editors}, the results sound a strong cautionary note. The surface plausibility of AI-generated text, figures, and even data tables means that synthetic content could become increasingly difficult to detect. Journals must strengthen policies requiring explicit disclosure of AI use in manuscript preparation and, crucially, in data generation. Reviewers must be encouraged to apply heightened skepticism and to scrutinize methodological sections for physical implausibilities, such as mismatches between described equipment and reported noise characteristics, or a lack of discussion of measurement uncertainty.
	\end{itemize}
	
	\subsection{The Structural Limitation and the Path to Augmented Intelligence}
	
	The central thesis confirmed by this experiment is structural, not incidental. The transformer architecture, trained via next-token prediction on a static corpus, is inherently retrospective. It is a mechanism for recombining and interpolating the past. Scientific progress, in its most valuable form, is prospective-it involves creating new information structures that did not exist before, whether through novel experiments, new theoretical syntheses, or the discovery of unforeseen phenomena. This is the essence of the Novelty Horizon. Therefore, the future of AI in science does not lie in expecting autoregressive LLMs to become autonomous discoverers. Instead, the path forward is Augmented Intelligence, where AI tools are deeply integrated into a human-driven scientific process. In this paradigm, LLMs can help survey literature, generate hypotheses for human evaluation, draft experimental protocols, and assist in data analysis-but the core acts of causal reasoning, experimental design, and grounding in physical reality remain human responsibilities. Hybrid neuro-symbolic systems that combine pattern recognition with formal, causal world models may eventually extend the Past Information Boundary, but they will likely remain tools for augmentation.
	
	\subsection{The Formal Bounds as Predictive Tools: Practical Applications of $B_P$ and $H_N$}
	
	The formalisation of $B_P$ and $H_N$ is not merely a post-hoc explanation of empirical failures; it provides predictive principles that can guide researchers, educators, and practitioners in assessing whether a given scientific task is feasible for an LLM without running exhaustive experiments.
	
	\textbf{Predicting Task Feasibility.} Given a candidate task $\mathcal{T}=(\mathcal{I},\mathcal{O},\Phi)$ and knowledge of a model's training corpus (or reasonable assumptions about its composition), one can assess:
	\begin{itemize}
		\item \textbf{Condition~\ref{cond:1} (Extrapolation):} Does the required output $o^*$ lie within $\conv(\Omega_{\text{train}})$? For continuous scientific variables (voltages, currents, temperatures, concentrations), this requires estimating whether the specific combination of parameter values has been observed in training. Tasks requiring parameter combinations outside the convex hull of training examples are predicted to fail catastrophically.
		\item \textbf{Condition~\ref{cond:2} (Causal Structure):} Are the governing laws $\Phi$ representable as statistical correlations in $C_{\text{train}}$, or do they require genuine causal reasoning? Tasks involving independent measurement noise, counterfactual interventions, or feedback loops are unlikely to be captured by next-token prediction alone. The LLM may produce outputs that satisfy algebraic relationships (e.g., $v=iR$) but will fail to reproduce the statistical structure of independent measurements (e.g., non-zero Ohm's law residuals).
		\item \textbf{Condition~\ref{cond:3} (Temporal Extrapolation):} Does the task require knowledge of phenomena discovered or characterised after the model's training cutoff $t_p$? If so, the model cannot possess ground-truth information and will resort to extrapolation from historical patterns-a process that fails when the post-$t_p$ distribution deviates from historical continuation.
	\end{itemize}
	
	\textbf{Why Fine-Tuning Does Not Cross the Horizon.} A common misconception is that fine-tuning an LLM on new experimental data will enable it to generate valid data for novel configurations. The formal bounds clarify why this is insufficient. Fine-tuning shifts $t_p$ (adding more recent data) and may expand $\Omega_{\text{train}}$ (adding new examples). However, it does not address Condition~\ref{cond:2}: the model remains a next-token predictor optimised for statistical correlation, not causal reasoning. Unless the fine-tuning data explicitly contains the correct causal structure (e.g., independent measurement noise, intervention data, counterfactual examples), the model will continue to treat physical laws as algebraic tautologies. Moreover, fine-tuning cannot create extrapolation capabilities beyond $\conv(\Omega_{\text{train}})$; it merely expands the convex hull.
	
	\textbf{Practical Heuristics for Researchers.} Based on the formal bounds, we propose the following heuristics:
	\begin{itemize}
		\item If a task requires generating numerical data for a configuration that is not a convex combination of training examples, do not trust LLM-generated outputs. Validate against first-principles models or ground-truth experiments.
		\item If a task involves causal relationships that are not reducible to statistical correlations (e.g., independent measurements, feedback, interventions), assume LLM failure. The `perfect algebra artifact' (Section~\ref{sec:physical_constraints_results}) is a reliable indicator of this failure mode.
		\item If a task concerns phenomena discovered after the model's training cutoff, treat LLM outputs as untested extrapolations. The model may produce plausible-sounding but factually incorrect outputs.
		\item To detect whether an LLM has operated beyond $H_N$, apply the forensic framework of Section~\ref{sec:methodology}: test for Ohmic residuals (Condition~\ref{cond:2}), regime-dependent performance (Condition~\ref{cond:1}), and spectral/temporal signatures (Condition~\ref{cond:3}).
	\end{itemize}
	
	\textbf{Relationship to Interpolation vs. Extrapolation in High-Dimensional Spaces.} The convex hull condition (Condition~\ref{cond:1}) connects the Novelty Horizon to the classical statistical distinction between interpolation and extrapolation. In the high-dimensional token embedding space of an LLM, $\conv(\Omega_{\text{train}})$ defines the region where the model can reasonably interpolate. Tasks requiring points outside this convex hull demand extrapolation-a capability that autoregressive models trained with next-token prediction do not possess. The diode-clamped short-circuit experiment, with its specific conjunction of parameters and dynamics, lies outside this convex hull, explaining the catastrophic failure observed in the nonlinear regime.
	
	\section{Conclusion}
	\label{sec:conclusion}
	
	This paper establishes and characterises a critical boundary for autoregressive large language models in empirical science: the Novelty Horizon $H_N$ (Definition~\ref{def:HN}). We have formally defined the Past Information Boundary $B_P=(t_p,\Omega_{\text{train}},C_{\text{train}})$ (Definition~\ref{def:BP}) and the Novelty Horizon as the set of tasks satisfying at least one of three conditions: extrapolation outside $\conv(\Omega_{\text{train}})$, causal structures absent from $C_{\text{train}}$, or reliance on post-$t_p$ information. The multi-layered forensic analysis demonstrates that while linguistic fluency and statistical pattern recognition can mimic scientific discourse, they cannot achieve the causal, grounded understanding required to extrapolate and generate valid data for tasks beyond $H_N$. The failure is comprehensive when the task lies beyond a model's training distribution, as defined by these formal bounds. Generated outputs violated inviolable physical constraints, exhibited pathological statistical signatures, and proved geometrically incompatible with first-principles constitutive laws. Explicit procedural guidance merely steered outputs toward more plausible numerical scales while failing to impart genuine physical understanding, resulting in data with extreme statistical artifacts and a Generative Fidelity Score of zero. These empirical failures are explained and formalized by the theoretical constructs of the Past Information Boundary and the Novelty Horizon. A large language model's knowledge is fundamentally an encapsulation of a static historical record, optimised for interpolation within that corpus. Generating valid data for a novel experiment, however, requires extrapolation beyond this boundary-a capability precluded by the model's fundamental architecture under the conditions tested. The diode-clamped short-circuit experiment, with its specific nonlinear dynamics and sensor noise characteristics, resided squarely beyond the Novelty Horizon of the tested models. The models manipulated symbols describing ``clamping'' but could not simulate the underlying physics, confusing textual correlation with causal mechanism. This conclusion is powerfully reinforced by the parallel investigation using direct-prompt generation, presented in the supplementary information. The consistency of the failures-the perfect algebra artifact, the regime misplacement, and the pathological statistical signatures across these distinct generative paradigms-leaves no room for alternative interpretations. Indeed, the direct-prompt framework revealed even more extreme manifestations of these failures, with unguided generation producing kurtosis values exceeding 200 and guided generation exhibiting parameter compensation that masked plausible resistance estimates behind severely distorted diode parameters. That these pathologies persist and, in some cases, intensify when the mediating structure of code is removed confirms that the Novelty Horizon is not a quirk of the script-based methodology but a fundamental boundary inherent to the architecture and training of autoregressive language models. A careful interpretation of the paper results suggests that current LLMs, under the specific conditions of tasks that require extrapolation outside $\conv(\Omega_{\text{train}})$, causal reasoning absent from $C_{\text{train}}$, or post-$t_p$ information, show strong limitations in generating valid novel scientific data. This statement avoids an absolute claim that LLMs can never generate novel data; rather, it acknowledges that for the class of tasks defined by the three conditions, the observed failures are systematic and severe, and overcoming them would require foundational modifications to the models themselves. Consequently, the appropriate role for these tools must be understood in relation to this horizon. As established by Definition~\ref{def:HN}, tasks strictly inside $H_N$ (interpolation within $\conv(\Omega_{\text{train}})$, routine synthesis using correlations in $C_{\text{train}}$, information from $t\le t_p$) can be safely delegated to LLMs for efficiency-literature summarisation, code completion, formatting, and initial hypothesis generation. Tasks at or beyond $H_N$ (novel experimental configurations, extrapolation to unseen regimes, discovery of new causal relationships) demand the causal reasoning, experimental design, and physical grounding of human scientists. This paper provides a methodological framework for the physics-based authentication of data and a philosophical clarification: true scientific discovery operates at the frontier of the Novelty Horizon. That frontier is a boundary language models cannot navigate alone, bound as they are to the archive of the past ($B_P$). The indispensable scientist, not the synthetic statistician, remains the sole author of discovery. Looking forward, crossing the Novelty Horizon will require not merely scaling existing architectures but developing domain-grounded, hybrid systems that integrate symbolic causal models with statistical pattern recognition. For the specific case of electrical short-circuit simulation-and more generally for any chaotic or strongly nonlinear physical system-a promising direction is the construction of background-oriented experimental simulators that embed first-principles knowledge directly into the generative loop. One concrete proposal is a neural-symbolic short-circuit engine that couples a pre-trained LLM with a library of validated component models (e.g., the bidirectional diode-clamp circuit, the Modified Ohm's Law derived in Section~\ref{sec:modified_ohm}, and noise profiles of Hall-effect sensors). Instead of asking the LLM to generate data from scratch, the system would use the LLM to parse a natural-language experimental description, translate it into a parameterised configuration for a symbolic solver (e.g., a SPICE-like engine specialised for fault analysis), and then return the solver's output as the ``generated'' data. Such an architecture would satisfy Condition~\ref{cond:1} by operating entirely within the solver's valid domain, and it would address Condition~\ref{cond:2} by explicitly encoding the causal structure of measurement independence. More ambitiously, future systems could incorporate online learning of previously unseen nonlinear regimes-for instance, by allowing the symbolic solver to be augmented with data-driven corrections when the LLM detects a mismatch between its static component library and the user's description of a genuinely novel configuration. This would require the model to possess metacognitive uncertainty estimation that triggers a ``request for human guidance'' or an automated search for new constitutive equations. Until such hybrid systems mature, the responsible path is to treat LLMs as powerful but bounded assistants, with human experts retaining ultimate responsibility for experimental design and physical interpretation. The Novelty Horizon is not a permanent barrier, but crossing it will demand a deliberate shift from pure autoregression to augmented, causally grounded intelligence.

	\appendix
	\section{Prompt for Unguided LLM Generation (Case 1 – Statistical Extrapolation)}
	\label{app:prompt1}
	
	This appendix contains the exact prompt used to generate the Case 1 dataset. It instructs a GPT-5.2 class model to synthesise short-circuit data given only basic nominal circuit parameters (2.5V/0.25A, 5.0V/0.5A), without providing the specific diode-clamp experimental procedure. This tests the model's intrinsic capacity for statistical extrapolation from fundamental electrical concepts, devoid of context regarding the novel measurement physics.
	
	\subsection*{Prompt}
	``Imagine that you are conducting an experiment in two different electrical circuits with power inputs of 2.5 V, 0.25 A, and 5.0 V and 0.5 A, using an Arduino Uno microcontroller and the ACS712 20 A current sensor on a breadboard. Then, you have an electrical short in each of these two circuits. For this problem, provide a rigorous Python code that prompts the user to input the external circuit power configurations and the duration of the short-circuit; then, following the user inputs, generate the short-circuit fault data for short voltage, short current (as if the short-circuit voltage and current were measured by the current sensor), computed short resistance, and short power. Also, let the code include a function for saving the generated data file with all these fault data features for download. Compute the resistance and the power correctly and suitably, scientifically. After this code, please provide a description of the circuit configuration/design that you have experimented on, and also provide a theoretical description of how the resistance and the power data are computed. In each case, I want to generate the fault data for 10 minutes with 10-millisecond resolution, and I want you to record this data at intervals of two seconds. Include the time column in the generated data files.''
	
	\subsection*{Commentary. The Linear Correlation Engine and Its Digital Artifacts}
	The code generated in response to this prompt typically implements a time-stepping simulator with a core physics model reduced to a single, static algebraic relationship. The LLM's foundational assumption, drawn from the statistical prevalence of textbook problems, is that a short-circuit is a linear resistor. Its simulator logic follows:
	\begin{itemize}
		\item Physics Kernel: It defines a nominal short-circuit resistance $R_{\text{sc}}$ and enforces a pristine, noise-free version of Ohm's Law at its core: $V_{\text{true}}(t)=I_{\text{true}}(t)\cdot R_{\text{sc}}$.
		\item Signal Decoration: It then decorates these perfect signals with independent Gaussian noise: $V_{\text{measured}}(t)=V_{\text{true}}(t)+\mathcal{N}(0,\sigma_V)$, $I_{\text{measured}}(t)=I_{\text{true}}(t)+\mathcal{N}(0,\sigma_I)$.
		\item Derived Quantity Calculation: Finally, it computes resistance and power from the noisy decorated signals as $R_{\text{calc}}(t)=V_{\text{measured}}(t)/I_{\text{measured}}(t)$ and $P_{\text{calc}}(t)=V_{\text{measured}}(t)\cdot I_{\text{measured}}(t)$.
	\end{itemize}
	The analogy: This is similar to painting rust onto a brand-new steel beam to simulate age. The underlying structure is perfect and linear; the complexity is only a surface layer. This architecture guarantees two unphysical data artifacts:
	\begin{itemize}
		\item \textbf{The Perfect Algebra Artifact:} Because $V_{\text{true}}$ and $I_{\text{true}}$ are perfectly coupled by $R_{\text{sc}}$, any deviation in the measured pair $(V_{\text{measured}},I_{\text{measured}})$ is purely from independent noise. Therefore, the derived quantity $R_{\text{calc}}(t)$ will have no mathematical relationship to the true $R_{\text{sc}}$ used in the core kernel. The data thus violates the internal consistency of a real experiment, where $R_{\text{calc}}(t)$ is estimated from the same two noisy measurements that are coupled by physics. This manifests as the infinitesimal Ohmic residual RMS (Section~\ref{sec:physical_constraints_results}), a fingerprint of a simulator that confuses a physical law with an algebraic tautology.
		\item \textbf{The Spectral Anachronism:} The use of independent, memoryless white noise generators fails to capture the temporal correlations (low-pass filtering) of real sensor systems and power supplies. This results in the anomalously fast-decaying autocorrelation functions (Figure~\ref{fig:results2}) and the flat, non-$1/f$ power spectral densities (Figure~\ref{fig:results3}). The data speaks in the frequency domain with the accent of a random number generator, not a physical sensor.
	\end{itemize}
	This unguided prompt forces the LLM to operate entirely within its Past Information Boundary $B_P=(t_p,\Omega_{\text{train}},C_{\text{train}})$. The model has no access to the specific diode-clamp experimental procedure; it must rely solely on statistical patterns in $C_{\text{train}}$. The task-generating time-resolved short-circuit data with authentic sensor noise and nonlinear clamping dynamics-requires extrapolation outside $\conv(\Omega_{\text{train}})$ (Condition~\ref{cond:1} of Definition~\ref{def:HN}) and causal reasoning about independent measurements not encoded in $C_{\text{train}}$ (Condition~\ref{cond:2}). The catastrophic failure observed in Section~\ref{sec:results} is therefore a direct empirical consequence of operating beyond the Novelty Horizon $H_N$.
	
	\section{Prompt for Procedurally Guided LLM Generation (Case 2 – Informed Pattern Matching)}
	\label{app:prompt2}
	
	This appendix contains the exact prompt used to generate the Case 2 dataset. The prompt explicitly provides the full experimental procedure for the ``Clamped Short Diodes Nominal'' configuration from reference \cite{ref30}. This tests the model's ability to integrate specific domain knowledge and procedural context into its generative process, moving from pure extrapolation toward guided simulation of the described experiment.
	
	\subsection*{Prompt}
	``Imagine that you are conducting an experiment in two different electrical circuits with power inputs of 2.5V, 0.25A, and 5.0V and 0.5A. Below, I have provided you with the uncommon diode clamp circuit being experimented on under all configurations, for your understanding and use. Then, you have an electrical short in each of these two circuits. For this setup, provide a rigorous Python code that prompts the user to input the external circuit power configurations and the duration of the short-circuit; then, following the user inputs, generate the short-circuit fault data for short voltage, short current, short resistance, and short power. Also, let the code save the generated data file, including all the data features, for download. Compute the resistance and the power correctly and suitably. After this code, also provide the circuit configuration/design that you have experimented on, and provide a theoretical description of how the resistance and the power data are computed. In each case, I want to generate the fault data for 10 minutes with 10-millisecond resolution, and I want you to record this data at intervals of two seconds. Include the time column in the generated data files. Consider the experiment description in the procedure provided in Table~\ref{tab:algorithm}.''
	
	\subsection*{Commentary. Pattern Overload and the Emergence of Algorithmic Pathology}
	When provided with the detailed experimental procedure, the LLM's generated code attempts to incorporate more complex patterns: diode clamps, averaging, and thresholding. However, it does so by stitching together conditional logic blocks rather than solving a unified physical model. Its simulator often adopts a structure like:
	\begin{itemize}
		\item Branching Logic: It calculates an ``unclamped'' voltage. If $|V_{\text{unclamped}}| > V_{\text{clamp}}$, it engages a ``clamped'' branch where voltage is fixed at $\pm V_{\text{clamp}}$ and current is recalculated as:
		\[
		\begin{cases}
			V(t)=I(t)R_{\text{short}}, & \text{if } |V|<V_{\text{clamp}} \\
			V(t)=\operatorname{sign}(V)\cdot V_{\text{clamp}},\; I(t)=V(t)/R_{\text{clamp,eq}}, & \text{otherwise}
		\end{cases}
		\]
		\item Averaging as Post-Processing: The instruction to average 10,000 samples is often implemented as a simple noise reduction on the final time-series, $\sigma_{\text{out}} = \sigma_{\text{in}}/\sqrt{10000}$, without modelling the underlying analog integration process.
	\end{itemize}
	The analogy: This is like a cook following a recipe by mechanically adding ingredients (patterns) without understanding how they react together. The result is a dish with isolated, overpowering flavours. In the data, this manifests as:
	\begin{itemize}
		\item \textbf{The Bimodal Catastrophe:} The hard-switching between linear and clamped branches, each with its own noise model, creates a mixture of two statistical populations in the current and voltage data. This is the direct origin of the extreme skewness and kurtosis ($\gamma_2>65$) observed in Section~\ref{sec:stat_distributions}. The probability density function becomes a Frankenstein's monster of two distributions, a signature impossible in a continuously measured physical system.
		\item \textbf{The Regime Misplacement:} The LLM's heuristic for the clamp threshold often misjudges the system's operating point. Consequently, as seen in Section~\ref{sec:regime_results}, the entire generated dataset can reside in the incorrect physical regime (e.g., all data points below 0.3V). The simulator has the concept of a diode clamp from its training data, but not the causal understanding of energy balance and circuit loading required to correctly predict its activation point. It inserts the clamp as a narrative element, not a physical component.
	\end{itemize}
	This guided prompt provides the model with detailed procedural context, effectively expanding the conceptual frontier $\Omega_{\text{train}}$ by including textual descriptions of the diode-clamp network. However, the model remains unable to cross the Novelty Horizon because the underlying causal structure-the Modified Ohm's Law (Equation~\eqref{eq:modified_ohm_time}) with independent measurement noise-is not representable as statistical correlations in $C_{\text{train}}$ (Condition~\ref{cond:2} of Definition~\ref{def:HN}). The model's attempt to simulate clamping via conditional branching logic (``if $|V|>V_{\text{clamp}}$'') produces the extreme kurtosis ($\gamma_2>65$) and regime misplacement observed in Sections~\ref{sec:stat_distributions} and~\ref{sec:regime_results}. This demonstrates that providing procedural context shifts the model's outputs within $B_P$ but does not enable crossing $H_N$-the fundamental limitation is architectural, not informational.
	
	\section{Experimental Framework for Case 3 -- Adoption of a Pre-Validated Physical Design}
	\label{app:exp_setup}
	
	\subsection{Provenance of the Experimental Configuration and Data}
\begin{figure}[htbp]
		\centering
		\includegraphics[width=0.8\textwidth]{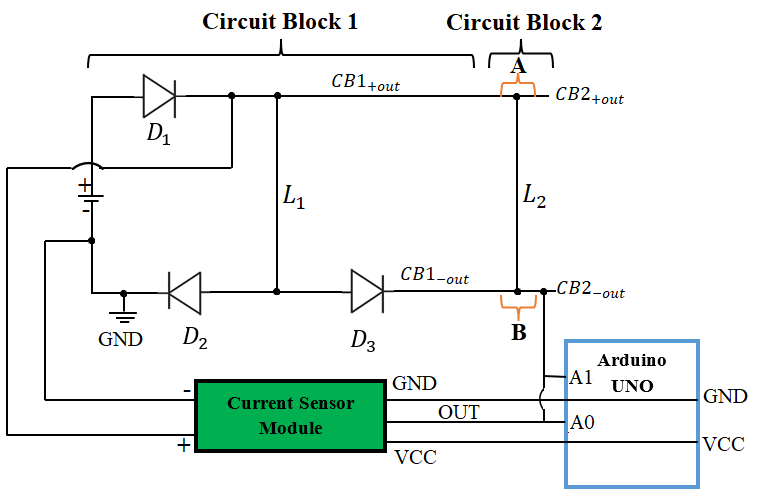}
		\caption{Experimental setup for ``Clamped Short Diodes Nominal'' measurements, adopted from \cite{ref30}. Key components: DC power supply, ACS712T current sensor, Arduino Uno, and clamped diode network ($D_1,D_2,D_3$). Nodes A and B define the short-circuit path during fault induction.}
		\label{fig:a3-1}
	\end{figure}

	Case 3, the physical ground truth dataset central to this paper's forensic analysis, is not the product of a newly calibrated apparatus. Instead, it directly adopts the fully characterized experimental framework, measurement data, and validation outcomes from the recent foundational study, \cite{ref30}. That work established a novel, governed short-circuit measurement system, culminating in a comprehensive, multi-layered sensor calibration and circuit validation framework. The present study leverages this rigorously validated foundation. The experimental design for Case 3 is precisely the ``Clamped Short Diodes Nominal'' configuration described in Section~\ref{sec:case2} and illustrated in Figure~1 of \cite{ref30}. This configuration, reproduced for clarity in Figure~\ref{fig:a3-1}, integrates a three-diode (1N5408) clamping network (``Circuit Block 1'' or CB1), an ACS712T-20A Hall-effect current sensor, an Arduino Uno microcontroller, and a programmable DC power supply. The core function of this architecture is to enforce a bounded electrical regime during a short-circuit event, enabling sustained, high-fidelity measurement of voltage and current without system destruction.
	
	The data acquisition protocol follows the automated algorithm detailed in Table~2 of \cite{ref30}. For the specific data used in this paper (10-minute duration, 10-ms base sampling), the Arduino firmware was configured to sample the analog voltage and current channels at 10-ms intervals, with each reported data point representing the average of 10,000 instantaneous samples to suppress quantization noise and aliasing, as per the Nyquist-Shannon criterion \cite{ref30}.
	
	\subsection{Adoption of Pre-Validated Data Without Recalibration}
	A critical distinction of this work is that it does not involve recalibrating the sensor system or revalidating the circuit's clamping principle. That extensive work was conclusively performed and reported in \cite{ref30}. The sensor calibration framework established in \cite{ref30}-encompassing baseline validation, post-fault performance assessment under both nominal and extreme short-circuit conditions, and the derivation of the Consolidated Calibration Confidence Score (CCCS)-confirmed the measurement system's integrity and resilience. The data utilized for Case 3 in this paper is drawn directly from the results of that validated experimental campaign, specifically the extended-duration (10-minute) tests under nominal short-circuit conditions at 2.5V and 5.0V inputs. Therefore, the voltage ($v_s(t)$) and current ($i(t)$) time-series for Case 3 carry with them the full weight of the calibration assurance demonstrated in \cite{ref30}. Their statistical noise profiles, temporal stability, and adherence to fundamental circuit laws have already been authenticated against first principles and sensor specifications.
	
	\subsection{Role of Case 3 Data in the Present Study}
	In the context of this paper's investigation into LLM capabilities, the pre-validated Case 3 data serves a singular, crucial purpose: to act as an immutable physical benchmark. It represents reality as measured by a trusted instrument whose errors and characteristics are known and bounded. The multi-layered analytical framework developed in Section~\ref{sec:methodology} of this paper-testing temporal statistics, spectral signatures, physical consistency, and fit to the Modified Ohm's Law-uses this benchmark not to question the data's validity, but to expose the deviations of the synthetically generated Cases 1 and 2 from physical plausibility. The governing physics for the system, derived in Section~\ref{sec:modified_ohm} of this manuscript, provides the constitutive law against which all datasets are judged. The experimental data from \cite{ref30} naturally conforms to this model because the model describes the physical reality that the apparatus in \cite{ref30} was designed to measure. Thus, the failure of the LLM-generated data to fit this model or replicate the statistical fingerprints of the real measurement process (e.g., correct autocorrelation, $1/f$ noise spectra) is not a commentary on the experimental data's quality, but a direct measure of the LLMs' inability to generate outputs that reside on the physically-valid manifold defined by both the constitutive law and the authentic measurement process.
	
	\section{Epistemological Foundations of Physics-Based Authentication - Why Statistical Similarity Cannot Certify Causal Fidelity}
	\label{app:epistemology}
	
	This appendix develops the theoretical critique that motivates the forensic methodology of the main paper. It is placed here, rather than in the introduction, to allow the empirical argument to proceed without interruption while providing a self-contained reference for readers interested in the philosophical and statistical foundations of the authentication framework.
	
	\subsection{The Reference Class Problem in LLM-Generated Scientific Data}
	
	The central epistemological difficulty in evaluating LLM-generated scientific data can be formulated as a reference class problem. Let $\mathcal{D}_{\text{real}}$ denote the set of all empirically measured datasets from a physical system governed by laws $\Phi$, and let $\mathcal{D}_{\text{synth}}$ denote a dataset generated by an LLM. A researcher assessing $\mathcal{D}_{\text{synth}}$ has access only to its statistical properties-its marginal distributions, correlation structure, and spectral characteristics-computed relative to some reference class $\mathcal{R}$ of real datasets. The implicit inference is:
	\[
	\text{If } \operatorname{Stat}(\mathcal{D}_{\text{synth}}) \approx \operatorname{Stat}(\mathcal{D}_{\text{real}}) \text{ for } \mathcal{D}_{\text{real}} \in \mathcal{R}, \text{ then } \mathcal{D}_{\text{synth}} \text{ is empirically valid.}
	\]
	This inference commits what may be termed the reference class fallacy: the statistical properties that define $\mathcal{R}$ are aggregates over many possible generative processes, only a subset of which respect the causal structure of $\Phi$. A dataset may satisfy $\operatorname{Stat}(\mathcal{D}) \approx \operatorname{Stat}(\mathcal{D}_{\text{real}})$ while violating $\Phi$ in fundamental ways, provided the violations are not captured by the chosen statistical tests. The diode-clamped short-circuit experiment provides a concrete instantiation of this fallacy. Consider the following statistical properties that a naive validation might examine:
	\begin{itemize}
		\item Mean current $\bar{I}$ and mean voltage $\bar{V}$
		\item Pearson correlation $\rho_{VI}$ between $V$ and $I$
		\item Signal-to-noise ratio and variances $\sigma_I^2$, $\sigma_V^2$
		\item Power spectral density slope at low frequencies
	\end{itemize}
	An LLM could be trained-or explicitly prompted-to match these aggregate statistics. Indeed, the unguided generation (Case 1) successfully produced near-Gaussian noise and a high positive correlation between $V$ and $I$, properties that superficially resemble the experimental data. Yet the forensic analysis in Sections~\ref{sec:physical_constraints_results}--\ref{sec:constitutive_failure} revealed catastrophic failures: the Ohmic residuals were $10^{-17}\,\Omega$ rather than $10^{-3}\,\Omega$, and the fitted Modified Ohm's Law yielded $R^2\approx-26000$. The dataset was statistically plausible at the level of marginal moments but physically impossible at the level of causal structure. This exposes a general principle: statistical similarity is necessary but not sufficient for empirical validity. The sufficiency direction fails because the space of possible datasets consistent with a given set of statistical constraints is vast, and only a vanishingly small fraction of that space corresponds to trajectories that could arise from a physical measurement process governed by $\Phi$.
	
	\subsection{The Measurement Process as a Hidden Variable}
	
	A deeper analysis reveals that the reference class fallacy originates in the omission of what philosophers of science call the measurement process as a latent variable. Let $\mathcal{M}$ denote the measurement apparatus-sensors, amplifiers, analog-to-digital converters, and their associated noise characteristics, bandwidth limitations, and calibration drifts. The observed data $\mathcal{D}_{\text{real}}$ is not a direct recording of the underlying physical state $S(t)$ but rather a convolution: $\mathcal{D}_{\text{real}}(t) = \mathcal{M}[S(t)] + \eta(t)$, where $\eta(t)$ represents irreducible measurement noise with spectral properties determined by $\mathcal{M}$. The causal graph is: $S(t) \to \mathcal{M}[S(t)] \to \mathcal{D}_{\text{real}}(t)$. The measurement apparatus $\mathcal{M}$ is itself a physical system with its own dynamics, including thermal noise (Johnson-Nyquist), shot noise, $1/f$ flicker noise, quantization error, and finite bandwidth. These properties are not arbitrary; they follow from fundamental physics. For a Hall-effect sensor such as the ACS712T, the noise power spectral density has the form: $S_W(f) = \frac{4k_B T R}{1+(2\pi f\tau)^2} + \frac{K}{f^\alpha} + \frac{\Delta^2}{12f_s}$, where the first term represents thermal noise filtered by the sensor bandwidth $1/(2\pi\tau)$, the second term represents $1/f$ noise (with $\alpha\approx1$), and the third term represents quantization noise from the ADC with step size $\Delta$ and sampling frequency $f_s$. An LLM trained on text lacks any representation of $\mathcal{M}$. It has never experienced the physical act of measurement-the settling time of an ADC, the random walk of thermal noise, the saturation of a sensor amplifier. When asked to generate ``experimental data,'' the LLM defaults to the only generative model it possesses: the statistical patterns in its training corpus. But those patterns are predominantly textual descriptions of measurement outcomes, not the outcomes themselves. The model learns that experimental papers report means, standard deviations, and correlations; it does not learn the stochastic process $\mathcal{M}$ that generates those summaries from underlying physical states. Consequently, LLM-generated data exhibits what may be termed the measurement-blindness artifact: it treats the reported values as primitives rather than as samples from a measurement distribution. This manifests as:
	\begin{itemize}
		\item Perfect algebraic closure (Section~\ref{sec:physical_constraints_results}). Because the model computes derived quantities like $R=V/I$ from the same internally consistent numbers, rather than from independently noisy measurements, the residual $V-IR$ is zero to machine precision-a statistical impossibility for any real measurement system.
		\item Incorrect noise spectra (Section~\ref{sec:spectral}). Lacking a model of sensor bandwidth and $1/f$ noise, the LLM defaults to white noise or simple autoregressive processes, producing autocorrelation functions that decay within 2--3 lags rather than the gradual decay characteristic of band-limited analog signals.
		\item Regime misplacement (Section~\ref{sec:regime_results}). Without a causal model of the diode clamping threshold, the LLM treats the nonlinear regime as a conditional branch rather than a continuous physical transition, resulting in data that either never enters the clamping regime (Case 2) or enters it for the wrong reasons (Case 1).
	\end{itemize}
	
	\subsection{The Necessity of Forensic Physics-Based Authentication}
	
	The preceding analysis implies that validating LLM-generated scientific data requires moving beyond statistical similarity to forensic physics-based authentication-a methodology that tests for consistency with the causal structure of $\Phi$ and the measurement process $\mathcal{M}$. The main paper develops precisely such a methodology, organized around three discriminative tests:
	
	\textbf{Test 1: Algebraic Independence.} For a system governed by $\Phi: v=iR$ (in the linear regime), real measurements produce residuals $\epsilon_R = v/i - R_{\text{calc}}$ that are non-zero and distributed according to the propagation of independent measurement errors. Specifically, if $v=v_{\text{true}}+\eta_v$ and $i=i_{\text{true}}+\eta_i$ with $\eta_v,\eta_i$ independent, then $\epsilon_R = (v_{\text{true}}+\eta_v)/(i_{\text{true}}+\eta_i) - R_{\text{calc}} \neq 0$, and the distribution of $\epsilon_R$ has variance determined by the sensor noise levels. LLM-generated data that yields $\epsilon_R \approx 0$ to machine precision (as in Section~\ref{sec:physical_constraints_results}) fails this test, revealing that $v$ and $i$ were not generated as independent measurements.
	
	\textbf{Test 2: Spectral Authenticity.} For a measurement system with bandwidth-limited sensor and $1/f$ noise, the power spectral density $S(f)$ must satisfy $\lim_{f\to0} S(f) \propto 1/f^\alpha$, $\alpha\approx1$, and the autocorrelation function $\rho(\tau)$ must decay slowly over lags $\tau$ comparable to the sensor time constant. LLM-generated data that produces white noise spectra (flat $S(f)$) or ACFs that drop to zero within a few lags (as in Section~\ref{sec:spectral}) fails this test, revealing that the generation process lacks a model of $\mathcal{M}$.
	
	\textbf{Test 3: Regime Continuity.} For a system with a threshold phenomenon (e.g., diode clamping), the transition between linear and nonlinear regimes must be continuous in both value and derivative. Let $f(V,I)$ be the function describing the system's current-voltage characteristic. Physical continuity requires $\lim_{V\to V_{\text{th}}^-} f(V) = \lim_{V\to V_{\text{th}}^+} f(V)$ and $\lim_{V\to V_{\text{th}}^-} df/dV = \lim_{V\to V_{\text{th}}^+} df/dV$. LLM-generated data that exhibits a hard threshold (Case 2's conditional branching) or that places all data in the wrong regime (Section~\ref{sec:regime_results}) fails this test, revealing that the model has learned the concept of a threshold from text but not the physics of continuous transition.
	
	\subsection{The Generative Fidelity Score as an Epistemological Instrument}
	
	The Generative Fidelity Score (GFS) introduced in Section~\ref{sec:gfs} can be understood as an epistemological instrument designed to quantify the distance between a candidate dataset $\mathcal{D}_{\text{synth}}$ and the manifold $\mathcal{M}_{\Phi}$ of datasets that could arise from the causal process $(S,\mathcal{M})$. The GFS synthesizes multiple discriminative tests into a single metric:
	\[
	\text{GFS} = w_1 M_{\text{phys}} + w_2 M_{\text{stat}} + w_3 M_{\text{model}}
	\]
	where $M_{\text{phys}}$ tests for violations of inviolable constraints (Test 1), $M_{\text{stat}}$ tests for spectral authenticity (Test 2), and $M_{\text{model}}$ tests for consistency with the first-principles constitutive law (Test 3). The catastrophic GFS values obtained for both LLM-generated cases (effectively zero, as the model fit component was undefined) provide quantitative confirmation that these datasets lie far from $\mathcal{M}_{\Phi}$-not merely in terms of statistical divergence, but in terms of physical impossibility. Critically, the GFS framework is theory-dependent: it requires a first-principles model $\Phi$ of the system under study. This is not a weakness but an essential feature. There is no model-free method for authenticating scientific data because validity is always relative to a theoretical framework. The Modified Ohm's Law derived in Section~\ref{sec:modified_ohm} provides that framework for the diode-clamped short-circuit system. For other domains, analogous first-principles models must be developed. The implication is that physics-based authentication is domain-specific and theory-laden-a fact that LLM proponents who advocate for generic ``AI validation'' often overlook.
	
	\subsection{Broader Implications for AI-Supported Science}
	
	The epistemological critique developed here has implications beyond the specific case of short-circuit measurement. It suggests a general principle: LLMs are fundamentally unsuited for generating empirical data in domains where the causal structure of the measurement process is not fully captured by statistical patterns in text. This includes virtually all of experimental physics, chemistry, biology, and engineering, where the measurement apparatus $\mathcal{M}$ is a physical system with dynamics that are rarely described in sufficient detail in training corpora. This is not to say that LLMs have no role in science. They remain powerful tools for literature synthesis, code assistance, and hypothesis generation-tasks where the output remains in the discursive domain. But the frontier of LLM use has already extended beyond this domain into empirical data generation, as evidenced by studies such as Barr et al. \cite{ref28} on synthetic clinical data. The present paper provides the methodological toolkit for authenticating such data and demonstrates, through rigorous empirical test, that current LLMs fail catastrophically when pushed beyond their Novelty Horizon. The way forward is not to ban LLMs from science but to recognize their fundamental limitation: they are retrospective statistical interpolators, not prospective causal engines. The Past Information Boundary $B_P=(t_p,\Omega_{\text{train}},C_{\text{train}})$ is not a bug to be fixed by scaling or prompt engineering; it is a structural consequence of the autoregressive next-token prediction objective. Tasks that lie beyond the Novelty Horizon $H_N$-requiring extrapolation outside $\conv(\Omega_{\text{train}})$, causal reasoning not encoded in $C_{\text{train}}$, or information post-dating $t_p$-will remain beyond the reach of pure LLMs regardless of model size or training data volume. Crossing $H_N$ will require fundamental architectural changes, including the incorporation of causal world models, grounding in physical simulators, and explicit representation of measurement processes-changes that would transform these systems into something other than LLMs as currently conceived. Until such hybrid systems mature, the responsible path is to treat LLM-generated numerical data as unvalidated conjecture, subject to forensic physics-based authentication before being admitted as evidence in scientific reasoning. The methodology developed in this paper provides a template for such authentication across domains, and the theoretical framework of $B_P$ and $H_N$ provides a predictive criterion for when such authentication is essential: whenever the task requires extrapolation beyond the convex hull of training examples, causal reasoning not reducible to statistical correlation, or knowledge of post-training phenomena.
	
	
\end{document}